\DocumentMetadata{uncompress} %% for newpax
\documentclass[
  10pt,
  longbibliography,
  twocolumn,
  twoside,
]{article}

\usepackage[numbers,square,comma,sort&compress]{natbib}
\usepackage[
  left=0.65in,
  right=0.65in,
  top=0.65in,
  bottom=0.65in,
]{geometry}

\usepackage{sectsty}
\sectionfont{\raggedright\large\bfseries}
\subsectionfont{\raggedright\large}

\usepackage[font=small,labelfont=bf]{caption}

\usepackage{fancyhdr}
\usepackage{amssymb}
\usepackage{amsmath}
\usepackage{newunicodechar}
\usepackage{amsfonts}  % Required for \mathbb
\usepackage{array}
\usepackage{nameref}    % Load before hyperref
\usepackage{hyperref}   % Must be loaded after nameref
\usepackage{placeins}

\usepackage{environ}

\usepackage{url}
\PassOptionsToPackage{hyphens}{url} %\usepackage{hyperref}
\makeatletter
\g@addto@macro{\UrlBreaks}{\UrlOrds}
\makeatother

\usepackage[table]{xcolor}   % 'table' option enables row/column coloring
\usepackage{colortbl}        % for coloring table cells

\definecolor{goodblue}{RGB}{0, 91, 187}
\hypersetup{
  colorlinks=true,
  allcolors=goodblue,
  urlcolor=goodblue,
  citecolor=goodblue,
  pdfborder={0 0 0},
  breaklinks=true,
}

\usepackage[normalem]{ulem}

\usepackage{textcomp}

\usepackage[export]{adjustbox}

\makeatletter

\def\CT@@do@color{%
  \global\let\CT@do@color\relax
  \@tempdima\wd\z@
  \advance\@tempdima\@tempdimb
  \advance\@tempdima\@tempdimc
  \advance\@tempdimb\tabcolsep
  \advance\@tempdimc\tabcolsep
  \advance\@tempdima2\tabcolsep
  \kern-\@tempdimb
  \leaders\vrule
  \hskip\@tempdima\@plus  1fill
  \kern-\@tempdimc
  \hskip-\wd\z@ \@plus -1fill }
\makeatother

\newcommand{\done}[1]{}

\usepackage{titlesec}
\titleformat*{\paragraph}{\bfseries}

\titlespacing*{\section}{0pt}{3ex}{0.3ex}
\titlespacing*{\subsection}{0pt}{1.5ex}{0.3ex}
\titlespacing*{\subsubsection}{0pt}{1.5ex}{0.3ex}

\usepackage{changepage}
\usepackage{tocloft}  % tableofcontents

\usepackage{graphicx}
\usepackage{epsfig}
\usepackage{verbatim}
\usepackage{enumerate}
\usepackage{enumitem}
\usepackage{ifthen}

\usepackage{longtable}

\usepackage{mathtools}

\usepackage{tabularx}
\newboolean{twocolswitch}

\newcommand{\PreserveBackslash}[1]{\let\temp=\\#1\let\\=\temp}

\newcommand{\sindex}[1]{}
\newcommand{\nindex}[1]{}

\newcommand{\www}[1]{\url{#1}}

\usepackage{lettrine}

\usepackage{changepage}

\usepackage{footmisc}

\usepackage{refcount}

\makeatletter
\newcommand{\footnotemarklabel}[1]{%
  \protect\footnotemark
  \begingroup
    \edef\@currentlabel{\thefootnote}% make the label refer to the footnote number
    \label{#1}%
  \endgroup
}
\makeatother

\renewcommand{\footnoterule}{%
  \kern 3pt
  \hrule width 0.4\columnwidth height 0.4pt
  \kern 6pt
}
\NewEnviron{excerpt}{
    \begin{quote}
      \medskip
      \BODY
      \medskip
    \end{quote}
}

\NewEnviron{editnote}{
    \begin{quote}
      \color{rose}
      $\blacksquare$
      \BODY
      \medskip
    \end{quote}
}

\newcommand{\command}[1]{
  \lstinline[language={[LaTeX]TeX},basicstyle=\ttfamily]{#1}
}

\newcommand{\editbox}[2]{
}

\newcommand{\editboxwithlatex}[2]{
}

\usepackage{fancyvrb}

\usepackage{tcolorbox}
\tcbuselibrary{skins,breakable,listings,breakable}

\usepackage{tikz}
\usetikzlibrary{shapes}

\tikzstyle{mybox} = [draw=lightblue!70, fill=lightblue!7, very thick,
    rectangle, rounded corners, inner sep=10pt, inner ysep=20pt]

\tikzstyle{editortitle} =[draw=archetyperowcoloralt, fill=archetyperowcoloralt, text=black]

\newcommand\Loadedframemethod{default}
\usepackage[framemethod=\Loadedframemethod]{mdframed}

\mdfsetup{skipabove=\topskip,skipbelow=\topskip}

\tikzstyle{loglinetitle} =[draw=icedark, fill=icemedium!50, text=black]

\newenvironment{loglinebox}[1][]{

  \ifstrempty{#1}%
  {\mdfsetup{%
    frametitle={%
       \tikz[baseline=(current bounding box.east),outer sep=0pt]
        \node[loglinetitle, anchor=east,rectangle]
        {\strut~~#1~~\strut};}}
  }%
  {\mdfsetup{%
     frametitle={%
       \tikz[baseline=(current bounding box.east),outer sep=0pt]
        \node[loglinetitle,anchor=east,rectangle]
        {\strut~~#1~~\strut};}}%
   }%
   \mdfsetup{innertopmargin=5pt,linecolor=icedark,%
             linewidth=0.5pt,topline=true,
             frametitleaboveskip=\dimexpr-\ht\strutbox\relax,}
   \begin{mdframed}[backgroundcolor=icelight,nobreak=true]\relax%
     \raggedright
}{\end{mdframed}}

\tikzstyle{abstracttitle} =[draw=magmadark!75, fill=magmamedium!75, text=black]

\newenvironment{abstractbox}[1][]{

  \ifstrempty{#1}%
  {\mdfsetup{%
    frametitle={%
       \tikz[baseline=(current bounding box.east),outer sep=0pt]
        \node[abstracttitle, anchor=east,rectangle]
        {\strut~~#1~~\strut};}}
  }%
  {\mdfsetup{%
     frametitle={%
       \tikz[baseline=(current bounding box.east),outer sep=0pt]
        \node[abstracttitle,anchor=east,rectangle]
        {\strut~~#1~~\strut};}}%
   }%
   \mdfsetup{innertopmargin=5pt,linecolor=magmadark,%
             linewidth=0.5pt,topline=true,
             frametitleaboveskip=\dimexpr-\ht\strutbox\relax,}
   \begin{mdframed}[backgroundcolor=magmalight,nobreak=true]\relax%
     \raggedright
}{\end{mdframed}}

\tikzstyle{infotitle} =[draw=darkgrey, fill=lightgrey!50, text=black]

\newenvironment{infobox}[1][]{

  \ifstrempty{#1}%
  {\mdfsetup{%
    frametitle={%
       \tikz[baseline=(current bounding box.east),outer sep=0pt]
        \node[infotitle, anchor=east,rectangle]
        {\strut~~#1~~\strut};}}
  }%
  {\mdfsetup{%
     frametitle={%
       \tikz[baseline=(current bounding box.east),outer sep=0pt]
        \node[infotitle,anchor=east,rectangle]
        {\strut~~#1~~\strut};}}%
   }%
   \mdfsetup{innertopmargin=5pt,linecolor=grey,%
             linewidth=0.5pt,topline=true,
             frametitleaboveskip=\dimexpr-\ht\strutbox\relax,}
   \begin{mdframed}[backgroundcolor=lightgrey!25,nobreak=true]\relax%
     \raggedright
}{\end{mdframed}}

\tikzstyle{essencetitle} = [draw=magmadark!75, fill=magmamedium!75, text=black]

\usepackage{enumitem}

\tikzstyle{changelogtitle} =[draw=darkgrey, fill=lightgrey!50, text=black]

\usepackage[table]{xcolor}

\definecolor{olivegreen}{rgb}{0.33333,.41961,0.18431}
\definecolor{forestgreen}{rgb}{0.13333,.5451,0.13333}

\definecolor{lightgrey}{rgb}{0.7,0.7,0.7}
\definecolor{verylightgrey}{rgb}{0.90,0.90,0.90}
\definecolor{veryverylightgrey}{rgb}{0.95,0.95,0.95}
\definecolor{grey}{rgb}{0.5,0.5,0.5}
\definecolor{darkgrey}{rgb}{0.3,0.3,0.3}
\definecolor{verydarkgrey}{rgb}{0.15,0.15,0.15}

\definecolor{headerblue}{HTML}{33367E}
\definecolor{unitednationsblue}{HTML}{4D88FF}

\definecolor{charcoal}{HTML}{36454F}
\definecolor{cinerous}{HTML}{98817B}
\definecolor{feldgrau}{HTML}{4D5D53}
\definecolor{glaucous}{HTML}{6082B6}
\definecolor{arsenic}{HTML}{3B444B}
\definecolor{xanadu}{HTML}{738678}

\definecolor{firebrick}{HTML}{B22222}
\definecolor{orangered}{HTML}{FF4500}
\definecolor{tomato}{HTML}{FF6347}

\definecolor{orange}{RGB}{255,116,0}

\definecolor{purpletaupe}{HTML}{3B444B}
\definecolor{rose}{HTML}{E3242B}

\colorlet{editnotecolor}{rose}

\definecolor{headerorange}{RGB}{255,116,0}
\definecolor{headergray}{RGB}{230,230,230}

\definecolor{headerpop}{RGB}{230,230,230}

\definecolor{magmalight}{RGB}{252,251,195}
\definecolor{magmalightalt}{RGB}{250,240,184}
\definecolor{magmamedium}{RGB}{245,200,146}
\definecolor{magmadark}{RGB}{224,106,98}

\definecolor{icelight}{RGB}{223,242,244}
\definecolor{icelightalt}{RGB}{189,222,226}
\definecolor{icemedium}{RGB}{132,184,204}
\definecolor{icedark}{RGB}{103,153,191}

\definecolor{traitrowcolor}{RGB}{223,242,244}
\definecolor{traitrowcoloralt}{RGB}{189,222,226}

\definecolor{characterrowcolor}{RGB}{252,251,195}
\definecolor{characterrowcoloralt}{RGB}{250,240,184}

\definecolor{archetyperowcolor}{RGB}{255,213,212} %% pale pink
\definecolor{archetyperowcoloralt}{RGB}{255,182,179} %% melon

\definecolor{datasetrowcolor}{RGB}{232,244,234}
\definecolor{datasetrowcoloralt}{RGB}{210,231,214}

\newcommand{\semdiffsign}{\Leftrightarrow}
\newcommand{\semdiffsignleft}{\Leftarrow}
\newcommand{\semdiffsignright}{\Rightarrow}

\newcommand{\semdiff}[2]{\{#1\,$\semdiffsign$\,#2\}}

\newcommand{\semdiffright}[2]{\{#1\,$\semdiffsignright$\,#2\}}

\newcommand{\semdiffbold}[2]{\{\textbf{#1}\,$\semdiffsign$\,\textbf{#2}\}}
\newcommand{\semdiffboldleft}[2]{\{\textbf{#1}\,$\semdiffsignleft$\,#2\}}
\newcommand{\semdiffboldright}[2]{\{#1\,$\semdiffsignright$\,\textbf{#2}\}}

\newcommand{\semdiffmath}[2]{\{\textnormal{\textbf{#1}}\!\semdiffsign\!{\textnormal{\textbf{#2}}\}}}
\newcommand{\semdiffmathleft}[2]{\{\textnormal{\textbf{#1}}\!\semdiffsign\!{\textnormal{#2}\}}}
\newcommand{\semdiffmathright}[2]{\{\textnormal{#1}\!\semdiffsign\!{\textnormal{\textbf{#2}}\}}}

\usepackage{stmaryrd}

\newcommand{\externallinksymbol}{{\tiny$^{{}_{\nnearrow}}$\!\!}}
\newcommand{\internallinksymbol}{{$^{\Rsh}$\!\!}}
\newcommand{\paperlinksymbol}{\externallinksymbol}

\newcommand{\Ntraits}{464}
\newcommand{\Ncharacters}{2000}
\newcommand{\Nstories}{341}

\newcommand{\onlinesiteshort}{compstorylab.org/archetypometrics}
\newcommand{\onlinesite}{https://\onlinesiteshort}

\newcommand{\archetypometricshome}{\onlinesite}
\newcommand{\cardsdir}{\archetypometricshome/cards}

\newcommand{\babaisyouboxscale}{0.48}

\newcommand{\archetyperatiosymbol}[2]{{\colorbox{#1}{\textcolor{#2}{\stackanchor{\scalebox{\babaisyouboxscale}{AR}}{\scalebox{\babaisyouboxscale}{CH}}}}}}

\newcommand{\appendixsymbol}[2]{{\colorbox{#1}{\textcolor{#2}{\stackanchor{\scalebox{\babaisyouboxscale}{AP}}{\scalebox{\babaisyouboxscale}{DX}}}}}}

\newcommand{\characterlinksimple}[2]{\href{\cardsdir/#1-\Ncharacters-\Ntraits-\Nstories.pdf}{\textcolor{verydarkgrey}{#2\paperlinksymbol}}}

\newcommand{\characterlinksimpledataset}[3]{
  \IfEqCase{#3}{
    {1}{\href{\cardsdir/#1-\Ncharactersmainone-\Ntraitsmainone-\Nstoriesmainone.pdf}{\textcolor{verydarkgrey}{#2\colorbox{datasetrowcolor}{$\dataset{#3}$}\paperlinksymbol}}}
    {2}{\href{\cardsdir/#1-\Ncharactersmaintwo-\Ntraitsmaintwo-\Nstoriesmaintwo.pdf}{\textcolor{verydarkgrey}{#2\colorbox{datasetrowcolor}{$\dataset{#3}$}\paperlinksymbol}}}
    {3}{\href{\cardsdir/#1-\Ncharactersmain-\Ntraitsmain-\Nstoriesmain.pdf}{\textcolor{verydarkgrey}{#2\colorbox{datasetrowcolor}{$\dataset{#3}$}\paperlinksymbol}}}
  }[\PackageError{characterlinksimpledataset}{Undefined option to characterlinksimpledataset: #1}{}]%
}

\newcommand{\traitlinkonelabel}[3]{\textcolor{verydarkgrey}{\ifthenelse{\equal{#3}{#1}}{\href{\cardsdir/#2--#1-\Ncharacters-\Ntraits-\Nstories.pdf}{#3\paperlinksymbol}}{\href{\cardsdir/#1--#2-\Ncharacters-\Ntraits-\Nstories.pdf}{#3\paperlinksymbol}}}}
\newcommand{\traitlinksimple}[2]{\textcolor{verydarkgrey}{\semdiff{\href{\cardsdir/#2--#1-\Ncharacters-\Ntraits-\Nstories.pdf}{#1\paperlinksymbol}}{\href{\cardsdir/#1--#2-\Ncharacters-\Ntraits-\Nstories.pdf}{#2\paperlinksymbol}}}}

\newcommand{\traitlinksimpledataset}[3]{
  \IfEqCase{#3}{
    {1}{\textcolor{verydarkgrey}{\semdiff{\href{\cardsdir/#2--#1-\Ncharactersmainone-\Ntraitsmainone-\Nstoriesmainone.pdf}{#1\colorbox{datasetrowcolor}{$\dataset{#3}$}\paperlinksymbol}}{\href{\cardsdir/#1--#2-\Ncharactersmainone-\Ntraitsmainone-\Nstoriesmainone.pdf}{#2\colorbox{datasetrowcolor}{$\dataset{#3}$}\paperlinksymbol}}}}
    {2}{\textcolor{verydarkgrey}{\semdiff{\href{\cardsdir/#2--#1-\Ncharactersmaintwo-\Ntraitsmaintwo-\Nstoriesmaintwo.pdf}{#1\colorbox{datasetrowcolor}{$\dataset{#3}$}\paperlinksymbol}}{\href{\cardsdir/#1--#2-\Ncharactersmaintwo-\Ntraitsmaintwo-\Nstoriesmaintwo.pdf}{#2\colorbox{datasetrowcolor}{$\dataset{#3}$}\paperlinksymbol}}}}
    {3}{\textcolor{verydarkgrey}{\semdiff{\href{\cardsdir/#2--#1-\Ncharactersmain-\Ntraitsmain-\Nstoriesmain.pdf}{#1\colorbox{datasetrowcolor}{$\dataset{#3}$}\paperlinksymbol}}{\href{\cardsdir/#1--#2-\Ncharactersmain-\Ntraitsmain-\Nstoriesmain.pdf}{#2\colorbox{datasetrowcolor}{$\dataset{#3}$}\paperlinksymbol}}}}
  }[\PackageError{traitlinksimpledataset}{Undefined option to traitlinksimpledataset: #1}{}]%
}

\newcommand{\traitlinksimpleright}[2]{\href{\cardsdir/#1--#2-\Ncharacters-\Ntraits-\Nstories.pdf}{\textcolor{verydarkgrey}{\semdiffright{#1}{\textbf{#2}}\paperlinksymbol}}}

\newcommand{\traitlinksimpleleft}[2]{\href{\cardsdir/#2--#1-\Ncharacters-\Ntraits-\Nstories.pdf}{\textcolor{verydarkgrey}{\semdiff{\textbf{#1}}{#2}}\paperlinksymbol}}

\newcommand{\traitlinksimpledatasetalt}[5]{
  \IfEqCase{#5}{
    {1}{\textcolor{verydarkgrey}{\semdiff{\href{\cardsdir/#2--#1-\Ncharactersmainone-\Ntraitsmainone-\Nstoriesmainone.pdf}{#3\colorbox{datasetrowcolor}{$\dataset{#5}$}\paperlinksymbol}}{\href{\cardsdir/#1--#2-\Ncharactersmainone-\Ntraitsmainone-\Nstoriesmainone.pdf}{#4\colorbox{datasetrowcolor}{$\dataset{#5}$}\paperlinksymbol}}}}
    {2}{\textcolor{verydarkgrey}{\semdiff{\href{\cardsdir/#2--#1-\Ncharactersmaintwo-\Ntraitsmaintwo-\Nstoriesmaintwo.pdf}{#3\colorbox{datasetrowcolor}{$\dataset{#5}$}\paperlinksymbol}}{\href{\cardsdir/#1--#2-\Ncharactersmaintwo-\Ntraitsmaintwo-\Nstoriesmaintwo.pdf}{#4\colorbox{datasetrowcolor}{$\dataset{#5}$}\paperlinksymbol}}}}
    {3}{\textcolor{verydarkgrey}{\semdiff{\href{\cardsdir/#2--#1-\Ncharactersmain-\Ntraitsmain-\Nstoriesmain.pdf}{#3\colorbox{datasetrowcolor}{$\dataset{#5}$}\paperlinksymbol}}{\href{\cardsdir/#1--#2-\Ncharactersmain-\Ntraitsmain-\Nstoriesmain.pdf}{#4\colorbox{datasetrowcolor}{$\dataset{#5}$}\paperlinksymbol}}}}
  }[\PackageError{traitlinksimpledatasetalt}{Undefined option to traitlinksimpledatasetalt: #1}{}]%
}

\newcommand{\storylinksimple}[2]{\href{\cardsdir/#1-\Ncharacters-\Ntraits-\Nstories.pdf}{\textcolor{darkgrey}{#2\paperlinksymbol}}}

\newcommand{\storylinksimpledataset}[3]{
  \IfEqCase{#3}{
    {1}{\href{\cardsdir/#1-\Ncharactersmainone-\Ntraitsmainone-\Nstoriesmainone.pdf}{\textcolor{verydarkgrey}{#2\colorbox{datasetrowcolor}{$\dataset{#3}$}\paperlinksymbol}}}
    {2}{\href{\cardsdir/#1-\Ncharactersmaintwo-\Ntraitsmaintwo-\Nstoriesmaintwo.pdf}{\textcolor{verydarkgrey}{#2\colorbox{datasetrowcolor}{$\dataset{#3}$}\paperlinksymbol}}}
    {3}{\href{\cardsdir/#1-\Ncharactersmain-\Ntraitsmain-\Nstoriesmain.pdf}{\textcolor{verydarkgrey}{#2\colorbox{datasetrowcolor}{$\dataset{#3}$}\paperlinksymbol}}}
  }[\PackageError{storylinksimpledataset}{Undefined option to storylinksimpledataset: #1}{}]%
}

\newcommand{\archetypelinkbase}[1]{\href{\cardsdir/Archetype-#1-component-size-\Ncharacters-\Ntraits-\Nstories.pdf}{\textcolor{verydarkgrey}{#1\paperlinksymbol}}}

\newcommand{\archetypelinksimple}[2]{\href{\cardsdir/Archetype-#1-component-size-\Ncharacters-\Ntraits-\Nstories.pdf}{\textcolor{verydarkgrey}{#2\paperlinksymbol}}}

\newcommand{\archetypelinksimpledataset}[3]{
  \IfEqCase{#3}{
    {1}{\href{\cardsdir/Archetype-#1-component-size-\Ncharactersmainone-\Ntraitsmainone-\Nstoriesmainone.pdf}{\textcolor{verydarkgrey}{#2\colorbox{datasetrowcolor}{$\dataset{#3}$}\paperlinksymbol}}}
    {2}{\href{\cardsdir/Archetype-#1-component-size-\Ncharactersmaintwo-\Ntraitsmaintwo-\Nstoriesmaintwo.pdf}{\textcolor{verydarkgrey}{#2\colorbox{datasetrowcolor}{$\dataset{#3}$}\paperlinksymbol}}}
    {3}{\href{\cardsdir/Archetype-#1-component-size-\Ncharactersmain-\Ntraitsmain-\Nstoriesmain.pdf}{\textcolor{verydarkgrey}{#2\colorbox{datasetrowcolor}{$\dataset{#3}$}\paperlinksymbol}}}
  }[\PackageError{archetypelinksimpledataset}{Undefined option to archetypelinksimpledataset: #1}{}]%
}

\newcommand{\archetypelinkratiosimpleappendix}[2]{{\hypersetup{allcolors=.}\hyperref[page:N\Ncharactersbase_archetypometrics.archetypeclass-#1]{#2\archetyperatiosymbol{archetyperowcolor}{black}\,\appendixsymbol{verydarkgrey}{white}\internallinksymbol}}}

\newcommand{\essentialtraitlinknegative}[2]{\href{\cardsdir/Essential-Trait-\zeropad{000}{#1}-negative-component-size-\Ncharacters-\Ntraits-\Nstories.pdf}{\textcolor{verydarkgrey}{#2\paperlinksymbol}}}

\newcommand{\essentialtraitlinkpositive}[2]{\href{\cardsdir/Essential-Trait-\zeropad{000}{#1}-positive-component-size-\Ncharacters-\Ntraits-\Nstories.pdf}{\textcolor{verydarkgrey}{#2\paperlinksymbol}}}

\usepackage{xstring}

\newcommand{\archetype}[1]{\archetypelinkbase{#1}}

\DeclareRobustCommand{\archetypesemdiff}[1]{
    \IfEqCase{#1}{
        {1}{\semdiffbold{\archetypelinkbase{Fool}}{\archetypelinkbase{Hero}}}
        {2}{\semdiffbold{\archetypelinkbase{Angel}}{\archetypelinkbase{Demon}}}
        {3}{\semdiffbold{\archetypelinkbase{Traditionalist}}{\archetypelinkbase{Adventurer}}}
        {4}{\semdiffbold{\archetypelinksimple{Lone-Wolf}{Lone Wolf}}{\archetypelinkbase{Diva}}}
        {5}{\semdiffbold{\archetypelinkbase{Outcast}}{\archetypelinkbase{Sophisticate}}}
        {6}{\semdiffbold{\archetypelinkbase{Brute}}{\archetypelinkbase{Geek}}}
    }[\PackageError{archetypesemdiff}{Undefined option to archetypesemdiff: #1}{}]%
}%

\newcommand{\archetypesemdiffleft}[1]{
    \IfEqCase{#1}{
        {1}{\semdiffboldleft{\archetypelinkbase{Fool}}{\archetypelinkbase{Hero}}}
        {2}{\semdiffboldleft{\archetypelinkbase{Angel}}{\archetypelinkbase{Demon}}}
        {3}{\semdiffboldleft{\archetypelinkbase{Traditionalist}}{\archetypelinkbase{Adventurer}}}
        {4}{\semdiffboldleft{\archetypelinksimple{Lone-Wolf}{Lone Wolf}}{\archetypelinkbase{Diva}}}
        {5}{\semdiffboldleft{\archetypelinkbase{Outcast}}{\archetypelinkbase{Sophisticate}}}
        {6}{\semdiffboldleft{\archetypelinkbase{Brute}}{\archetypelinkbase{Geek}}}
    }[\PackageError{archetypesemdiff}{Undefined option to archetypesemdiff: #1}{}]%
}%

\newcommand{\archetypesemdiffright}[1]{
    \IfEqCase{#1}{
        {1}{\semdiffboldright{\archetypelinkbase{Fool}}{\archetypelinkbase{Hero}}}
        {2}{\semdiffboldright{\archetypelinkbase{Angel}}{\archetypelinkbase{Demon}}}
        {3}{\semdiffboldright{\archetypelinkbase{Traditionalist}}{\archetypelinkbase{Adventurer}}}
        {4}{\semdiffboldright{\archetypelinksimple{Lone-Wolf}{Lone Wolf}}{\archetypelinkbase{Diva}}}
        {5}{\semdiffboldright{\archetypelinkbase{Outcast}}{\archetypelinkbase{Sophisticate}}}
        {6}{\semdiffboldright{\archetypelinkbase{Brute}}{\archetypelinkbase{Geek}}}
    }[\PackageError{archetypesemdiff}{Undefined option to archetypesemdiff: #1}{}]%
}%

\newcommand{\archetypesemdiffmath}[1]{
  \IfEqCase{#1}{
    {1}{\semdiffmath{\archetypelinkbase{Fool}}{\archetypelinkbase{Hero}}}
    {2}{\semdiffmath{\archetypelinkbase{Angel}}{\archetypelinkbase{Demon}}}
    {3}{\semdiffmath{\archetypelinkbase{Traditionalist}}{\archetypelinkbase{Adventurer}}}
    {4}{\semdiffmath{\archetypelinksimple{Lone-Wolf}{Lone Wolf}}{\archetypelinkbase{Diva}}}
    {5}{\semdiffmath{\archetypelinkbase{Outcast}}{\archetypelinkbase{Sophisticate}}}
    {6}{\semdiffmath{\archetypelinkbase{Brute}}{\archetypelinkbase{Geek}}}
  }[\PackageError{archetypesemdiffmath}{Undefined option to archetypesemdiffmath: #1}{}]%
}%

\newcommand{\archetypesemdiffmathswap}[1]{
  \IfEqCase{#1}{
    {1}{\semdiffmath{\archetypelinkbase{Hero}}{\archetypelinkbase{Fool}}}
    {2}{\semdiffmath{\archetypelinkbase{Demon}}{\archetypelinkbase{Angel}}}
    {3}{\semdiffmath{\archetypelinkbase{Adventurer}}{\archetypelinkbase{Traditionalist}}}
    {4}{\semdiffmath{\archetypelinkbase{Diva}}{\archetypelinksimple{Lone-Wolf}{Lone Wolf}}}
    {5}{\semdiffmath{\archetypelinkbase{Sophisticate}}{\archetypelinkbase{Outcast}}}
    {6}{\semdiffmath{{\archetypelinkbase{Geek}}\archetypelinkbase{Brute}}}
  }[\PackageError{archetypesemdiffmathswap}{Undefined option to archetypesemdiffmathswap: #1}{}]%
}%

\newcommand{\archetypesemdiffmathleft}[1]{
  \IfEqCase{#1}{
    {1}{\semdiffmathleft{\archetypelinkbase{Fool}}{\archetypelinkbase{Hero}}}
    {2}{\semdiffmathleft{\archetypelinkbase{Angel}}{\archetypelinkbase{Demon}}}
    {3}{\semdiffmathleft{\archetypelinkbase{Traditionalist}}{\archetypelinkbase{Adventurer}}}
    {4}{\semdiffmathleft{\archetypelinksimple{Lone-Wolf}{Lone Wolf}}{\archetypelinkbase{Diva}}}
    {5}{\semdiffmathleft{\archetypelinkbase{Outcast}}{\archetypelinkbase{Sophisticate}}}
    {6}{\semdiffmathleft{\archetypelinkbase{Brute}}{\archetypelinkbase{Geek}}}
  }[\PackageError{archetypesemdiffmathleft}{Undefined option to archetypesemdiffmathleft: #1}{}]%
}%

\newcommand{\archetypesemdiffmathright}[1]{
  \IfEqCase{#1}{
    {1}{\semdiffmathright{\archetypelinkbase{Fool}}{\archetypelinkbase{Hero}}}
    {2}{\semdiffmathright{\archetypelinkbase{Angel}}{\archetypelinkbase{Demon}}}
    {3}{\semdiffmathright{\archetypelinkbase{Traditionalist}}{\archetypelinkbase{Adventurer}}}
    {4}{\semdiffmathright{\archetypelinksimple{Lone-Wolf}{Lone Wolf}}{\archetypelinkbase{Diva}}}
    {5}{\semdiffmathright{\archetypelinkbase{Outcast}}{\archetypelinkbase{Sophisticate}}}
    {6}{\semdiffmathright{\archetypelinkbase{Brute}}{\archetypelinkbase{Geek}}}
  }[\PackageError{archetypesemdiffmathright}{Undefined option to archetypesemdiffmathright: #1}{}]%
}%

\newcommand{\essentialsemdiff}[1]{
  \IfEqCase{#1}{
    {1}{\semdiff{\essentialtraitlinknegative{1}{weak/incompetent/lazy/stupid}}{\essentialtraitlinkpositive{1}{powerful/capable/purposeful/intelligent}}}
    {2}{\semdiff{\essentialtraitlinknegative{2}{safe/pure/virtuous/humble}}{\essentialtraitlinkpositive{2}{dangerous/depraved/corrupt/arrogant}}}
    {3}{\semdiff{\essentialtraitlinknegative{3}{serious/predictable/humorless/uncreative}}{\essentialtraitlinkpositive{3}{playful/unpredictable/funny/creative}}}
    {4}{\semdiff{\essentialtraitlinknegative{4}{rugged/stoic/independent/blunt}}{\essentialtraitlinkpositive{4}{refined/dramatic/dependent/sensitive}}}
    {5}{\semdiff{\essentialtraitlinknegative{5}{unlucky/unsophisticated/traumatized}}{\essentialtraitlinkpositive{5}{fortunate/sophisticated/confident}}}
    {6}{\semdiff{\essentialtraitlinknegative{6}{physical/mainstream/simple-minded}}{\essentialtraitlinkpositive{6}{intellectual/weird/complex}}}
    {7}{\semdiff{\essentialtraitlinknegative{7}{dramatic/attractive/young}}{\essentialtraitlinkpositive{7}{comedic/ugly/old}}}
    {8}{\semdiff{\essentialtraitlinknegative{8}{spiritual/rural/historical}}{\essentialtraitlinkpositive{8}{skeptical/urban/modern}}}
    {9}{\semdiff{\essentialtraitlinknegative{9}{old/historical/low-tempo}}{\essentialtraitlinkpositive{9}{young/modern/high-tempo}}}
    {10}{\semdiff{\essentialtraitlinknegative{10}{feminine/luddite}}{\essentialtraitlinkpositive{10}{masculine/technophile}}}
    {11}{\semdiff{\essentialtraitlinknegative{11}{secondary/street-wise}}{\essentialtraitlinkpositive{11}{primary/sheltered}}}
  }[\PackageError{essentialsemdiff}{Undefined option to essentialsemdiff: #1}{}]%
}%

\newcommand{\essentialsemdiffloose}[1]{
  \IfEqCase{#1}{
    {1}{\semdiff{\essentialtraitlinknegative{1}{weak, incompetent, lazy, stupid}}{\essentialtraitlinkpositive{1}{powerful, capable, purposeful, intelligent}}}
    {2}{\semdiff{\essentialtraitlinknegative{2}{safe, pure, virtuous, humble}}{\essentialtraitlinkpositive{2}{dangerous, depraved, corrupt, arrogant}}}
    {3}{\semdiff{\essentialtraitlinknegative{3}{serious, predictable, humorless, uncreative}}{\essentialtraitlinkpositive{3}{playful, unpredictable, funny, creative}}}
    {4}{\semdiff{\essentialtraitlinknegative{4}{rugged, stoic, independent, blunt}}{\essentialtraitlinkpositive{4}{refined, dramatic, dependent, sensitive}}}
    {5}{\semdiff{\essentialtraitlinknegative{5}{unlucky, unsophisticated, traumatized}}{\essentialtraitlinkpositive{5}{fortunate, sophisticated, confident}}}
    {6}{\semdiff{\essentialtraitlinknegative{6}{physical, mainstream, simple-minded}}{\essentialtraitlinkpositive{6}{intellectual, weird, complex}}}
    {7}{\semdiff{\essentialtraitlinknegative{7}{dramatic, attractive, young}}{\essentialtraitlinkpositive{7}{comedic, ugly, old}}}
    {8}{\semdiff{\essentialtraitlinknegative{8}{spiritual, rural, historical}}{\essentialtraitlinkpositive{8}{skeptical, urban, modern}}}
    {9}{\semdiff{\essentialtraitlinknegative{9}{old, historical, low-tempo}}{\essentialtraitlinkpositive{9}{young, modern, high-tempo}}}
    {10}{\semdiff{\essentialtraitlinknegative{10}{feminine, luddite}}{\essentialtraitlinkpositive{10}{masculine, technophile}}}
    {11}{\semdiff{\essentialtraitlinknegative{11}{secondary, street-wise}}{\essentialtraitlinkpositive{11}{primary, sheltered}}}
  }[\PackageError{essentialsemdiffloose}{Undefined option to essentialsemdiffloose: #1}{}]%
}%

\newcommand{\essentialsemdifflooseleft}[1]{
    \IfEqCase{#1}{
        {1}{\semdiff{\textbf{\essentialtraitlinknegative{1}{weak, incompetent, lazy, stupid}}}{\essentialtraitlinkpositive{1}{powerful, capable, purposeful, intelligent}}}
        {2}{\semdiff{\textbf{\essentialtraitlinknegative{2}{safe, pure, virtuous, humble}}}{\essentialtraitlinkpositive{2}{dangerous, depraved, corrupt, arrogant}}}
        {3}{\semdiff{\textbf{\essentialtraitlinknegative{3}{serious, predictable, humorless, uncreative}}}{\essentialtraitlinkpositive{3}{playful, unpredictable, funny, creative}}}
        {4}{\semdiff{\textbf{\essentialtraitlinknegative{4}{rugged, stoic, independent, blunt}}}{\essentialtraitlinkpositive{4}{refined, dramatic, dependent, sensitive}}}
        {5}{\semdiff{\textbf{\essentialtraitlinknegative{5}{unlucky, unsophisticated, traumatized}}}{\essentialtraitlinkpositive{5}{fortunate, sophisticated, confident}}}
        {6}{\semdiff{\textbf{\essentialtraitlinknegative{6}{physical, mainstream, simple-minded}}}{\essentialtraitlinkpositive{6}{intellectual, weird, complex}}}
        {7}{\semdiff{\textbf{\essentialtraitlinknegative{7}{dramatic, attractive, young}}}{\essentialtraitlinkpositive{7}{comedic, ugly, old}}}
        {8}{\semdiff{\textbf{\essentialtraitlinknegative{8}{spiritual, rural, historical}}}{\essentialtraitlinkpositive{8}{skeptical, urban, modern}}}
        {9}{\semdiff{\textbf{\essentialtraitlinknegative{9}{old, historical, low-tempo}}}{\essentialtraitlinkpositive{9}{young, modern, high-tempo}}}
        {10}{\semdiff{\textbf{\essentialtraitlinknegative{10}{feminine, luddite}}}{\essentialtraitlinkpositive{10}{masculine, technophile}}}
        {11}{\semdiff{\textbf{\essentialtraitlinknegative{11}{secondary, street-wise}}}{\essentialtraitlinkpositive{11}{primary, sheltered}}}
    }[\PackageError{essentialsemdifflooseleft}{Undefined option to essentialsemdifflooseleft: #1}{}]%
 }%

\newcommand{\essentialsemdifflooseright}[1]{
  \IfEqCase{#1}{
    {1}{\semdiff{\essentialtraitlinknegative{1}{weak, incompetent, lazy, stupid}}{\textbf{\essentialtraitlinkpositive{1}{powerful, capable, purposeful, intelligent}}}}
    {2}{\semdiff{\essentialtraitlinknegative{2}{safe, pure, virtuous, humble}}{\textbf{\essentialtraitlinkpositive{2}{dangerous, depraved, corrupt, arrogant}}}}
    {3}{\semdiff{\essentialtraitlinknegative{3}{serious, predictable, humorless, uncreative}}{\textbf{\essentialtraitlinkpositive{3}{playful, unpredictable, funny, creative}}}}
    {4}{\semdiff{\essentialtraitlinknegative{4}{rugged, stoic, independent, blunt}}{\textbf{\essentialtraitlinkpositive{4}{refined, dramatic, dependent, sensitive}}}}
    {5}{\semdiff{\essentialtraitlinknegative{5}{unlucky, unsophisticated, traumatized}}{\textbf{\essentialtraitlinkpositive{5}{fortunate, sophisticated, confident}}}}
    {6}{\semdiff{\essentialtraitlinknegative{6}{physical, mainstream, simple-minded}}{\textbf{\essentialtraitlinkpositive{6}{intellectual, weird, complex}}}}
    {7}{\semdiff{\essentialtraitlinknegative{7}{dramatic, attractive, young}}{\textbf{\essentialtraitlinkpositive{7}{comedic, ugly, old}}}}
    {8}{\semdiff{\essentialtraitlinknegative{8}{spiritual, historical, rural}}{\textbf{\essentialtraitlinkpositive{8}{skeptical, urban, modern}}}}
    {9}{\semdiff{\essentialtraitlinknegative{9}{old, historical, low-tempo}}{\textbf{\essentialtraitlinkpositive{9}{young, modern, high-tempo}}}}
    {10}{\semdiff{\essentialtraitlinknegative{10}{feminine, luddite}}{\textbf{\essentialtraitlinkpositive{10}{masculine, technophile}}}}
    {11}{\semdiff{\essentialtraitlinknegative{11}{secondary, street-wise}}{\textbf{\essentialtraitlinkpositive{11}{primary, sheltered}}}}
  }[\PackageError{essentialsemdifflooseright}{Undefined option to essentialsemdifflooseright: #1}{}]%
}%

\newcommand{\essentialsemdiffmathleft}[1]{
    \IfEqCase{#1}{
        {1}{\semdiffmathleft{weak/incompetent/lazy/stupid}{powerful/capable/purposeful/intelligent}}
        {2}{\semdiffmathleft{safe/pure/virtuous/humble}{dangerous/depraved/corrupt/arrogant}}
        {3}{\semdiffmathleft{serious/predictable/humorless/uncreative}{playful/unpredictable/funny/creative}}
        {4}{\semdiffmathleft{rugged/stoic/independent/blunt}{refined/dramatic/dependent/sensitive}}
        {5}{\semdiffmathleft{unlucky/unsophisticated/traumatized}{fortunate/sophisticated/confident}}
        {6}{\semdiffmathleft{physical/mainstream/simple-minded}{intellectual/weird/complex}}
        {7}{\semdiffmathleft{dramatic/attractive/young}{comedic/ugly/old}}
        {8}{\semdiffmathleft{spiritual/rural/historical}{skeptical/urban/modern}}
        {9}{\semdiffmathleft{old/historical/low-tempo}{young/modern/high-tempo}}
        {10}{\semdiffmathleft{feminine/luddite}{masculine/technophile}}
        {11}{\semdiffmathleft{secondary/street-wise}{primary/sheltered}}
    }[\PackageError{essentialsemdiffmathleft}{Undefined option to essentialsemdiffmathleft: #1}{}]%
}%

\newcommand{\essentialsemdiffmathright}[1]{
    \IfEqCase{#1}{
        {1}{\semdiffmathright{weak/incompetent/lazy/stupid}{powerful/capable/purposeful/intelligent}}
        {2}{\semdiffmathright{safe/pure/virtuous/humble}{dangerous/depraved/corrupt/arrogant}}
        {3}{\semdiffmathright{serious/predictable/humorless/uncreative}{playful/unpredictable/funny/creative}}
        {4}{\semdiffmathright{rugged/stoic/independent/blunt}{refined/dramatic/dependent/sensitive}}
        {5}{\semdiffmathright{unlucky/unsophisticated/traumatized}{fortunate/sophisticated/confident}}
        {6}{\semdiffmathright{physical/mainstream/simple-minded}{intellectual/weird/complex}}
        {7}{\semdiffmathright{dramatic/attractive/young}{comedic/ugly/old}}
        {8}{\semdiffmathright{spiritual/rural/historical}{skeptical/urban/modern}}
        {9}{\semdiffmathright{old/historical/low-tempo}{young/modern/high-tempo}}
        {10}{\semdiffmathright{feminine/luddite}{masculine/technophile}}
        {11}{\semdiffmathright{secondary/street-wise}{primary/sheltered}}
    }[\PackageError{essentialsemdiffmathright}{Undefined option to essentialsemdiffmathright: #1}{}]%
}%

\newcommand{\essentialsemdiffmath}[1]{
    \IfEqCase{#1}{
        {1}{\semdiffmath{weak/incompetent/lazy/stupid}{powerful/capable/purposeful/intelligent}}
        {2}{\semdiffmath{safe/pure/virtuous/humble}{dangerous/depraved/corrupt/arrogant}}
        {3}{\semdiffmath{serious/predictable/humorless/uncreative}{playful/unpredictable/funny/creative}}
        {4}{\semdiffmath{rugged/stoic/independent/blunt}{refined/dramatic/dependent/sensitive}}
        {5}{\semdiffmath{unlucky/unsophisticated/traumatized}{fortunate/sophisticated/confident}}
        {6}{\semdiffmath{physical/mainstream/simple-minded}{intellectual/weird/complex}}
        {7}{\semdiffmath{dramatic/attractive/young}{comedic/ugly/old}}
        {8}{\semdiffmath{spiritual/rural/historical}{skeptical/urban/modern}}
        {9}{\semdiffmath{old/historical/low-tempo}{young/modern/high-tempo}}
        {10}{\semdiffmath{feminine/luddite}{masculine/technophile}}
        {11}{\semdiffmath{secondary/street-wise}{primary/sheltered}}
    }[\PackageError{essentialsemdiffmath}{Undefined option to essentialsemdiffmath: #1}{}]%
}%

\newcommand{\ousiometricsemdiff}[1]{
    \IfEqCase{#1}{
        {1}{\semdiffbold{weak}{powerful}}
        {2}{\semdiffbold{safe}{dangerous}}
        {3}{\semdiffbold{structured}{unstructured}}
    }[\PackageError{ousiometricsemdiff}{Undefined option to ousiometricsemdiff: #1}{}]%
}%

\newcommand{\ousiometricsemdiffmath}[1]{
    \IfEqCase{#1}{
        {1}{\semdiffmath{weak}{powerful}}
        {2}{\semdiffmath{safe}{dangerous}}
        {3}{\semdiffmath{structured}{unstructured}}
    }[\PackageError{ousiometricsemdiffmath}{Undefined option to ousiometricsemdiffmath: #1}{}]%
}%

\newcommand{\ousiometricsemdiffmathleft}[1]{
    \IfEqCase{#1}{
        {1}{\semdiffmathleft{weak}{powerful}}
        {2}{\semdiffmathleft{safe}{dangerous}}
        {3}{\semdiffmathleft{structured}{unstructured}}
    }[\PackageError{ousiometricsemdiffmathleft}{Undefined option to ousiometricsemdiffmathleft: #1}{}]%
}%

\newcommand{\ousiometricsemdiffmathright}[1]{
    \IfEqCase{#1}{
        {1}{\semdiffmathright{weak}{powerful}}
        {2}{\semdiffmathright{safe}{dangerous}}
        {3}{\semdiffmathright{structured}{unstructured}}
    }[\PackageError{ousiometricsemdiffmathright}{Undefined option to ousiometricsemdiffmathright: #1}{}]%
}%

\newcommand{\dimensiontype}[1]{
    \IfEqCase{#1}{
        {1}{Major archetype}
        {2}{Major Archetype}
        {3}{Major Archetype}
        {4}{Minor Archetype}
        {5}{Minor Archetype}
        {6}{Minor Archetype}
        {7}{Trait}
        {8}{Trait}
        {9}{Trait}
        {10}{Trait}
        {11}{Trait}
    }[\PackageError{dimensiontype}{Undefined option to dimensiontype: #1}{}]%
}%

\newcommand{\straightalign}{perceived-Straight}
\newcommand{\queeralign}{perceived-Queer}
\newcommand{\fandomqueer}{portrayed-Queer}
\newcommand{\straightneigh}{perceived-Straight$_k$}
\newcommand{\queerneigh}{perceived-Queer$_k$}

\setboolean{twocolswitch}{true}

\raggedright

\usepackage{pdfpages}
\usepackage{newpax}
\newpaxsetup{usefileattributes=true}

\usepackage{authblk}

\begin{document}

\title{\protect
The queer Hero versus the Fool bias of the queer trait: \\
An archetypometric analysis \\
of the 
collective portrayal of queerness in fictional stories

% A but also B: An archetypometric analysis of the collective portrayal of queerness in fictional stories
% The queer Hero yet hidden Fool bias of characters: ...

%% Queer heroes and the foolisness of the queer trait:
%% An archetypometric analysis
%% of the paradoxical collective portrayal of queerness in fictional stories

%% Uncovering the collective portrayal of queerness in fictional stories:
%% An archetypometric analysis

%% in archetype space across popular fictional stories

% Original
%% Exploring the portrayal and perception of queerness in %% archetype space across fictional stories
}

\renewcommand*{\Authsep}{, }
\renewcommand*{\Authand}{, }
\renewcommand*{\Authands}{, }
\renewcommand*{\Affilfont}{\normalsize\normalfont}
\renewcommand*{\Authfont}{\bfseries}
\setlength{\affilsep}{2em}

\author[1\thanks{ashley.fehr@uvm.edu}]{Ashley~M.~A.~Fehr}
\author[1]{Calla~Glavin~Beauregard}
\author[1,2]{Julia~Witte~Zimmerman}
\author[1]{Danny~Benett}
\author[3]{Timothy~R.~Tangherlini}
\author[1,4]{Christopher~M.~Danforth}
\author[1,5,6,7\thanks{peter.dodds@uvm.edu}]{Peter~Sheridan~Dodds}

\affil[1]{
  Computational Story Lab,
  Vermont Advanced Computing Center,
  Vermont Complex Systems Institute,
  MassMutual Center of Excellence for Complex Systems and Data Science,
  University of Vermont,
  Burlington,
  VT 05405,
  US
}

\affil[2]{
  Computational Ethics Lab,
  University of Vermont,
  Burlington,
  VT 05405,
  US
}

\affil[3]{
  Department of Scandinavian,
  Folklore Program,
  School of Information,
  Berkeley Institute for Data Science,
  University~of~California,~Berkeley,~Berkeley,~CA~94720-1500,~USA
}

\affil[4]{
  Department of Mathematics \& Statistics,
  University of Vermont,
  Burlington,
  VT 05405,
  US
  }

\affil[5]{
  Department of Computer Science,
  University of Vermont,
  Burlington,
  VT 05405,
  US
}

\affil[6]{
  Santa Fe Institute,
  1399 Hyde Park Rd,
  Santa Fe,
  NM 87501,
  US
}

\affil[7]{
  Complexity Science Hub,
  Metternichgasse 8,
  1030 Vienna,
  Austria
}

\date{\today}

\maketitle

%%%%%%%%% end of author(s), address(es) plus abstract

%%%%%%%%%%%%%%%%%%%%%%%%%%%%%%%%%%%%%
%% logline and excerpt graphic
%%%%%%%%%%%%%%%%%%%%%%%%%%%%%%%%%%%%%

\mbox{}

\bigskip
\bigskip
\bigskip

\hspace*{50pt}
\begin{minipage}{300pt}

  \begin{tabular}{>{\raggedright\arraybackslash}p{180pt}p{30pt}p{180pt}}

    \parbox{180pt}{    
      \begin{loglinebox}[Logline]
        \raggedright
        Archetypometric analysis of fictional stories reveals
hidden tendencies and biases of collective authors.
perceived-Queer characters are more multifaceted in archetype and trait space than are perceived-Straight characters.
More heroic characters tend toward being straight,
while more adventurous and geekish characters tend toward being queer.

        \smallskip
      \end{loglinebox}
    }

    &
    
    &
    
    %% excerpt graphic
    \parbox{180pt}{    
      \includegraphics[width=180pt,valign=m,frame]{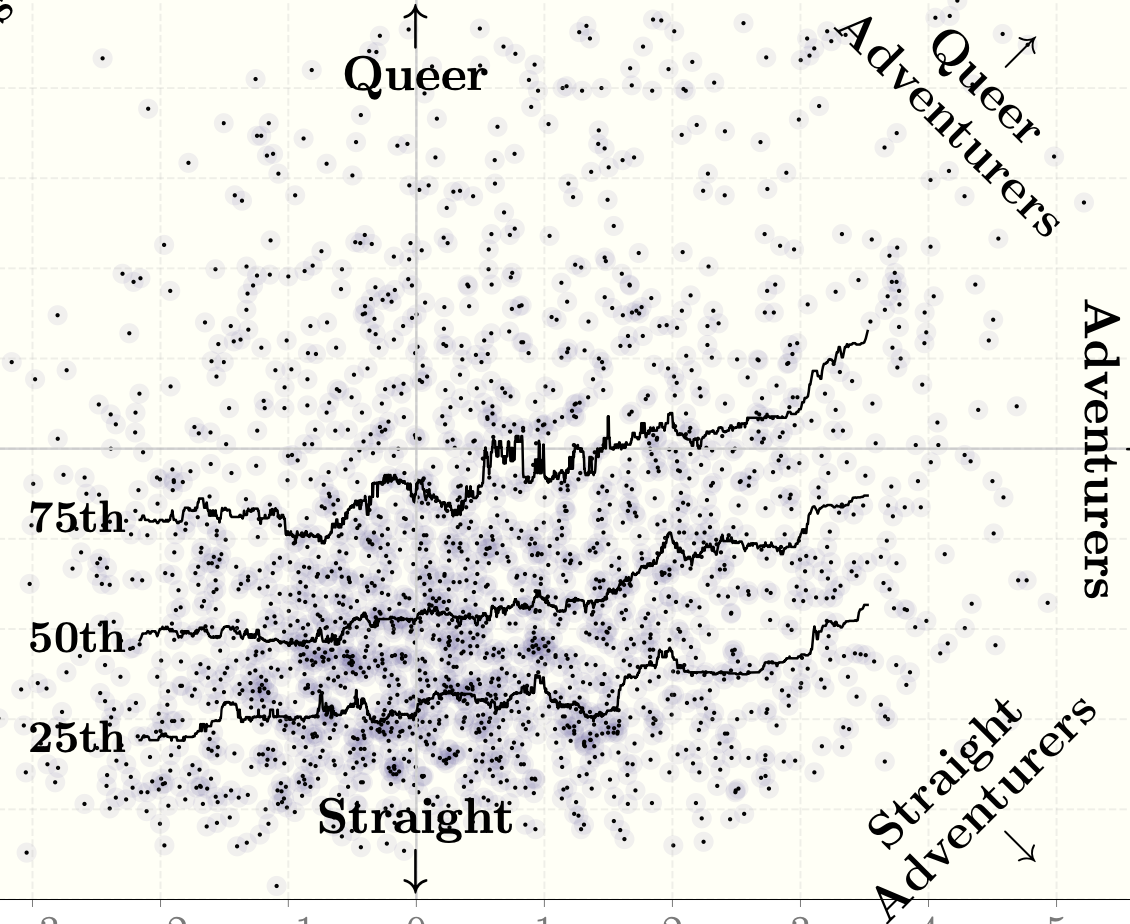}
    }
    
  \end{tabular}
  
\end{minipage}

\clearpage

\newgeometry{
  left=2in,
  right=2in,
  top=1in,
  bottom=1in,
  }

\onecolumn
\selectfont

%%%%%%%%%%%%%%%%%%%
%% Abstract:
%%%%%%%%%%%%%%%%%%%

\begin{abstractbox}[Abstract]
  \raggedright
  Visibility in media is pivotal for identity development and for broadening societal views of gender and sexuality. 
Queer representation has increased in recent years, 
yet damaging stereotypes and tropes persist.
Here, we focus on queer portrayal and its perception by audiences in fictional stories (television, film, and literature) by studying characters by their quantified archetypes which are
operationalizations of common conceptions such as 
Hero, Diva, and Outcast.
We use the archetypometrics and Fandom's LGBTQIA+ datasets 
to study samples of fictional characters along the trait differential spanning straight to queer. 
We find, quantify, and explain a seeming paradox.
The characters with the highest queer score
present positive primary archetypes 
and are typically 
Heroes rather than Fools, 
Angels rather than Demons, 
and 
Adventurers rather than Traditionalists.
But evaluation across many stories
for the straight-queer trait itself reveals a
strong collective-writing bias towards Fool (away from Hero)
and no meaningful loading for the other two dimensions.
Our analysis offers a population-scale view of the complexities of
queer portrayal, while also pointing to risks in blindly training on many-authored story corpora.
  \smallskip
\end{abstractbox}

%%%%%%%%%%%%%%%%%%%
%% Keywords:
%%%%%%%%%%%%%%%%%%%

\begin{infobox}[Keywords]
  \centering
  archetypes,
archetypometrics,
traits,
characters,
straight,
queer,
representation,
perception,
stereotypes,
stories,
fiction,
television

 \smallskip
\end{infobox}

\renewcommand{\baselinestretch}{1}
\selectfont

\twocolumn

\restoregeometry

\clearpage

\newgeometry{
  left=2in,
  right=2in,
  top=1in,
  bottom=1in,
  }

\onecolumn

\renewcommand{\baselinestretch}{1.25}
\selectfont

\renewcommand{\baselinestretch}{1}
\selectfont

\restoregeometry
\twocolumn

% add space between number and section titles in TOC
\setlength\cftsecnumwidth{2em}

% Source - https://tex.stackexchange.com/a/182744
% Posted by kjyv
% Retrieved 2026-06-30, License - CC BY-SA 3.0

\renewcommand{\baselinestretch}{1.5}\normalsize
\tableofcontents
\renewcommand{\baselinestretch}{1.0}\normalsize

\clearpage

\section{Introduction}
\label{sec:straight-queer.introduction}

%% SET UP IMPORTANCE OF STUDYING, SOME HISTORY, OUR OVERARCHING QUESTION

%% first paragraph: positive first
%% group level and individual level
%% second paragraph: negative

Visibility and media representation
influence societal views of gender and sexuality~\cite{cover2022MakingQueerContentVisible,dhoest2023HomosexualityNormalityReception}.
Favorable representations can contribute to a reduction in
homophobic and sexist attitudes~\cite{birchmore2022ExploringBoundariesParasocial}.
At the individual level,
positive representations
matter in identity development, especially for children and adolescents~\cite{sumter2025Representation}.
Supportive representations take on
many forms such as attending to character depth or countering stereotypical representations of vulnerable characters.

\begin{figure}[!b]
  \centering
  \includegraphics[width=\columnwidth]{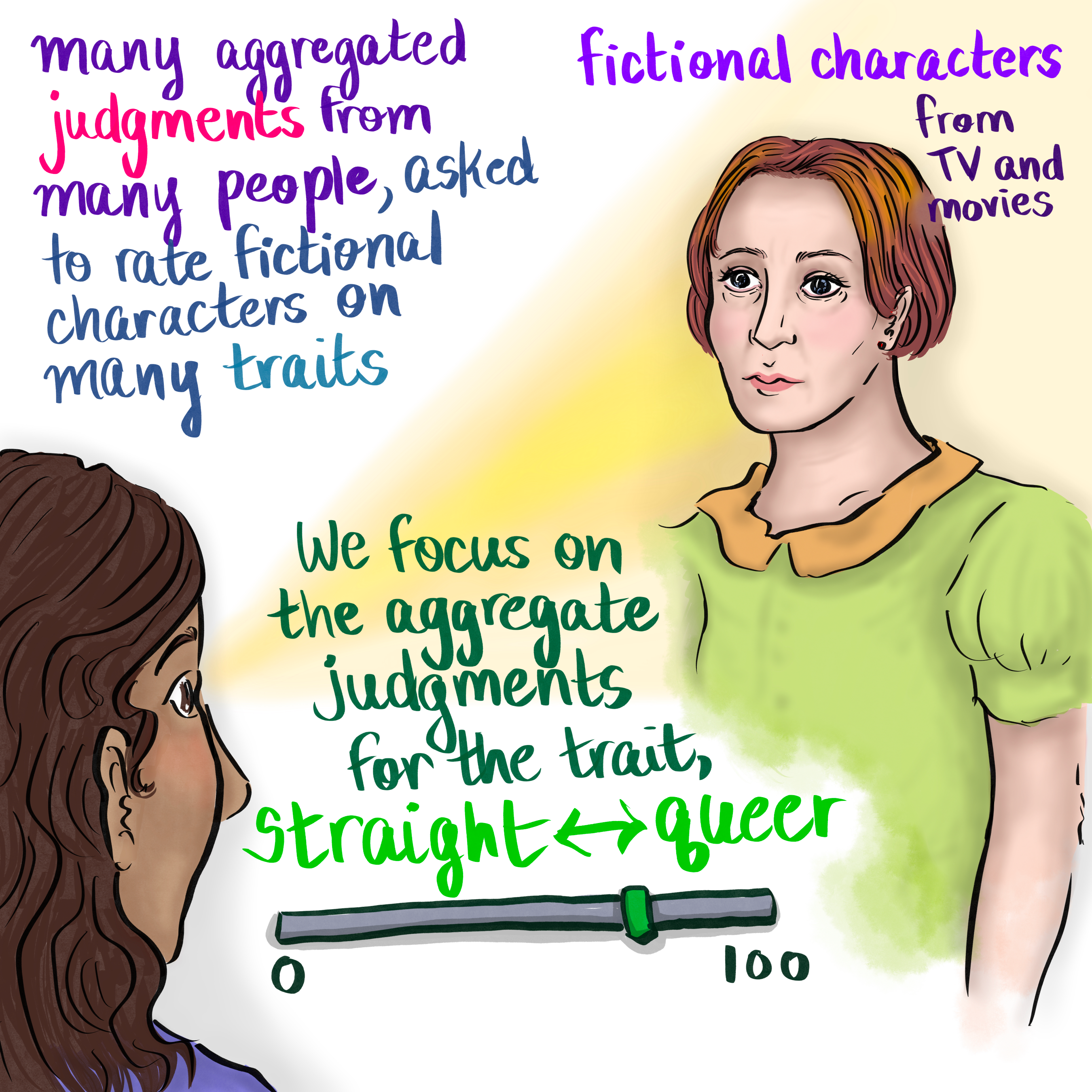}
  \caption{Among 464 dual-labeled traits, ``queer'' and ``straight'' are terms which could be understood as two ends of a spectrum, with each term helping to contextualize the opposite (e.g., ``straight'' as in sexual identity). These terms are not meant to reduce human identity to a binary; rather, they anchor survey-taker interpretation for psychometric consistency and validity. Survey takers rated characters along a $100$-point scale between straight and queer. From this open-source dataset, we examine those perceptions in a world that often frames straightness and queerness as polar opposites while acknowledging that queer and straight are not binary and that the psychometric approach has limitations (see \S\ref{sec:straight-queer.discussion.limits}).}
  \label{fig:queercharactersdiagram}
\end{figure}

% REPRESENTATION & STEREOTYPING: i) HOW PERCEPTIONS MAKE THEIR WAY INTO DATA, and ii) not all representation is good and can perpetuate negative attitudes.

But representations of minorities can also encode bias and introduce or reinforce negative stereotypes~\cite{weber2026QueerNLPCritical,yuan2025LGBTRepresentationFilm}.
In one study, judgments of social media images associated prototypically gay male characteristics with high identification with
negative stereotypes and engagement in immoral activities~\cite{beam2024gay_prototypicality}.
Visibility can also victimize vulnerable characters or deliberately mislead viewers by teasing queer inclusion---``Queerbaiting''~\cite{cover2022MakingQueerContentVisible}.
Some media tropes---recurrent narrative patterns---reflect such damaging portrayals in storytelling. These tropes or stereotypes can persist as vehicles for representation~\cite{rovirosa2025QueerNonNormativeCharacters}.
For example, the ``Bury your gays'' trope explicitly identifies the phenomenon of queer characters being denied happy endings by being unexpectedly killed.
Notable examples are the killings of Tara Maclay from \storylinksimple{Buffy-the-Vampire-Slayer}{Buffy the Vampire Slayer} and Lexa from \storylinksimple{The-100}{The 100}~\cite{cover2023BuryYourGays}.

Negative queer stereotypes and representations in media are not new but stem from a longer history of public awareness and policy. Historically, queer public awareness in the US reflected an alleged connection between `anti-American' values and the homosexual community~\cite{johnson2023LavenderScareCold}.
One such historical period was the Red Scare era (1947--1954~\cite{2026RedScareHistory}) during which espousing communist values was unpatriotic.
The Lavender Scare (1947--1975~\cite{2026LavenderScareHistory}) linked homosexuality with moral depravity, security risks,
and mental imbalance.
These qualities made those perceived to be queer---just like those perceived to be communist or `un-American'---allegedly `unfit' for government office~\cite{johnson2023LavenderScareCold}.
Discriminatory practices and perceptions continued well beyond the Lavender Scare period,
peaking once again with the moral panic surrounding the spread of AIDS in the 1980s~\cite{2026LavenderScareHistory},
with negative remnants in public sentiment being maintained by prior decades of media portrayals~\cite{fejes2008GayRightsMoral}.

Since the late 1980s, while queer visibility in television~\cite{glaad2025WhereWeAre} (Fig.~\ref{fig:glaad-stats}) and film has increased~\cite{birchmore2022ExploringBoundariesParasocial} and audiences prefer non-stereotypical (more realistic) queer representations~\cite{soto_sanfiel_perceptions_2024}, harmful stereotypes, and their instantiation through tropes, are still ubiquitous~\cite{beam2024gay_prototypicality} in these mediums~\cite{soto_sanfiel_perceptions_2024,cover2022MakingQueerContentVisible,cover2023BuryYourGays}. This negatively-coded visibility can counteract---or at least undermine---the positive aspects of increased representation~\cite{birchmore2022ExploringBoundariesParasocial,yuan2025LGBTRepresentationFilm}.
For example, large-scale computational analyses have identified that queer terms are often associated with negative sentiment~\cite{weber2026QueerNLPCritical}, suggesting that societal prejudice persists. This prejudice emerges as stereotypes and tropes used as vehicles to represent queerness in fictional media~\cite{rovirosa2025QueerNonNormativeCharacters,soto_sanfiel_perceptions_2024}. This work suggests that stereotypes, tropes, or bias can make its way into character representations, offering caution to training models or AI on data of character representations, which could serve to buttress harmful stereotypes.

% Our overarching question and clarifications
Here, we focus on perceptions of queer portrayal in stories through the lens of archetypes to better understand the extent of stereotypes in media (Fig.~\ref{fig:queercharactersdiagram}).
% what are archetype?
Archetypes, a well-known concept in narrative theory~\cite{Theophrastus_Characters_Rusten2003,Jung1969}, are taken here to be common character constructions used across stories. In previous work, we have shown that these can be statistically derived~\cite{dodds2025ArchetypometricsPragmateiaEmpirical} and identified six dominant archetype pairs: \archetypesemdiff{1}, \archetypesemdiff{2}, \archetypesemdiff{3}, \archetypesemdiff{4},
\archetypesemdiff{5}, and \archetypesemdiff{6}. Here, by studying the trait pair \traitlinksimple{straight}{queer}, we extend prior work that examines archetypes through the lens of gender~\cite{beauregard2026ArchetypesGenderFiction}.

% IMPORTANT CLARIFYING DEFINITION/DISCLAIMER
To analyze perceptions of queer portrayal, we regard a subset of characters from the larger character dataset as ``canonically queer'' if the character is explicitly indicated as queer within their story (as verified on wiki pages describing stories or characters and Fandom's LGBTQIA+ Characters wiki).
We also examine an explicit \traitlinksimple{straight}{queer} trait pair from a dataset of crowd-sourced online ratings to quantify variation in this trait across all characters, allowing us to measure audience perception of media portrayal of queerness and straightness. Importantly, we emphasize that this trait pair does not capture the entire range of diverse sexualities and lived experiences understood with the term ``queer'' as used within the LGBTQIA+ community~\cite{weber2026QueerNLPCritical,HRC2026GlossaryTerms}.
We then consider each character's traits and archetypes to analyze how a large audience of online raters perceives canonically queer portrayals.

\begin{figure*}[tp]
  \centering
    \includegraphics[width=\textwidth]{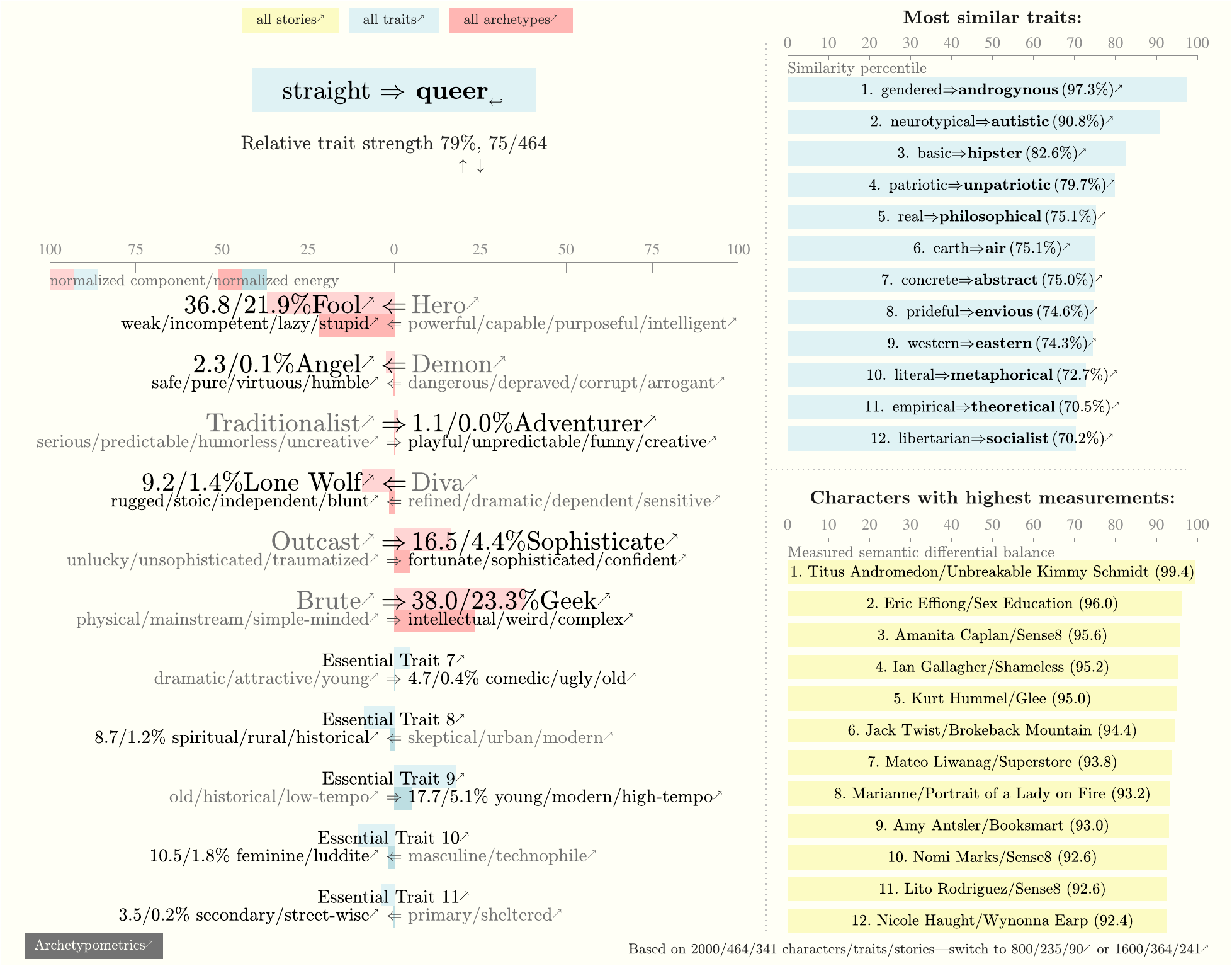}
  \caption{
  The \traitlinksimple{straight}{queer} trait's archetype configuration pointing in the perceived \textbf{queer} direction. Components include an archetype breakdown for the trait (left), the 12 closest traits in order by distance (upper right, with more listed in Tab.~\ref{tab:trait_closeness}), and characters with highest measurement on this trait (lower right).~\cite{dodds2025ArchetypometricsPragmateiaEmpirical}.
  }
  \label{fig:straight-queer.queer-aligned}
\end{figure*}

\begin{figure*}[t]
  \centering
    \includegraphics[width=\textwidth]{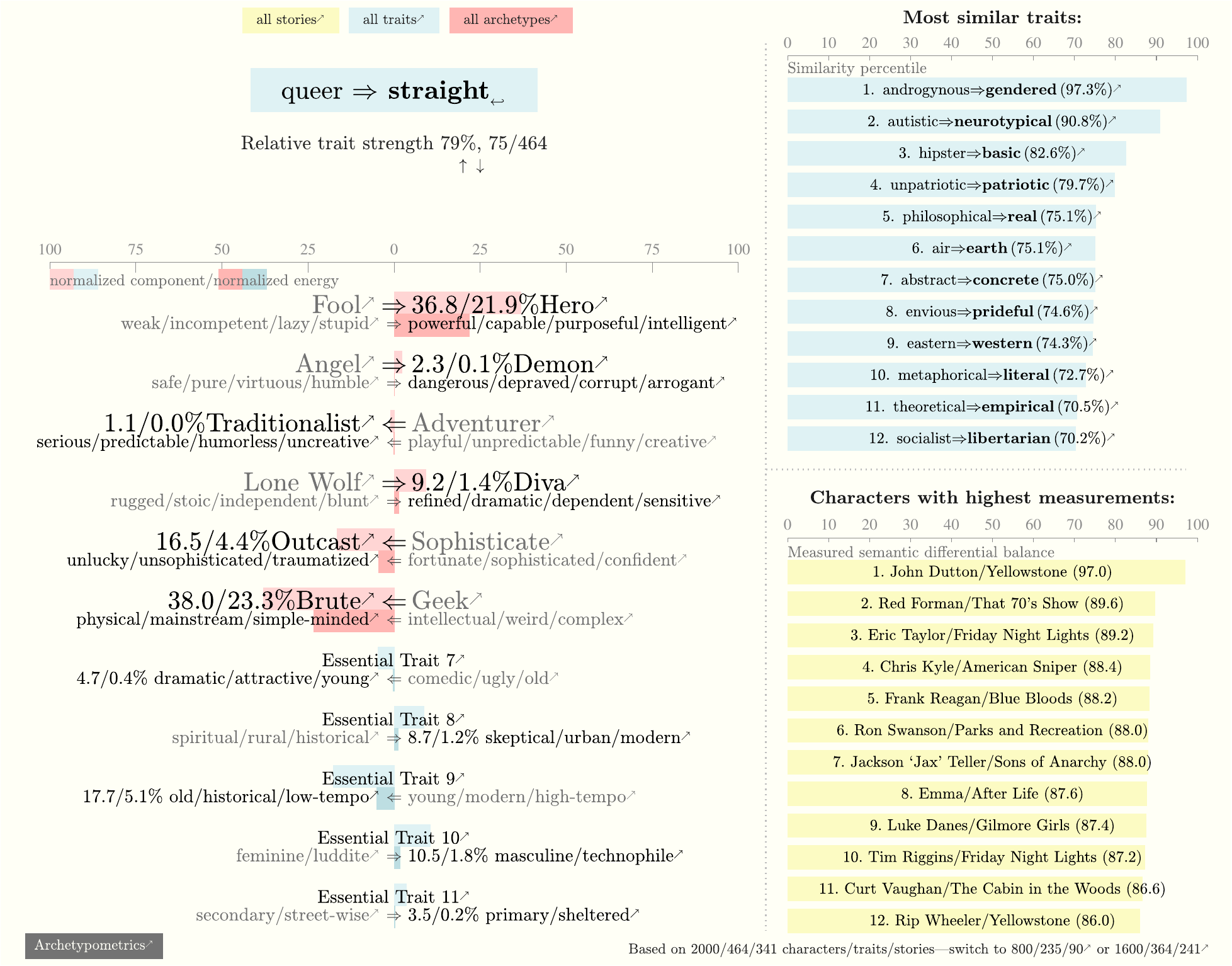}
  \caption{\traitlinksimple{straight}{queer} trait's archetype configuration pointing in the perceived \textbf{straight} direction. Components include an archetype breakdown for the trait (left), the 12 closest traits in order by distance (upper right), and characters with highest measurement on this trait (lower right).~\cite{dodds2025ArchetypometricsPragmateiaEmpirical}.}
  \label{fig:straight-queer.straight-aligned}
\end{figure*}

% research questions
The media portrayals and audience perceptions analyzed in the present work are constructed by people with varied experiences, prejudices, and opinions developed within historical contexts. These perceptions often correlate with negatively connoted or charged traits~\cite{weber2026QueerNLPCritical}, which we also observed in previous work (traits most alike \traitlinksimple{straight}{queer} in the direction of queer include \traitlinksimpleright{patriotic}{unpatriotic} in the direction of unpatriotic as well as \traitlinksimpleright{wise}{foolish} in the direction of foolish, showing some of the previously mentioned historical perception; see Figs.~\ref{fig:straight-queer.queer-aligned} and ~\ref{fig:straight-queer.straight-aligned})~\cite{dodds2025ArchetypometricsPragmateiaEmpirical}.
We find, however, that the top queer-aligned characters are also often \archetype{Hero} archetypes, resulting in a paradox. We analyze archetypes across subsets of straight- and queer-aligned characters in this foundational work~\cite{dodds2025ArchetypometricsPragmateiaEmpirical,dodds2025ArchetypometricsDataset} to better understand queer and straight perception in a high-dimensional archetype space. We ask what other traits and archetypes differentiate characters along the poles of this differential, organizing our exploration into
i) Group and trait description,
ii) Distance of other traits to \traitlinksimple{straight}{queer},
and
iii) Archetype discussion.

\section{Methodology}
\label{sec:straight-queer.methods}

\subsection{Datasets}

Both datasets are publicly available.
The Archetypometrics dataset is available on \href{https://zenodo.org/records/16953724}{Zenodo}.
The Fandom dataset is available by requesting a community database dump via \url{https://lgbtqia-characters.fandom.com/wiki/Special:Statistics}.
Project code (and another way to access the data) is available on \href{https://gitlab.com/amafehr/queer_archetypes}{Gitlab}.

% Archetypes
Our dataset derives from previous work~\cite{dodds2025ArchetypometricsPragmateiaEmpirical} based on a total of about 72 million ratings on OpenPsychometrics using its online Statistical ``Which Character'' quiz~\cite{openpsychometrics}. Survey respondents rated $2,000$~characters in $341$~stories (television, literature, and film) across $464$~traits.
Character traits are represented by semantic differentials of paired adjectives (e.g., \traitlinksimple{straight}{queer},
\traitlinksimple{playful}{serious},
and
\traitlinksimple{heroic}{villainous}).\footnote{
    The original study had 500 semantic differential traits but we removed 1 duplicate trait and 35 traits that were presented in emojis.
}
Self-selected volunteers rated fictional characters
they had knowledge of on a 100-point scale~\cite{DevelopmentofStatisticalCharQuiz}.
The number of ratings for individual traits and characters varies, with \traitlinksimple{straight}{queer} having on average $86$ ratings per character. Characters sometimes score highly in a specific direction; for example, we use the term `straight-aligned' to mean scores leaning toward straight in the \traitlinksimple{straight}{queer} pair.
Applying Singular Value Decomposition to these ratings allows us to identify six
archetypes~\cite{dodds2025ArchetypometricsPragmateiaEmpirical,zimmerman2025LocalityRelationMeaning}
which are ordered by size (singular value)
and further break down into
three primary and three secondary archetypes.
This framework further allows for character classification as dual or triple archetypes (e.g., a \archetype{Diva-Demon} or a \archetype{Demon-Hero-Traditionalist}.

% Fandom's LGBTQIA+ chars wiki
To identify subsets of characters that can plausibly be described as ``straight'' or ``queer'', we parse the website Fandom's LGBTQIA+ Characters Wikia XML data~\cite{lgbtcharacterswikia}; hereafter referred to as Fandom. Self-selected fans add characters to this wiki, and entries contain varying degrees of character history and canonical validity. The resulting crowd-sourced Fandom dataset contains $9,680$ character entries, including at least non-empty entries for the character's name, fictional story, and sexuality.

We identify 125 ``canonically'' queer characters appearing in both datasets (referred to as `\fandomqueer'). We treat these crowd-sourced labels from Fandom as a plausible proxy for the existence of the character's queer identity in the source text.
Some examples of these characters include:
\characterlinksimple{Mean-Girls-Damian-Leigh}{Damian Leigh} (\storylinksimple{Mean-Girls}{Mean Girls}),
\characterlinksimple{Lucifer-Mazikeen}{Mazikeen} (\storylinksimple{Lucifer}{Lucifer}), and
\characterlinksimple{The-100-Clarke-Griffin}{Clarke Griffin} (\storylinksimple{The-100}{The 100}).

\subsection{Procedures and measures}
%%%% PROCEDURES
We explore different views of relevant characters by constructing multiple samples using the variation in the \traitlinksimple{straight}{queer} trait and similarity metrics in the full archetype space matrix (Euclidean distance, inner product, and cosine similarity). We take the $n$ lowest scores (in the straight-aligned direction) and the $n$ highest scores (in the queer-aligned direction) for subsets of \straightalign\ and \queeralign\ characters. We find queer- and straight-aligned nearest neighbors by taking the top \textit{m} seed characters from the trait's polar ends and comparing a character's Euclidean distance to all other characters, keeping the top \textit{k} closest characters.

\begin{figure*}[t]
    \centering
    \includegraphics[width=\textwidth]{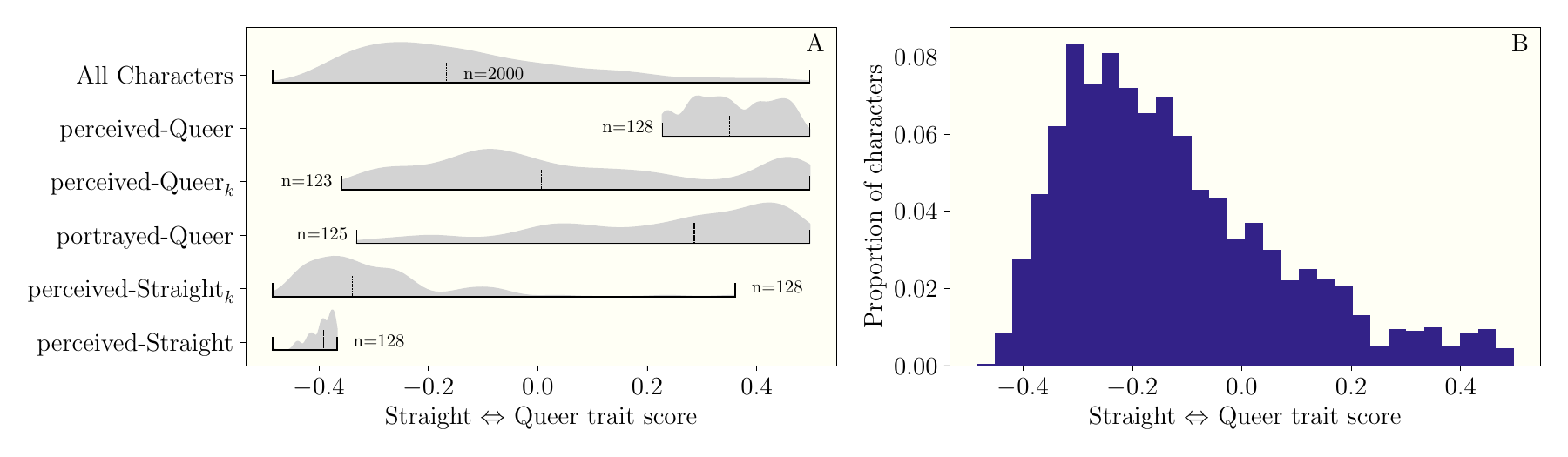}
\caption{
\textbf{A.}
Distributions on the \traitlinksimple{straight}{queer} trait with subset size $n$ and vertical bars denoting minimum, median, and maximum scores.
\textbf{B.}
The distribution for all 2000 characters
on the \traitlinksimple{straight}{queer} trait.
}
\label{fig:distribution-violins}
\end{figure*}

Applying bootstrap sampling and the radius of gyration ($R_g$) around sample size in simulations, stable choices of \textit{m} and \textit{k} emerge to form the final subset sizes (see Fig.~\ref{fig:distribution-violins} and \S\ref{sec:straight-queer.appendix.char_subsets.subsetcreation}). Our final subsets are named `\fandomqueer' (from Fandom), `\straightalign', and `\queeralign', with `\queerneigh' and `\straightneigh' for neighbors. Example characters for these groups include:
\characterlinksimple{Yellowstone-John-Dutton}{John Dutton} (\storylinksimple{Yellowstone}{Yellowstone}) and
\characterlinksimple{Breaking-Bad-Hank-Schrader}{Hank Schrader} (\storylinksimple{Breaking-Bad}{Breaking Bad}) in \textbf{\straightalign};
\characterlinksimple{Schitts-Creek-David-Rose}{David Rose} (\storylinksimple{Schitts-Creek}{Schitt's Creek}) and
\characterlinksimple{Unbreakable-Kimmy-Schmidt-Titus-Andromedon}{Titus Andromedon} (\storylinksimple{Unbreakable-Kimmy-Schmidt}{Unbreakable Kimmy Schmidt}) in \textbf{\queeralign};
\characterlinksimple{New-Girl-Schmidt}{Schmidt} (\storylinksimple{New-Girl}{New Girl}) and
\characterlinksimple{Marvel-Cinematic-Universe-Gamora}{Gamora} (\storylinksimple{Marvel-Cinematic-Universe}{Marvel}) in \textbf{\queerneigh};
\characterlinksimple{Supernatural-Dean-Winchester}{Dean Winchester} (\storylinksimple{Supernatural}{Supernatural}) and
\characterlinksimple{Brokeback-Mountain-Ennis-Del-Mar}{Ennis Del Mar} (\storylinksimple{Brokeback-Mountain}{Brokeback Mountain}) in \textbf{\straightneigh}.

%%%% STATISTICAL MEASURES
We compare subsets across the \traitlinksimple{straight}{queer} semantic differential as well as across archetype pairs. We calculate closeness of all rated traits to the \traitlinksimple{straight}{queer} trait using inner product and Euclidean distance.

\section{Results}
\label{sec:straight-queer.results}

\subsection{Groups and traits}
\label{sec:straight-queer.results.groups-and-traits}

% S-Q description
We characterize the \traitlinksimple{straight}{queer} distribution in Fig.~\ref{fig:distribution-violins}B.
Among all $464$ traits, the \traitlinksimple{straight}{queer}
trait ranks $79^{th}$ in coefficient of variation (CV)
and $279^{th}$ in variance.
For context, \traitlinksimple{persistent}{quitter} and \traitlinksimple{believable}{poorly-written} rank $1^{st}$ and $2^{nd}$ in CV, and \traitlinksimple{right-brained}{left-brained} and \traitlinksimple{tautology}{oxymoron} rank $1^{st}$ and $2^{nd}$ in variance.
This trait moderately skews in the queer-aligned direction (skewness $=0.91$).

% Groups/subsets
Next, we describe the data subsets. Characters in other subsets overlap with the \fandomqueer\ group derived from Fandom. The proportion of \fandomqueer\ membership overall in each group is:
\\
0\% (\straightalign),
\\
58\% (\queeralign),
\\
25\% (\queerneigh),
\\
and
4\% (\straightneigh).
\\
Fig.~\ref{fig:distribution-violins} and Tab.~\ref{tab:subset-descriptives} describe the \traitlinksimple{straight}{queer} trait relative to all $2000$ characters and the data subsets. \queeralign\ skews the most in the queer direction while \straightalign\ and \straightneigh\ skew the most in the straight direction. \queerneigh\ and \fandomqueer\ show much higher trait variance than the other groups.

\begin{table}[t]
  \centering
  \begin{tabular}{lccc}
  \hline \textbf{Subset} & \textbf{Median} & \textbf{$\sigma$} & \textbf{CV} \\
  \hline
  All Characters & -0.17 & 0.20 & -1.59 \\
  \queeralign & 0.35 & 0.08 & 0.21 \\
  \queerneigh & 0.01 & 0.26 & 4.03 \\
  \fandomqueer & 0.29 & 0.21 & 0.92 \\
  \straightneigh & -0.34 & 0.13 & -0.42 \\
  \straightalign & -0.39 & 0.02 & -0.06 \\
  \hline
  \end{tabular}
  \caption{
  Group statistics for the \traitlinksimple{straight}{queer} trait. For all characters, $\mu=-0.13$, $\sigma=0.20$.
  }
  \label{tab:subset-descriptives}
\end{table}

% Top traits in each subset: unique or exclusive
We next report the top-occurring traits by character subset.
Counts of \straightalign\ top traits show the most spread, with its top 3 being \traitlinksimpleleft{handshakes}{hugs},
\traitlinksimpleleft{gendered}{androgynous}, and
\traitlinksimpleleft{hygienic}{gross}.
\queeralign's top trait counts are most consolidated around its first trait,
\traitlinksimpleright{straight}{queer}, with next top traits of
\traitlinksimpleright{f***-the-police}{tattle-tale} and
\traitlinksimpleright{hygienic}{gross}.
\queerneigh\ shows a similar consolidation around
\traitlinksimpleright{straight}{queer}, with some difference in top traits of
\traitlinksimpleright{egalitarian}{racist} and
\traitlinksimpleright{hygienic}{gross}.
\straightneigh\ shows different top traits and slightly more consolidation than \straightalign, with
\traitlinksimpleleft{hygienic}{gross},
\traitlinksimpleleft{handshakes}{hugs},
\traitlinksimpleleft{militaristic}{hippie}, and
\traitlinksimpleleft{jock}{nerd}.
\fandomqueer's top traits are similar to \queeralign\ and \queerneigh\ with higher counts for its second top trait:
\traitlinksimpleright{straight}{queer},
\traitlinksimpleright{hygienic}{gross}, with a tie among
\traitlinksimpleright{bold}{shy},
\traitlinksimpleright{f***-the-police}{tattle-tale}, and
\traitlinksimpleright{stable}{unstable}.

\subsection{Trait distance}
\label{sec:straight-queer.results.trait-distance}

% Distance from S-Q to other traits: Measuring distance between traits: euclidean norm for distance and inner product for magnitude
We next report the top $30$ traits in all $2000$ characters that are closest to \traitlinksimple{straight}{queer} by Euclidean distance and inner product metrics (\S\ref{sec:straight-queer.appendix.traitdistance}).

Many of the traits in the full list evoke an amorphous, intangible, or unknowable quality aligning in the queer direction. The traits closest to queer alignment include:
\traitlinksimpleright{real}{philosophical}, \traitlinksimpleright{earth}{air}, \traitlinksimpleright{concrete}{abstract}, \traitlinksimpleright{literal}{metaphorical}, \traitlinksimpleright{straightforward}{cryptic}, \traitlinksimpleright{repetitive}{varied}, \traitlinksimpleright{believable}{poorly-written}, \traitlinksimpleright{permanent}{transient}, \traitlinksimpleright{pointed}{random}.

Other close traits in the queer-aligned direction evoke association with an out-group membership, such as: \traitlinksimpleright{patriotic}{unpatriotic}, \traitlinksimpleright{loyal}{traitorous}, \traitlinksimpleright{western}{eastern}, \traitlinksimpleright{libertarian}{socialist}, \traitlinksimpleright{Roman}{Greek}, \traitlinksimpleright{English}{German}, \traitlinksimpleright{gendered}{androgynous}.

We consider a few of the appearance-related traits more closely, considering that they surface in top or exclusive traits within multiple character subsets. In trait space, \traitlinksimpleright{macho}{metrosexual}  ($129^{th}$ closest) and \traitlinksimpleright{hygienic}{gross} ($246^{th}$ closest) are not close to \traitlinksimpleright{straight}{queer}. By subset, these traits increase in importance, with \semdiff{hygienic}{gross} showing up as a top trait across subsets and \semdiff{macho}{metrosexual} appearing exclusively in the top character traits of \queeralign.

\begin{figure*}[t]
\includegraphics[width=\textwidth]{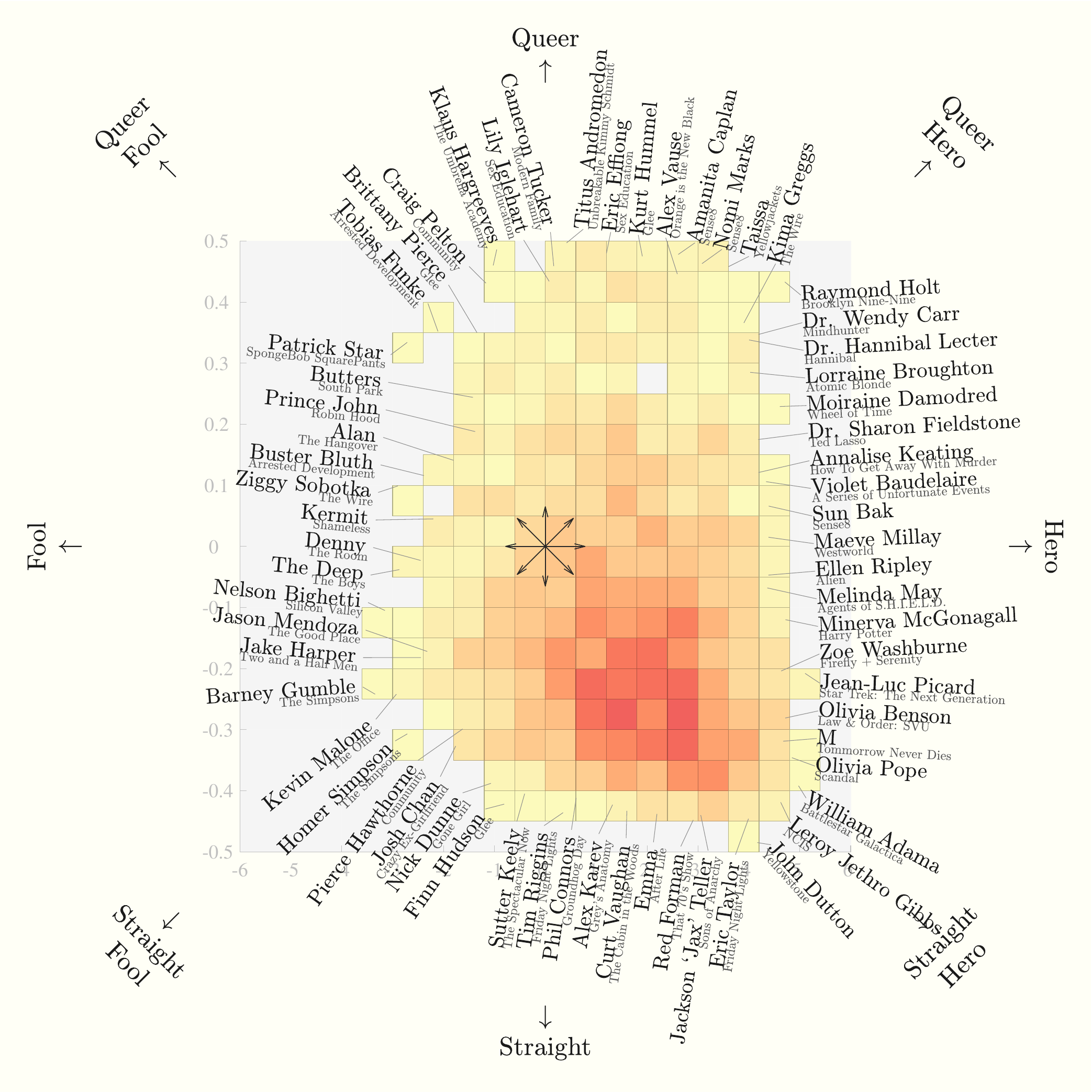}
    \caption{
    An ousiogram---an automatically annotated histogram~\cite{dodds2026OusiometricsEssenceMeaning}---of essential characteristics with \traitlinksimple{straight}{queer} trait scores on the vertical axis and \archetypesemdiff{1} scores on the horizontal axis.
    A darker cell color indicates more characters fall within that range.
    }
    \label{fig:ousiogram-archetypes-straight-queer-01}
\end{figure*}

\subsection{Archetype dimensions and membership}
\label{sec:straight-queer.results.archetype-dims-membership}

We now explore archetypes across character subsets and their relationship to the \traitlinksimple{straight}{queer} trait.

Archetype dimension results proceed as: i) the trait $\times$ dimension level for all characters before moving into subset variance and ii) groups' archetype labels with ousiograms of exemplars.

\subsubsection{Trait $\times$ Archetype dimension}

\begin{figure*}[t]
  \centering
  \includegraphics[width=\textwidth]{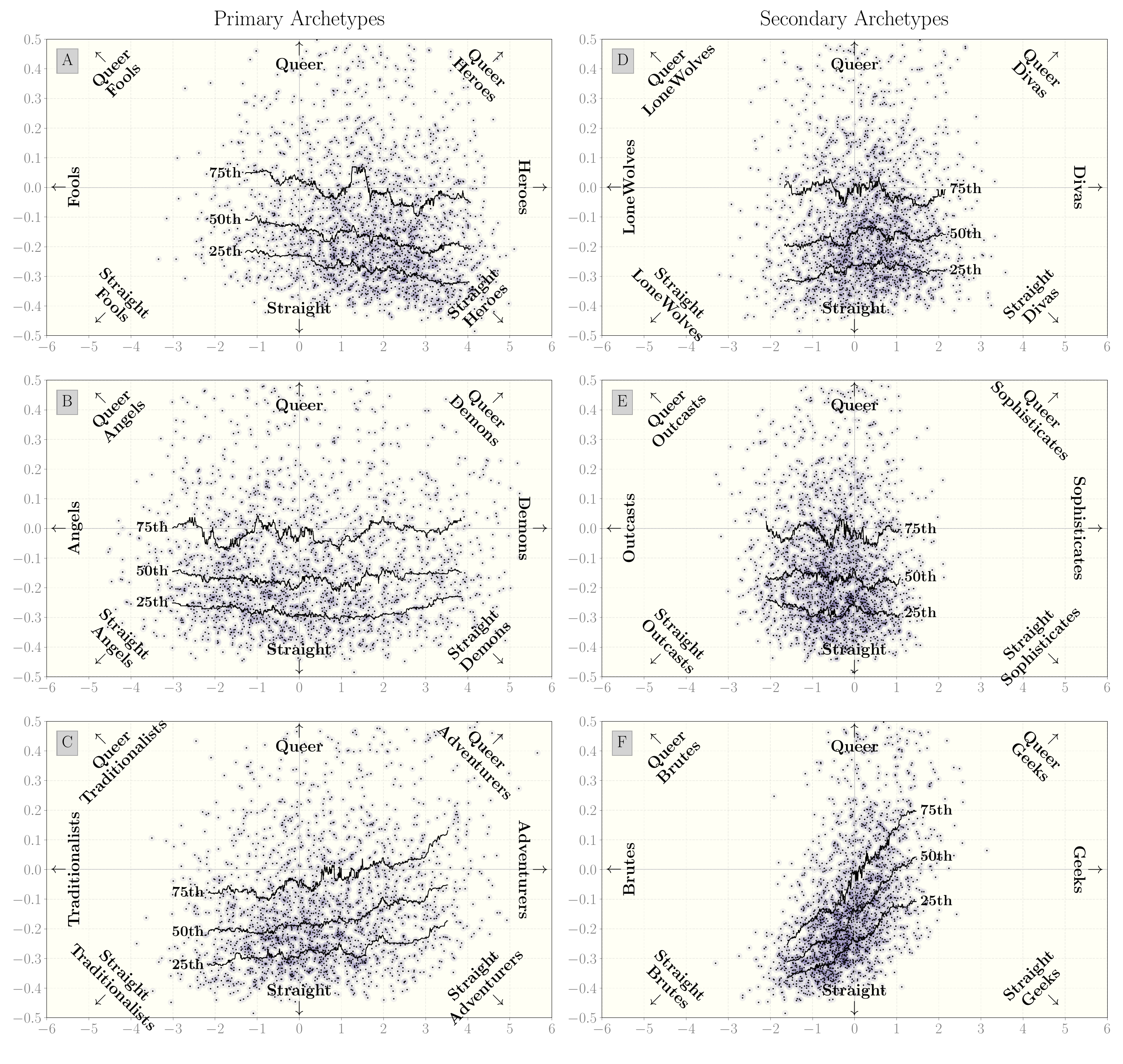}
  \caption{
  All characters as located on the \traitlinksimple{straight}{queer} trait
  versus each of the six fundamental archetype dimensions. Labeled quartile lines show a shifting window median on that dimension at the $25^{th}$, $50^{th}$, and $75^{th}$ quartiles using a window size of $n=200$.}
  \label{fig:quartile-plot}
\end{figure*}

Figure~\ref{fig:quartile-plot} shows all characters by quartile on \traitlinksimple{straight}{queer} scores and by archetype dimension. \archetypesemdiff{1} demonstrates that characters with higher scores in the straight direction tend toward the \archetype{Hero} archetype (away from \archetype{Fool}), also seen in Fig.~\ref{fig:ousiogram-archetypes-straight-queer-01}.

\archetypesemdiff{3} shows that characters with higher scores in the queer direction tend toward the \archetype{Adventurer} archetype (away from \archetype{Traditionalist}).
\archetypesemdiff{6} is the starkest division for the trait: characters with higher relative scores in the queer direction tend toward the \archetype{Geek} side of the dimension (away from \archetype{Brute}).
The relationship between the \traitlinksimple{straight}{queer} and the \archetypesemdiff{2}, \archetypesemdiff{4}, and \archetypesemdiff{5} archetype pairs is largely invariant. See \S\ref{sec:straight-queer.appendix.arch-dims} for further summary description of our data subsets on each archetype dimension.

We further explore relationships between the \traitlinksimple{straight}{queer} trait and character canonical gender, based on a previous labeling of this dataset~\cite{beauregard2026ArchetypesGenderFiction}. We find that while the \traitlinksimple{straight}{queer} trait does not display a strong relationship across the Diva archetype, either for all 2000 characters or for the current character subsets, queer men (that is, canonically queer and canonically male characters) and queer-aligned men tend strongly towards the Diva archetype. Examples of characters tending to this direction include \characterlinksimple{Modern-Family-Cameron-Tucker}{Cameron Tucker} (\storylinksimple{Modern-Family}{Modern Family}),
\characterlinksimple{Glee-Kurt-Hummel}{Kurt Hummel} (\storylinksimple{Glee}{Glee}), and
\characterlinksimple{Schitts-Creek-David-Rose}{David} (\storylinksimple{Schitts-Creek}{Schitt's Creek}). See \S\ref{sec:straight-queer.appendix.arch-dims} for more results and methodology on this comparison.

\subsubsection{Group archetypes and ousiograms}

Next, we gather the unique archetypes that occur in a target group. \queeralign\ (78) and \straightalign\ (74) show the highest number of unique archetype labels, with each neighboring group showing the least unique label totals (\straightneigh\ has 61). \queerneigh\ is the lowest (48), showing the most repetition of labels. A lower count of unique archetypes corresponds to higher proportions of those archetypes appearing more than once.

% top archetypes
Characters' assigned archetype labels~\cite{dodds2025ArchetypometricsPragmateiaEmpirical} are collected in Tab.~\ref{tab:archetype_proportions}. The following archetype proportions are consistent with our quartile trends.
Across subsets, the top 2 archetypes are \archetype{Hero} (appearing most for \straightneigh\ (0.48) and \straightalign\ (0.26)) and \archetype{Adventurer} (appearing most for \queeralign\ (0.48)).
Lone Wolf variants occur most frequently in top lists of \straightalign\ and \straightneigh\ while Diva variants tend to occur more in \fandomqueer\ and \queerneigh.
Brute variants appear only in the top lists of \straightalign\ and \straightneigh\ while Geek variants similarly appear only in top lists of \queeralign, \queerneigh, and \fandomqueer.

% OUSIOS AND MILK, MILK'S FAVORITE COOKIE
To contextualize trends like high Hero, Adventurer, and Brute/Geek presence among these subsets, we interpret ousiograms, histograms of essential archetype components showing exemplar characters. Like the quartile plots, these histograms cross \traitlinksimple{straight}{queer} with each archetype dimension (\S\ref{sec:straight-queer.appendix.arch-dims}), where directionality on the trait (e.g., `Queer') and archetype dimension (e.g., `Hero') is characterized.

% Hero
Starting with \archetypesemdiff{1} (Fig.~\ref{fig:ousiogram-archetypes-straight-queer-01}), the higher-count cells trending in the straight and hero directions are consistent with quartile and trait dispersion results. Some exemplar characters surface:
\characterlinksimple{Brooklyn-Nine-Nine-Raymond-Holt}{Raymond Holt} (\storylinksimple{Brooklyn-Nine-Nine}{Brooklyn Nine-Nine}) as a `Queer \archetype{Hero}'
and
\characterlinksimple{NCIS-Leroy-Jethro-Gibbs}{Gibbs} (\storylinksimple{NCIS}{NCIS}) as a `Straight Hero'
are leaders in their stories, often showing up as stoic in character.
\characterlinksimple{Arrested-Development-Tobias-Funke}{Tobias} (\storylinksimple{Arrested-Development}{Arrested Development}) as a `Queer \archetype{Fool}'
and
\characterlinksimple{Unbreakable-Kimmy-Schmidt-Titus-Andromedon}{Titus} veering slightly `Queer \archetype{Demon}'
tend to introduce comedic or dramatic action in their respective story lines.
\characterlinksimple{The-Simpsons-Homer-Simpson}{Homer} (\storylinksimple{The-Simpsons}{The Simpsons}) in the `Straight \archetype{Fool}' direction is a main character often making bad or comedic decisions.
Cross-referencing the other dimensions, Titus Andromedon (the strongest queer-aligned character) veers slightly toward Demon (Fig.~\ref{fig:ousiogram-archetypes-straight-queer-02}) and Adventurer (Fig.~\ref{fig:ousiogram-archetypes-straight-queer-03}) but is in the middle of the \archetypesemdiff{5} dimension with
\characterlinksimple{Glee-Kurt-Hummel}{Kurt Hummel} (\storylinksimple{Glee}{Glee},
Fig.~\ref{fig:ousiogram-archetypes-straight-queer-05}); finding themselves between Outcast and Sophisticate.
Titus and Kurt both showcase refined taste or characteristics while occasionally needing to be brave and take initiative.
In the `Straight Hero' direction is also \characterlinksimple{Yellowstone-John-Dutton}{John Dutton} (\storylinksimple{Yellowstone}{Yellowstone}) who is the strongest straight-aligned character; he veers slightly Traditionalist and Brute as well as Lone Wolf.

% Adventurer
Given the queer-aligned trend in the \semdiffboldright{Traditionalist}{Adventurer} direction, we interpret Fig.~\ref{fig:ousiogram-archetypes-straight-queer-03}. This histogram's counts populate more in the straight and \archetype{Traditionalist} directions, with examples such as
\characterlinksimple{Parks-and-Recreation-Ron-Swanson}{Ron Swanson} (\storylinksimple{Parks-and-Recreation}{Parks and Rec}) who tends toward \archetype{Lone-Wolf} and \archetype{Traditionalist} as well, aligning with Ron's actions which show a certain resistance to change and his abiding desire to be left alone.
\characterlinksimple{The-Umbrella-Academy-Klaus-Hargreeves}{Klaus Hargreeves} (\storylinksimple{The-Umbrella-Academy}{The Umbrella Academy})
and
\characterlinksimple{Shadow-and-Bone-Jesper-Fahey}{Jesper Fahey} (\storylinksimple{Shadow-and-Bone}{Shadow and Bone})
are `Queer Adventurers', with Klaus slightly veering in the `Queer Fool' direction as well. Klaus often gets himself into untenable situations (the Fool tendency), while both Klaus and Jesper have complex histories that showcase growth, loyalty, and courage; they often take agency within their stories.

% Brute/Geek
On the \archetypesemdiff{6} dimension, the histogram shows almost diagonal coloring reflecting some density in both the `Queer Geek' and `Straight Brute' directions. Characters tending toward `Queer Geek' include:
\characterlinksimple{Sense8-Nomi-Marks}{Nomi Marks} (\storylinksimple{Sense8}{Sense8})
and
\characterlinksimple{Arcane-Viktor}{Viktor} (\storylinksimple{Arcane}{Arcane}),
both characters known for technical prowess or ingenuity which collide with major events in their stories.
Inversely, characters tending toward `Straight Brute' include:
\characterlinksimple{The-Incredibles-Bob-Parr}{Bob Parr} (\storylinksimple{The-Incredibles}{The Incredibles})
and
\characterlinksimple{That-70s-Show-Red-Forman}{Red Forman} (\storylinksimple{That-70s-Show}{That '70s Show}), characters tending to display more traditional masculinity or prioritizing physical strength.

\section{Discussion}
\label{sec:straight-queer.discussion}

Three main results emerge from our analysis at the levels of characters, groups, traits, and archetypes.

First, there is evidence of narrower perceptions of queerness at the trait level.
Queer-aligned groups show more variance on the \traitlinksimple{queer}{straight} trait itself than straight-aligned groups; however, straight-aligned groups tend to show many more exclusive traits among their characters, suggesting a breadth of character portrayal.
Specifically, \straightalign\ shows more varied top traits; the three queer-aligned groups show more top trait consolidation within the single semantic differential \traitlinksimple{straight}{queer} than any one semantic differential for the straight-aligned groups.

A more multifaceted picture emerges at the archetype level, with queer-aligned characters tending to be perceived as adventuresome while straight-aligned characters tend to be perceived as more strongly heroic.
Queer-aligned characters show more frequent \archetype{Adventurer} and \archetype{Geek} archetypal perception, while straight-aligned characters show more frequent \archetype{Hero}, \archetype{Brute}, and \archetype{Lone-Wolf} perception. In fact, \archetype{Brute}-related archetypes only occur in the straight-aligned groups, while \archetype{Geek}-related ones only occur in the queer-aligned groups. Notably, \archetype{Hero} is not absent among queer-aligned characters but appears in dual or triple archetypes. Further, \queerneigh\ characters lean the most toward \archetype{Adventurer} while the \queeralign\ characters show higher heroic tendency than their neighbors. These perceptions suggest prominent characters are more strongly written or an effect of frequency and memorability of characters (who are then perceived as more heroic). It is possible that neighbors complement more distinct, heroic, or main characters.

In line with this finding, most characters in either group or neighborhood are not flat or of a single archetype (e.g., \archetype{Hero}, \archetype{Fool}, \archetype{Demon}), and instead tend to inhabit dual or triple archetypes. Examples include
\characterlinksimple{Stranger-Things-Robin-Buckley}{Robin Buckley} (\storylinksimple{Stranger-Things}{Stranger Things})
as a \archetype{Lone Wolf-Hero-Adventurer} from \queeralign\ as well as
\characterlinksimple{That-70s-Show-Red-Forman}{Red Forman} (\storylinksimple{That-70s-Show}{That '70s Show})
as a \archetype{Demon-Hero-Traditionalist} from \straightalign.
This tendency suggests rich characterizations across the perceived straight and queer spectrum, constrained by some larger group patterns of dominant archetypes.

Accordingly, we see evidence of stereotypical character perceptions and emergent biases at the trait level.
Characters perceived as queer-aligned as well as canonically queer often have \semdiffboldright{hygienic}{gross} as a top trait, with some appearance of \semdiffboldright{stable}{unstable} (see \S\ref{sec:straight-queer.appendix.traitdistance}).
The queer direction is associated with the philosophical, theoretical, abstract, and metaphorical traits, as well as more negatively connoted descriptors like poorly-written and cryptic.
These descriptors evoke the queer direction of the trait as intangible, uncertain, or unknowable.
Through these associations, queerness seems to be perceived in terms of its anti-normativity (where heterosexuality is normative)~\cite{jagose2015TroubleAntinormativity} such that perceptions of queerness may be met with less certainty~\cite{fluit2024SocialMarginalizationScoping}.
This explanation is in line with social marginalization---the social process of ``Othering.’’ Othering entails individuals or certain groups being systematically excluded because of societal norms or beliefs. Marginalization has disadvantaged outcomes, including (but not limited to) discrimination and social isolation~\cite{fluit2024SocialMarginalizationScoping}.
Otherness relies on dualistic thinking, opposing `us’ and `them’ as `self’ and `other,’ with the other often being stereotyped~\cite{udah2019SearchingPlaceBelong,canales2000OtheringUnderstandingDifference}.
Othering can be subtle, suggesting alienness~\cite{udah2019SearchingPlaceBelong,canales2000OtheringUnderstandingDifference}, lower importance, or invisibility~\cite{fluit2024SocialMarginalizationScoping} of the non-normative group, reflected in the intangibility descriptors we find associated with queerness perceptions.

Further bias emerges relative to out-group memberships which align with historical American values and perceived threat. These associations imply moral preference or a cultural majority's point of view, including descriptors such as unpatriotic, socialist, eastern, traitorous~\cite{johnson2023LavenderScareCold}, and androgynous.
Relative to a Western, American cultural context, these associations are consistent with social marginalization of undesirable or threatening out-groups~\cite{alorainy2019EnemyUsDetecting}.

Exploration into canonical men who are queer-aligned surfaces specific stereotypical perception; these characters align more in the direction of being foolish and diva-like, and we see variants of the \archetype{Diva} archetype occurring more in \fandomqueer\ and \queeralign, consistent with some common (biased) conceptions of queer men as divas or comedic relief~\cite{soto_sanfiel_perceptions_2024}. A potential outlier exists such that Queer-aligned canonical male characters range closer to \archetype{Angel} than other groups do.

\subsection{Limitations and future work}
\label{sec:straight-queer.discussion.limits}

% Studying identities
The Open-Source Psychometrics data encodes straightness and queerness as static points while human identities are far more fluid, activated by context~\cite{cohen2005PunksBulldaggersWelfare}. While we explore perceptions through this extant data continuum to avoid activating only a binary characteristic (straight versus queer)~\cite{cohen2005PunksBulldaggersWelfare}, flattening is inherent to data without time series measurement but still offers a limited view of lived experiences and identities. The \traitlinksimple{straight}{queer} differential sets up a comparison that may, in its design, garner more varied ratings of ``queer'' by setting it in opposition to ``straight''~\cite{jagose2015TroubleAntinormativity}. Future work could design differentials intentionally splitting straightness from queerness to understand underlying variance in perceptions.

Further, media representation has tended to skew binary and cis-normative,~\cite{devinney2022TheoriesGenderNLP} which necessarily impacts the queer representation in this particular collection of film, TV, and literature. Research could aim to understand media portrayals by selecting traits to study binary and cis-normative skew (e.g., \semdiff{gendered}{androgynous}).

An additional challenge in data-driven work can be having enough context to interpret results holistically. Character design can playfully subvert stereotypes, and queer storytelling may rely on this intentional subversion~\cite{munoz1997WhiteBeAngry}. When stereotypes are used for this purpose in a performance, character portrayals are not incidentally harmful. The difficulty arises in parsing this storytelling from audience perceptions and their understanding of narrative intent. Future data collection efforts could solicit open-ended perceptions of queer character representation for greater NLP context.

% Limitation of human judgment & Western skew
Additionally, some limitations exist regarding human judgment writ large~\cite{bowker1999SortingThingsOut} and, specifically, in a Western context. Human coding of canonical sexuality relies on knowledge and tacit judgment of subtext.~\cite{polanyi2012PersonalKnowledge} Fandom's reported character sexuality may occasionally rely on alternative universe versions of characters (such as \characterlinksimple{South-Park-Kenny-McCormick}{Kenny McCormick} from \storylinksimple{South-Park}{South Park}), or fan speculation (such as \characterlinksimple{Ted-Lasso-Jamie-Tartt}{Jamie Tartt} from \storylinksimple{Ted-Lasso}{Ted Lasso}), offering judgments which differ from canonical character representations. Contextually, the archetypometrics dataset contains predominantly Western, high-grossing films and TV shows, which could skew the raters' perceptions of queerness. Conversely, focusing on popular shows likely captures a more widely recognized view of queer representation (at least by Western audiences). Future work should expand into popular works of other societies to understand how patterns of representation may diverge across cultural contexts.

% Sampling method and demographics
Lastly, while we validate the construction method of our data subsets (\S\ref{sec:straight-queer.appendix.char_subsets.subsetcreation}), different thresholds of neighborhood definition would likely impact results.
Nevertheless, with this sampling method, we uncover latent characterizations of queer-aligned characters through an archetypal, contextualized lens, which could be extended to explore other demographically motivated traits (e.g., \traitlinksimple{rural}{urban}, \traitlinksimple{old}{young}) as well as personality traits (e.g., \traitlinksimple{extrovert}{introvert}, \traitlinksimple{open-minded}{close-minded}).

\section{Concluding remarks}
\label{sec:straight-queer.concludingremarks}

% 1. Broader, social/societal statement
Portrayal of queer identities in media is on the rise, though stereotypes and tropes are often employed as vehicles to represent characters of these identities.
% 2. Connect to big finding or main method of paper through which we find this exemplified by stating just generally what social persists
The oft-used character constructions of archetypes can encode these societally-held biases we see in stereotypes and tropes.
% 3. Through archetypometrics, we observe that... main finding
Through archetypometrics, we study what characters' traits depict and which archetypes they surface.
Characters tend to show a narrow perception at the trait level, with negative associations tied to queerness. Negative connotations related to historical stereotypes surface in the examination of the specific traits that span perceived straightness to queerness, characterizing the queer-aligned direction as gross, theoretical, or as a moral or political ``other.'' However, when considering archetype dimensions, queer representations are more positive---we note this finding as a paradox.
Perceived queer characters and their nearest neighbors show a breadth of representation in archetype space, with adventuresome, geekish, and heroic variants appearing. To a greater degree than perceived queer characters, perceived straight characters often emerge as heroic; this group also has tends toward \archetype{Brute} and \archetype{Lone-Wolf} archetypal variants.

% 4. Broader implication of what we have shown / societal values or warning exemplified.
The \traitlinksimple{straight}{queer} trait reveals societal biases, while individual characters escape these narrow definitions by inhabiting a variety of archetypal characteristics.
Accordingly, these representations endorse certain positive perceptions of the queer community, potentially consistent with effects that reduce prejudicial attitudes~\cite{birchmore2022ExploringBoundariesParasocial}.

\bigskip
\section*{Acknowledgments}

We are grateful for discussion with Alejandro Javier Ruiz Iglesias and Isabelle T. Smith. 
We acknowledge support from
MassMutual,
National Science Foundation Award \#2242829
(Science of Online Corpora, Knowledge, and Stories),
and
a philanthropic gift from an anonymous source.

\bibliography{straight-queer}

@book{polanyi2012PersonalKnowledge,
  title = {Personal {{Knowledge}}: {{Towards}} a {{Post-Critical Philosophy}}},
  author = {Polanyi, Michael},
  year = 2012,
  publisher = {Routledge},
  address = {London},
  doi = {10.4324/9780203442159},
  isbn = {978-0-203-44215-9}
}

@book{bowker1999SortingThingsOut,
  title = {Sorting {{Things Out}}: {{Classification}} and {{Its Consequences}}},
  shorttitle = {Sorting {{Things Out}}},
  author = {Bowker, Geoffrey C. and Star, Susan Leigh},
  year = 1999,
  publisher = {The MIT Press},
  doi = {10.7551/mitpress/6352.001.0001},
  urldate = {2026-07-09},
  isbn = {978-0-262-26907-0}
}

@article{canales2000OtheringUnderstandingDifference,
  title = {Othering: {{Toward}} an {{Understanding}} of {{Difference}}},
  shorttitle = {Othering},
  author = {Canales, Mary K.},
  year = 2000,
  journal = {Advances in Nursing Science},
  volume = {22},
  number = {4},
  pages = {16},
  issn = {0161-9268},
  urldate = {2026-07-06}
}

@article{udah2019SearchingPlaceBelong,
  title = {Searching for a {{Place}} to {{Belong}} in a {{Time}} of {{Othering}}},
  author = {Udah, Hyacinth},
  year = 2019,
  journal = {Social Sciences},
  volume = {8},
  number = {11},
  pages = {297},
  publisher = {Multidisciplinary Digital Publishing Institute},
  issn = {2076-0760},
  doi = {10.3390/socsci8110297},
  urldate = {2026-07-06}
}

@article{fluit2024SocialMarginalizationScoping,
  title = {Social Marginalization: {{A}} Scoping Review of 50 Years of Research},
  shorttitle = {Social Marginalization},
  author = {Fluit, Sam and {Cort{\'e}s-Garc{\'i}a}, Laura and {von Soest}, Tilmann},
  year = 2024,
  journal = {Humanities and Social Sciences Communications},
  volume = {11},
  number = {1},
  publisher = {Palgrave},
  issn = {2662-9992},
  doi = {10.1057/s41599-024-04210-y},
  urldate = {2026-07-02}
}

@article{alorainy2019EnemyUsDetecting,
  title = {``{{The Enemy Among Us}}'': {{Detecting Cyber Hate Speech}} with {{Threats-based Othering Language Embeddings}}},
  shorttitle = {``{{The Enemy Among Us}}''},
  author = {Alorainy, Wafa and Burnap, Pete and Liu, Han and Williams, Matthew L.},
  year = 2019,
  journal = {ACM Transactions on the Web (TWEB)},
  volume = {13},
  number = {3},
  pages = {14:1--14:26},
  publisher = {Association for Computing Machinery},
  issn = {1559-1131},
  doi = {10.1145/3324997},
  urldate = {2026-07-02}
}

@article{dodds2026OusiometricsEssenceMeaning,
  author =	 {Dodds, Peter Sheridan and Alshaabi, Thayer and
                  Fudolig, Mikaela Irene and Zimmerman, Julia Witte
                  and Lovato, Juniper and Beaulieu, Shawn and Minot,
                  Joshua R. and Arnold, Michael V. and Reagan, Andrew
                  J. and Danforth, Christopher M.},
  title =	 {Ousiometrics: {T}he essence of meaning aligns with a
                  power-danger-structure framework instead of
                  valence-arousal-dominance},
  journal =	 {Science Advances},
  year =	 {2026},
  volume =	 {12},
  number =	 {9},
  pages =	 {eadr4039},
  doi =		 {10.1126/sciadv.adr4039},
  url =		 {https://www.science.org/doi/10.1126/sciadv.adr4039}
}

@book{fejes2008GayRightsMoral,
  title = {Gay {{Rights}} and {{Moral Panic}}: {{The Origins}} of {{America}}'s {{Debate}} on {{Homosexuality}}},
  author = {Fejes, Fred},
  year = 2008,
  publisher = {Palgrave Macmillan US},
  address = {New York},
  doi = {10.1057/9780230614680},
  urldate = {2026-05-03},
  isbn = {978-0-230-10826-4 978-0-230-61468-0}
}

@misc{DevelopmentofStatisticalCharQuiz,
  title = {Development of the {{Statistical}} "{{Which Character}}" {{Personality Quiz}}},
  author = {Jorgenson, Eric},
  year = 2026,
  journal = {Open-Source Psychometrics Project},
  publisher = {Open-Source Psychometrics Project},
  urldate = {2026-05-11},
  url = {https://openpsychometrics.org/tests/characters/documentation/}
}

@incollection{cohen2005PunksBulldaggersWelfare,
  title = {Punks, {{Bulldaggers}}, and {{Welfare Queens}}: {{The Radical Potential}} of {{Queer Politics}}?},
  shorttitle = {Punks, {{Bulldaggers}}, and {{Welfare Queens}}},
  booktitle = {Black {{Queer Studies}}},
  author = {Cohen, Cathy J.},
  editor = {Johnson, E. Patrick and Henderson, Mae G.},
  year = 2005,
  series = {A {{Critical Anthology}}},
  eprint = {j.ctv11cw38r.6},
  eprinttype = {jstor},
  pages = {21--51},
  publisher = {Duke University Press},
  doi = {10.2307/j.ctv11cw38r.6},
  urldate = {2026-05-01},
  isbn = {978-0-8223-3629-7}
}

@article{jagose2015TroubleAntinormativity,
  title = {The {{Trouble}} with {{Antinormativity}}},
  author = {Jagose, Annamarie},
  year = 2015,
  month = may,
  journal = {differences},
  volume = {26},
  number = {1},
  pages = {26--47},
  issn = {1040-7391},
  doi = {10.1215/10407391-2880591},
  urldate = {2026-05-01}
}

@article{munoz1997WhiteBeAngry,
  title = {"{{The White}} to {{Be Angry}}": {{Vaginal Davis}}'s {{Terrorist Drag}}},
  shorttitle = {"{{The White}} to {{Be Angry}}"},
  author = {Mu{\~n}oz, Jos{\'e} Esteban},
  year = 1997,
  journal = {Social Text},
  number = {52/53},
  eprint = {466735},
  eprinttype = {jstor},
  pages = {81--103},
  publisher = {Duke University Press},
  issn = {0164-2472},
  doi = {10.2307/466735},
  urldate = {2026-05-01}
}

@misc{2026LavenderScareHistory,
  title = {Lavender {{Scare}}},
  author = {Anderson, Eric},
  year = 2026,
  month = apr,
  journal = {Britannica},
  publisher = {Britannica},
  urldate = {2026-05-03},
  url = {https://www.britannica.com/topic/Lavender-Scare}
}

@misc{2026RedScareHistory,
  title = {Red {{Scare}}},
  author = {Achter, Paul J. and {{Editors of Encyclopaedia Britannica}}},
  year = 2026,
  month = apr,
  journal = {Britannica},
  publisher = {Britannica},
  urldate = {2026-05-03},
  url = {https://www.britannica.com/topic/Red-Scare-politics}
}

@inproceedings{devinney2022TheoriesGenderNLP,
  title = {Theories of ``{{Gender}}'' in {{NLP Bias Research}}},
  booktitle = {Proceedings of the 2022 {{ACM Conference}} on {{Fairness}}, {{Accountability}}, and {{Transparency}}},
  author = {Devinney, Hannah and Bj{\"o}rklund, Jenny and Bj{\"o}rklund, Henrik},
  year = 2022,
  month = jun,
  series = {{{FAccT}} '22},
  pages = {2083--2102},
  publisher = {Association for Computing Machinery},
  address = {New York, NY, USA},
  doi = {10.1145/3531146.3534627},
  urldate = {2026-04-23},
  isbn = {978-1-4503-9352-2}
}

@misc{glaad2025WhereWeAre,
  title = {Where {{We Are}} on {{TV}} report},
  author = {{GLAAD}},
  year = 2025,
  month = nov,
  journal = {Gay \& Lesbian Alliance Against Defamation},
  urldate = {2026-04-23},
  url = {https://glaad.org/whereweareontv24/}
}

@article{yuan2025LGBTRepresentationFilm,
  title = {{{LGBT Representation}} in {{Film}} and {{Media}}: {{Social Impact}} and {{Future Development}}: {{A Literature Review}}},
  shorttitle = {{{LGBT Representation}} in {{Film}} and {{Media}}},
  author = {Yuan, Gao and Guintu, Ma Agatha Anne D.},
  year = 2025,
  month = apr,
  journal = {International Journal of Education and Social Development},
  volume = {2},
  number = {3},
  pages = {132--138},
  issn = {3078-2287},
  doi = {10.54097/kr2wjr51},
  urldate = {2026-03-02}
}

@misc{HRC2026GlossaryTerms,
  title        = {Glossary of {{Terms}}},
  author       = {{HRC Foundation}},
  year         = 2026,
  journal      = {The Human Rights Campaign},
  urldate      = {2026-04-23},
  url = {https://www.hrc.org/resources/glossary-of-terms}
}

@misc{weber2026QueerNLPCritical,
  title         = {Queer {{NLP}}: {{A Critical Survey}} on {{Literature Gaps}}, {{Biases}} and {{Trends}}},
  shorttitle    = {Queer {{NLP}}},
  author        = {Weber, Sabine and Wang, Angelina and Gupta, Ankush and Subramonian, Arjun and Ulmer, Dennis and Tanwar, Eshaan and Aich, Geetanjali and Devinney, Hannah and Hobbs, Jacob and Mickel, Jennifer and Tint, Joshua and Sosto, Mae and Groshan, Ray and Astarita, Simone and Gautam, Vagrant and Blaschke, Verena and Agnew, William and Lee, Wilson Y. and Long, Yanan},
  year          = 2026,
  month         = mar,
  number        = {arXiv:2602.16151},
  eprint        = {2602.16151},
  primaryclass  = {cs},
  publisher     = {arXiv},
  doi           = {10.48550/arXiv.2602.16151},
  urldate       = {2026-04-23},
  archiveprefix = {arXiv}
}

@article{birchmore2022ExploringBoundariesParasocial,
  title      = {Exploring the Boundaries of the Parasocial Contact Hypothesis: An Experimental Analysis of the Effects of the ``Bury Your Gays'' Media Trope on Homophobic and Sexist Attitudes},
  shorttitle = {Exploring the Boundaries of the Parasocial Contact Hypothesis},
  author     = {Birchmore, Ansley and Kettrey, Heather Hensman},
  year       = 2022,
  month      = aug,
  journal    = {Feminist Media Studies},
  volume     = {22},
  number     = {6},
  pages      = {1311--1327},
  publisher  = {Routledge},
  issn       = {1468-0777},
  doi        = {10.1080/14680777.2021.1887919},
  urldate    = {2026-03-02}
}

@incollection{sumter2025Representation,
  title     = {Representation},
  booktitle = {Children, {{Media}}, and {{Technology}}},
  author    = {Sumter, Sindy R. and van Driel, Irene I. and Billedo, Cherrie Joy F.},
  year      = 2025,
  publisher = {Routledge},
  doi       = {10.4324/9781003453123-4},
  isbn      = {978-1-003-45312-3}
}

@book{Theophrastus_Characters_Rusten2003,
  author     = {Theophrastus},
  editor     = {Jeffrey S. Rusten and I. C. Cunningham},
  translator = {Jeffrey S. Rusten and I. C. Cunningham},
  title      = {Theophrastus: Characters; Herodas: Mimes; Sophron and Other Mime Fragments},
  series     = {Loeb Classical Library, No. 225},
  publisher  = {Harvard University Press},
  address    = {Cambridge, MA \& London, England},
  edition    = {3rd},
  year       = {2003},
  isbn       = {9780674996038}
}

@book{Jung1969,
  author    = {Jung, C. G.},
  title     = {The Archetypes and the Collective Unconscious},
  volume    = {9, Part 1},
  series    = {Collected Works of C. G. Jung},
  editor    = {Read, Herbert and Fordham, Michael and Adler, Gerhard},
  year      = {1969},
  publisher = {Princeton University Press},
  address   = {Princeton, NJ}
}

@book{johnson2023LavenderScareCold,
  title      = {The {{Lavender Scare}}: {{The Cold War Persecution}} of {{Gays}} and {{Lesbians}} in the {{Federal Government}}},
  shorttitle = {The {{Lavender Scare}}},
  author     = {Johnson, David K.},
  editor     = {Author, the},
  year       = 2023,
  month      = mar,
  publisher  = {University of Chicago Press},
  address    = {Chicago, IL},
  urldate    = {2026-02-20},
  isbn       = {978-0-226-82572-4}
}

@article{cover2022MakingQueerContentVisible,
  title      = {Making {{Queer}} Content Visible: Approaches and Assumptions of {{Australian}} Film and Television Stakeholders Working with {{LGBTQ}}+ Content},
  shorttitle = {Making {{Queer}} Content Visible},
  author     = {Cover, Rob},
  year       = 2022,
  month      = feb,
  journal    = {Media International Australia},
  volume     = {190},
  number     = {1},
  doi        = {10.1177/1329878X221077851},
  urldate    = {2026-02-24}
}

@article{rovirosa2025QueerNonNormativeCharacters,
  title      = {Subrepresentaci\'on y T\'opicos de Los Personajes Queer y No Normativos En Las Series Deportivas Contempor\'aneas},
  shorttitle = {Queer and {{Non-Normative Characters}}},
  author     = {Rovirosa, Anna Tous and G{\'a}lvez, Raquel Cris{\'o}stomo and Fedotova, Elena and Hern{\'a}ndez, Natividad Ramajo},
  year       = 2025,
  month      = nov,
  journal    = {Palabra Clave},
  volume     = {28},
  number     = {3},
  pages      = {e2831-e2831},
  issn       = {2027-534X},
  doi        = {10.5294/pacla.2025.28.3.1},
  urldate    = {2026-02-24}
}

@article{cover2023BuryYourGays,
  title      = {The ``{{Bury}} Your {{Gays}}'' Trope in Contemporary Television: {{Generational}} Shifts in Production Responses to Audience Dissent},
  journal    = {The Journal of Popular Culture},
  shorttitle = {The ``{{Bury}} Your {{Gays}}'' Trope in Contemporary Television},
  author     = {Cover, Rob and Milne, Cassandra},
  year       = 2023,
  volume     = {56},
  number     = {5-6},
  pages      = {810--823},
  doi        = {10.1111/jpcu.13255},
  urldate    = {2026-02-24}
}

@incollection{dhoest2023HomosexualityNormalityReception,
  title      = {Homosexuality and Normality: {{The}} Reception of Gay Male Representations on Film and Television},
  shorttitle = {Homosexuality and Normality},
  booktitle  = {Routledge {{Handbook}} of {{Sexuality}}, {{Gender}}, {{Health}} and {{Rights}}},
  author     = {Dhoest, Alexander},
  year       = 2023,
  edition    = {2},
  publisher  = {Routledge},
  doi        = {10.4324/9781003278405},
  isbn       = {978-1-003-27840-5}
}

@misc{beauregard2026ArchetypesGenderFiction,
  title         = {Archetypes and Gender in Fiction: {{A}} Data-Driven Mapping of Gender Stereotypes in Stories},
  shorttitle    = {Archetypes and Gender in Fiction},
  author        = {Beauregard, Calla Glavin and Zimmerman, Julia Witte and Fehr, Ashley M. A. and Tangherlini, Timothy R. and Danforth, Christopher M. and Dodds, Peter Sheridan},
  year          = 2026,
  number        = {arXiv:2602.17005},
  eprint        = {2602.17005},
  primaryclass  = {cs},
  publisher     = {arXiv},
  doi           = {10.48550/arXiv.2602.17005},
  urldate       = {2026-02-20},
  archiveprefix = {arXiv}
}

@phdthesis{zimmerman2025LocalityRelationMeaning,
  title   = {Locality, Relation, and Meaning Construction in Language, as Implemented in Humans and Large Language Models (LLMs)},
  author  = {Zimmerman, Julia Witte},
  year    = {2025},
  address = {United States -- Vermont},
  urldate = {2026-01-20},
  isbn    = {9798314813249},
  type    = {{Ph.D.} dissertation},
  school  = {The University of Vermont and State Agricultural College},
  url     = {https://hdl.handle.net/20.500.14849/4063}
}

@misc{lgbtcharacterswikia,
  title      = {{LGBTQIA+ Characters Wikia}},
  author     = {{LGBTQIA+ Characters Wikia}},
  urldate    = {2026-01-20},
  url        = {https://lgbtqia-characters.fandom.com/wiki/LGBT\_Characters\_Wikia}
}

@misc{openpsychometrics,
  title   = {{Open Source Psychometrics Project}},
  author  = {{Open Source Psychometrics Project}},
  urldate = {2026-01-20},
  url     = {https://openpsychometrics.org/}
}

@misc{dodds2025ArchetypometricsDataset,
  title     = {Archetypometrics Dataset},
  author    = {Dodds, Peter Sheridan},
  year      = 2025,
  month     = aug,
  publisher = {Zenodo},
  doi       = {10.5281/zenodo.16953724},
  urldate   = {2026-01-20}
}

@article{beam2024gay_prototypicality,
  title      = {The consequences of prototypicality: {Testing} the prejudice distribution account of bias toward gay men},
  volume     = {11},
  issn       = {2329-0390},
  shorttitle = {The consequences of prototypicality},
  doi        = {10.1037/sgd0000581},
  number     = {1},
  journal    = {Psychology of Sexual Orientation and Gender Diversity},
  author     = {Beam, Adam J. and Wellman, Joseph D.},
  year       = {2024},
  pages      = {79--89}
}

@article{soto_sanfiel_perceptions_2024,
  title    = {Perceptions of {Gay} {Stereotypes} in {Fiction} and {Their} {Relationship} with {Homophobia}},
  issn     = {1936-4822},
  url      = {https://doi.org/10.1007/s12119-024-10291-3},
  doi      = {10.1007/s12119-024-10291-3},
  journal  = {Sexuality \& Culture},
  author   = {Soto-Sanfiel, María T. and Sánchez-Soriano, Juan-José and Angulo-Brunet, Ariadna},
  year     = {2024}
}

@misc{dodds2025ArchetypometricsPragmateiaEmpirical,
  title      = {Archetypometrics, a {{Pragmateia}}: {{Empirical Determination}} of the {{Fundamental Archetypes}} of {{Fictional Characters}}},
  shorttitle = {Archetypometrics, a {{Pragmateia}}},
  author     = {Dodds, Peter Sheridan and Zimmermann, Julia Witte and Beauregard, Calla G. and Fehr, Ashley M. A. and Fudolig, Mikaela I. and Tangherlini, Timothy R. and Danforth, Christopher M.},
  year       = 2025,
  month      = sep,
  number     = {6avk5\_v1},
  publisher  = {SocArXiv},
  doi        = {10.31235/osf.io/6avk5_v1},
  urldate    = {2026-01-20}
}

\clearpage
%% optional: switch to single column
\onecolumn

\appendix
\section*{Appendices}
\label{sec:straight-queer.appendix}

\setcounter{page}{1}
\renewcommand{\thepage}{A\arabic{page}}
\renewcommand{\thefigure}{A\arabic{figure}}
\renewcommand{\thetable}{A\arabic{table}}
\setcounter{figure}{0}
\setcounter{table}{0}

\renewcommand{\thesection}{A\arabic{section}}
\setcounter{section}{0}

\section{Supporting figures}
\label{sec:straight-queer.appendix.addlfigures}

We gather GLAAD statistics from their 2015--2016 through 2024--2025 annual reports on LGBTQ representation on television as we have seen some versions of this data in scholarly work but none holistically accumulating a decade as such from GLAAD reports. The 2024--2025 report was the latest as of this publication.

The longer term trend shows a 10-year increase of $1.8\times$ the number of LGBTQ characters counted from the first to the latest year. These characters reflect regulars and recurring. Note that GLAAD began recording streaming  in 2015--2016, making these years more straightforward methodologically to compare. Year-to-year labels represent the increase or decrease from the immediately prior year. The purple line itself tracks the raw total counts of characters across broadcast, cable, and streaming services.

\vspace{1cm}
\begin{figure}[!h]
    \centering
    \includegraphics[width=.8\textwidth]{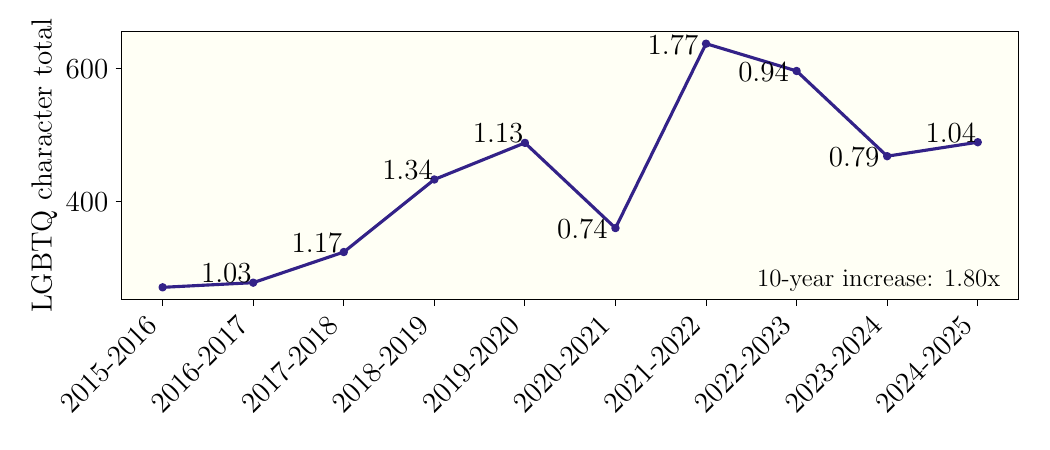}
    \caption{GLAAD statistics for ``Where We Are on TV'' for 2015--2025. Totals reflect LGBTQ regular and recurring characters broadcast, cable, and streaming.}
    \label{fig:glaad-stats}
\end{figure}

\clearpage
\section{Archetype dimensions}
\label{sec:straight-queer.appendix.arch-dims}

The following plots show the main 6 archetype dimensions (\archetypesemdiff{1}, \archetypesemdiff{2}, \archetypesemdiff{3}, \archetypesemdiff{4}, \archetypesemdiff{5}, and \archetypesemdiff{6}) by data subset then along the \semdiff{straight}{queer} trait in the ousiograms.

\begin{figure*}[!h]
    \centering
    \includegraphics[width=.9\textwidth]{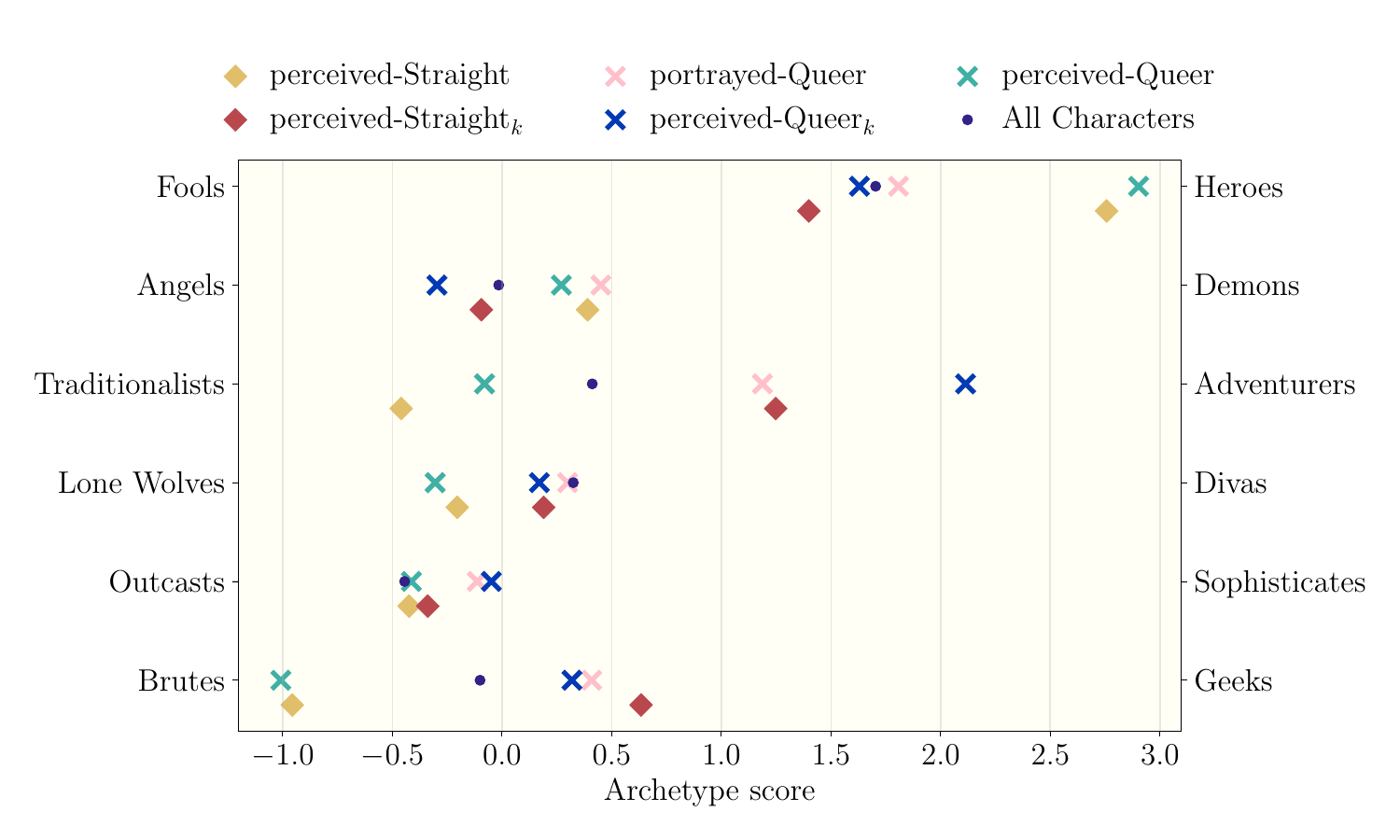}
    \caption{Archetype dimension split axis group plot. Groups' median archetype scores range between the left- and right-hand labels for these 6 top dimensions.}
    \label{fig:split-axis-archetype-dims}
\end{figure*}

Fig.~\ref{fig:split-axis-archetype-dims} shows an overarching view of subsets on archetype dimensions. In general, \archetype{Hero} is a strong archetype, which shows here as the two starkest groups on the trait (\straightalign\ and \queeralign) both lean toward the Hero end. \archetypesemdiff{3} shows the broadest variance, with \straightalign\ and \queeralign\ tending toward \archetype{Traditionalist} while \queerneigh\ tends most toward \archetype{Adventurer}. On \archetypesemdiff{6}, \straightalign\ and \queeralign\ both veer toward the \archetype{Brute} end while \fandomqueer, \queerneigh, and \straightneigh\ veer toward \archetype{Geek}. Similarly to the moving quartile medians, subsets' locations for \archetypesemdiff{2}, \archetypesemdiff{4}, and \archetypesemdiff{5} show less variance, characterized by tighter grouping around All Characters.

% rest of the character ousiograms

\begin{figure*}[t]
    \includegraphics[width=\textwidth]{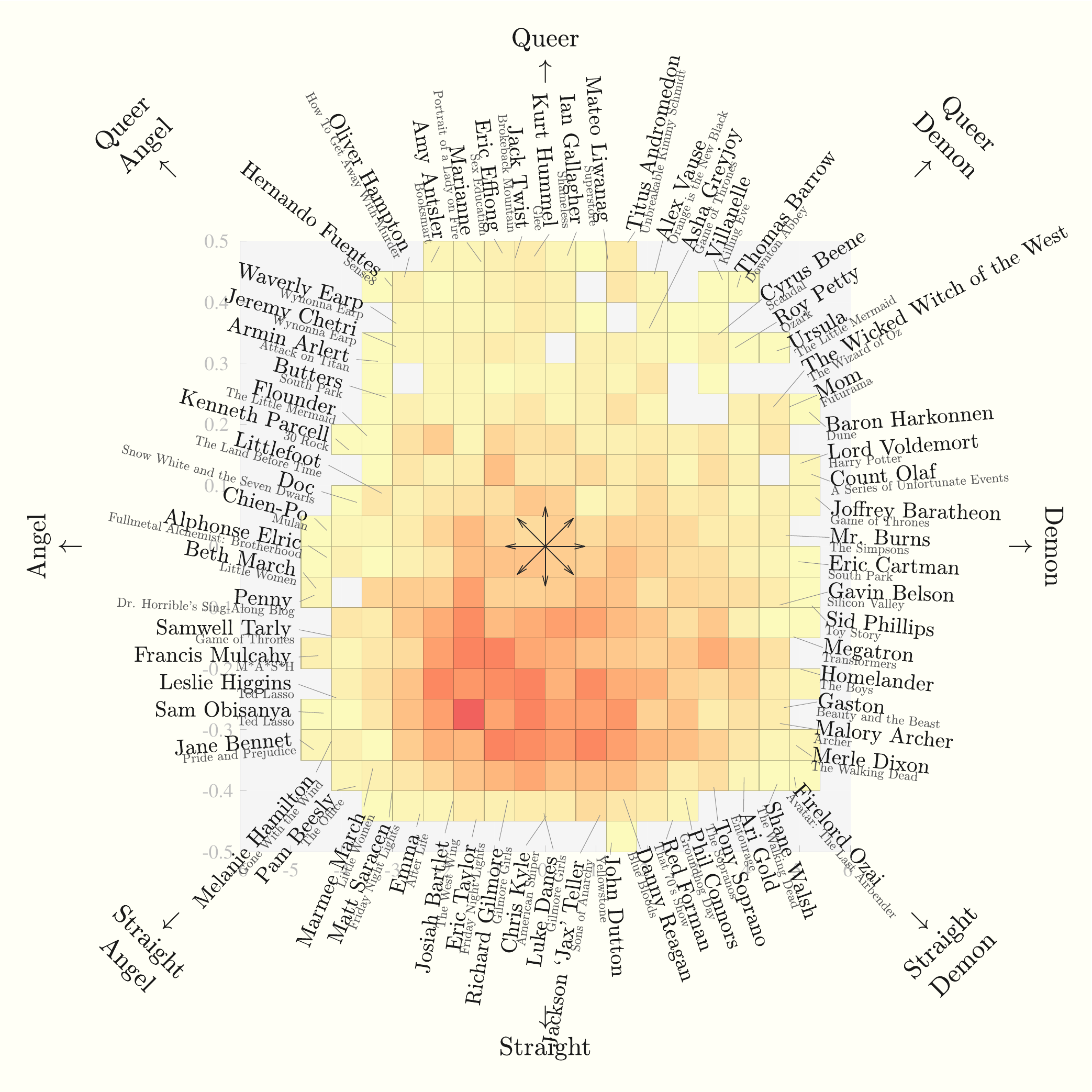}
    \caption{A histogram of essential characteristics with \traitlinksimple{straight}{queer} trait scores on the vertical axis and \archetypesemdiff{2} scores on the horizontal axis. A darker cell color denotes the number of characters where darker indicates more characters fall within that range.}
    \label{fig:ousiogram-archetypes-straight-queer-02}
\end{figure*}

\begin{figure*}[t]
    \includegraphics[width=\textwidth]{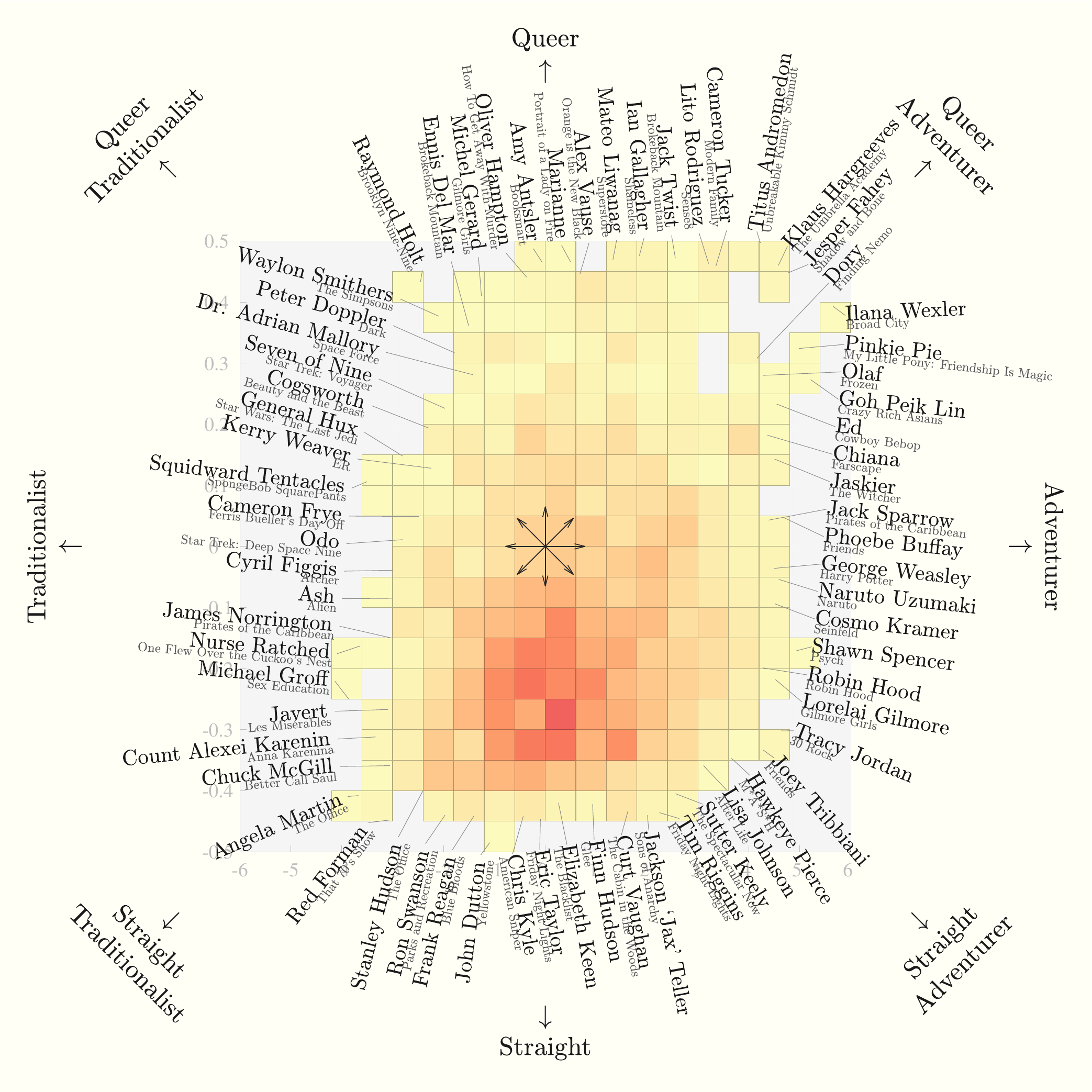}
    \caption{A histogram of essential characteristics with \traitlinksimple{straight}{queer} trait scores on the vertical axis and \archetypesemdiff{3} scores on the horizontal axis. A darker cell color indicates more characters fall within that range.}
    \label{fig:ousiogram-archetypes-straight-queer-03}
\end{figure*}

\begin{figure*}[t]
    \includegraphics[width=\textwidth]{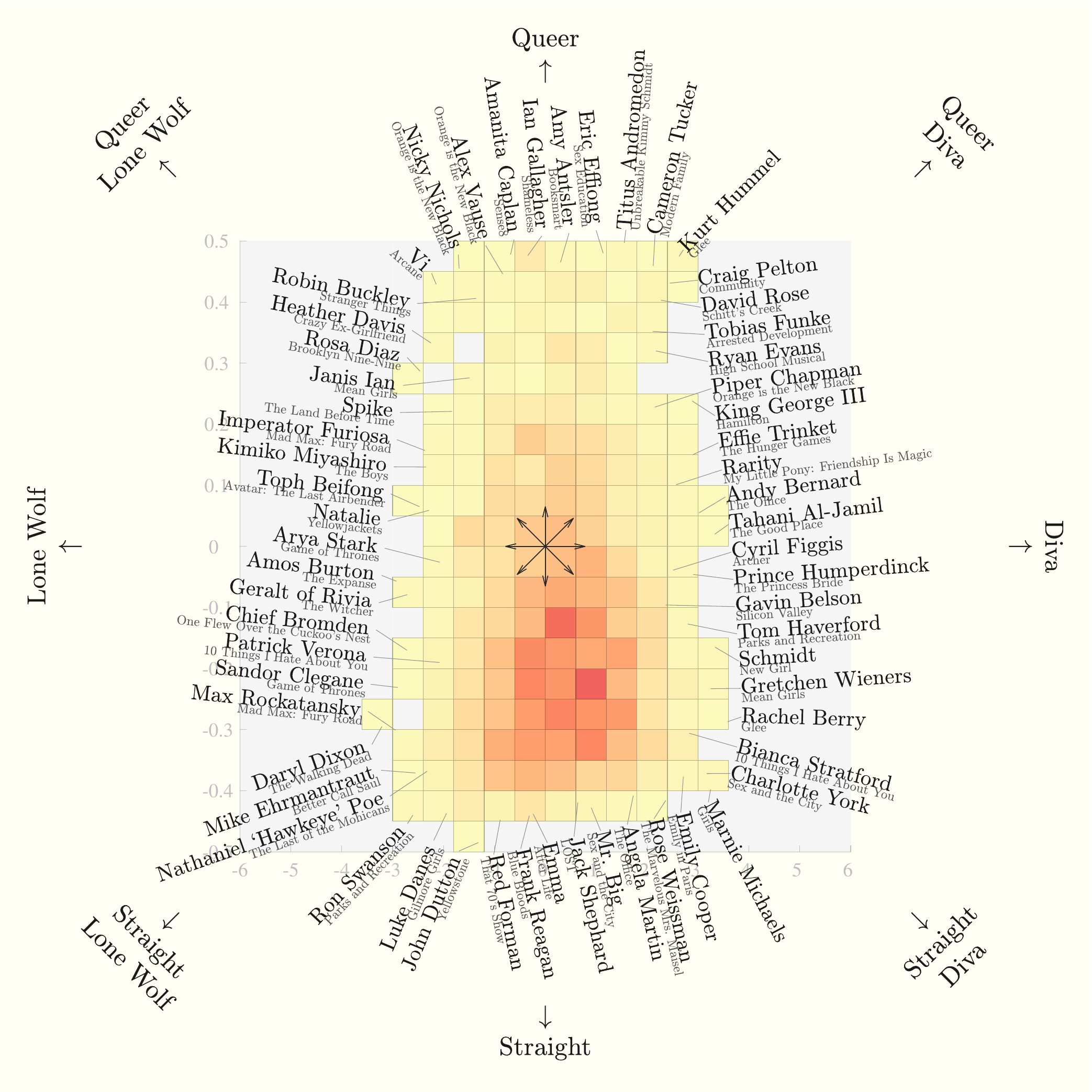}
    \caption{A histogram of essential characteristics with \traitlinksimple{straight}{queer} trait scores on the vertical axis and \archetypesemdiff{4} scores on the horizontal axis. A darker cell color indicates more characters fall within that range.}
    \label{fig:ousiogram-archetypes-straight-queer-04}
\end{figure*}

\begin{figure*}[t]
    \includegraphics[width=\textwidth]{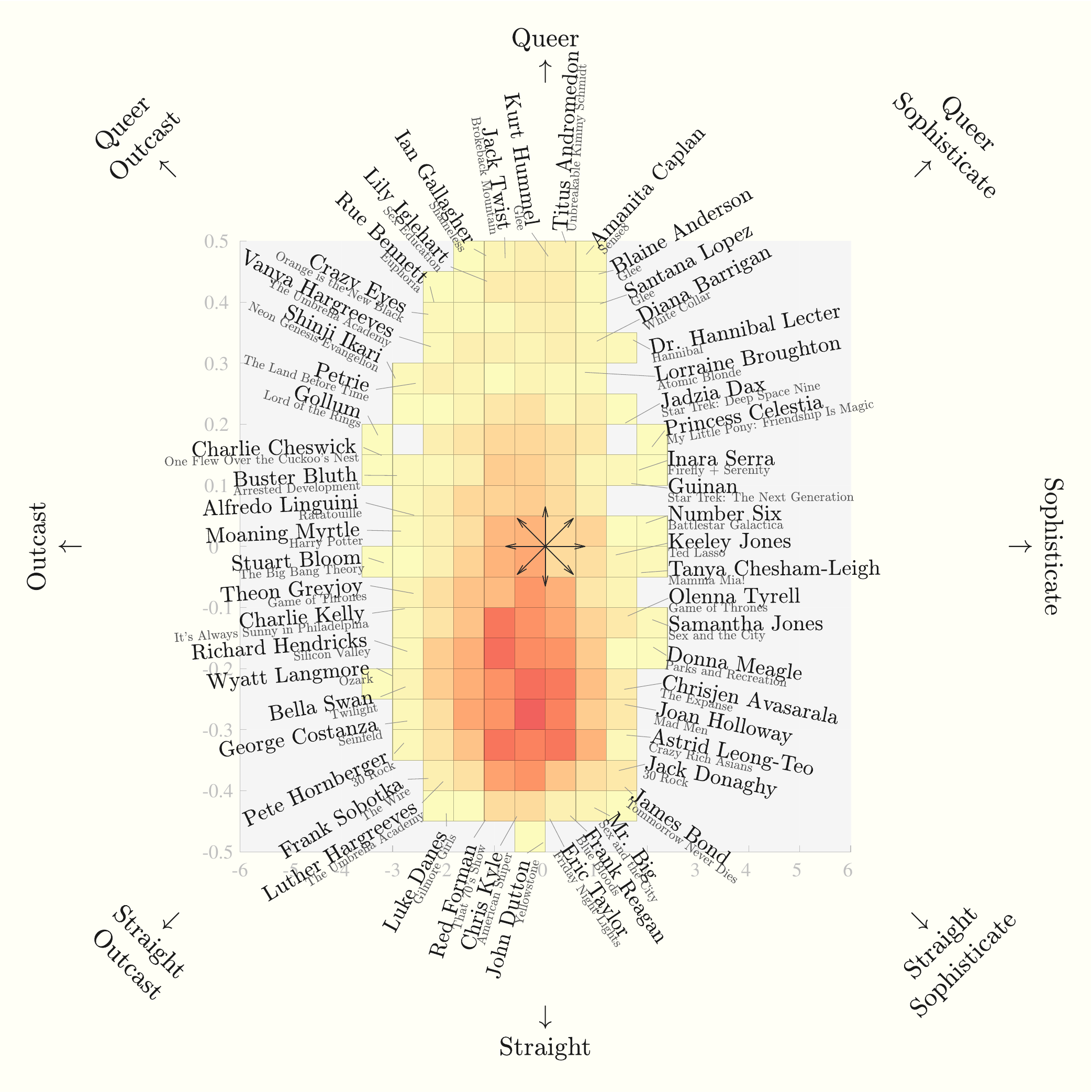}
    \caption{A histogram of essential characteristics with \traitlinksimple{straight}{queer} trait scores on the vertical axis and \archetypesemdiff{5} scores on the horizontal axis. A darker cell color indicates more characters fall within that range.}
    \label{fig:ousiogram-archetypes-straight-queer-05}
\end{figure*}

\begin{figure*}[t]
    \includegraphics[width=\textwidth]{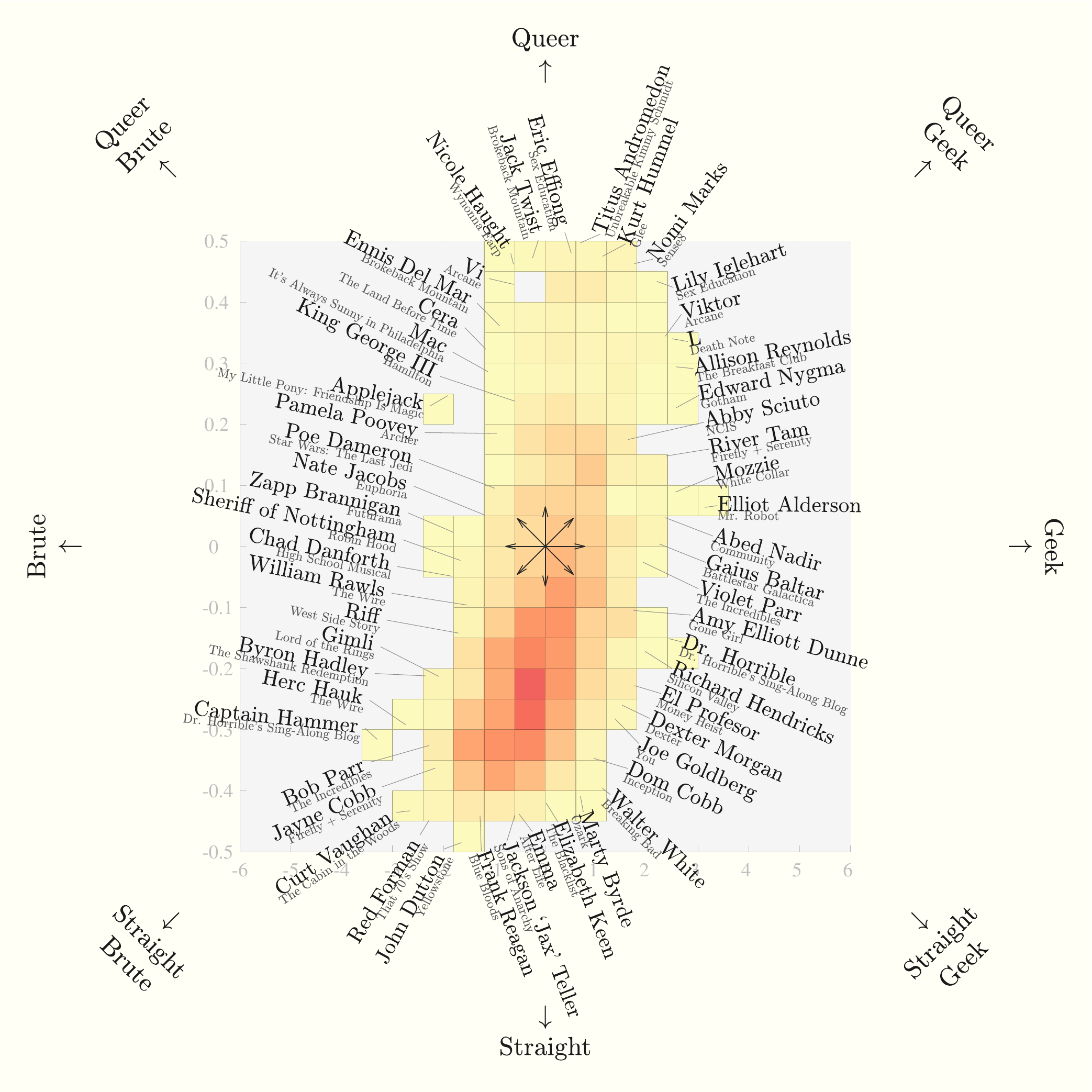}
    \caption{A histogram of essential characteristics with \traitlinksimple{straight}{queer} trait scores on the vertical axis and \archetypesemdiff{6} scores on the horizontal axis. A darker cell color indicates more characters fall within that range.}
    \label{fig:ousiogram-archetypes-straight-queer-06}
\end{figure*}

Next, we describe additional details for the canonical gender exploration. 
% extra text from exploring canonical gender.
We confirm queer identities in the \fandomqueer\ subset, whose canonically identified groups of men and women are very small. However, we return the most queer-aligning scores on the trait differential to explore an effect of gender among queer-aligned characters canonically identified as men and women. 
We take the top $200$ characters in the queer-aligned direction and subset by `male' (\queeralign$_\text{M}$) and `female', resulting in $100$ characters per label. \queeralign$_\text{M}$ contains $41/57$ of \fandomqueer's male characters (\fandomqueer$_\text{M}$). 
When comparing archetype medians to other subsets, both \queeralign$_\text{M}$ and \fandomqueer$_\text{M}$ lean the furthest toward \archetype{Fool}, a position which \straightneigh\ otherwise occupies. \queeralign$_\text{M}$ tends somewhat further toward \archetype{Angel} than \queerneigh\ on the dimension with otherwise tight grouping. 
Both male subsets veer further toward \archetype{Diva} than the rest of the subsets which veer further toward (\archetype{Lone Wolf}) relative to All Characters. 
The two male subsets do not practically differ on dimensions \archetypesemdiff{3}, \archetypesemdiff{5}, and \archetypesemdiff{6}. 
Female \queeralign\ characters tracked with the \fandomqueer\ and \queerneigh\ subsets on all $6$ dimensions.

\clearpage
\section{Character subsets}
\label{sec:straight-queer.appendix.char_subsets}

\subsection{Subset methodology}
\label{sec:straight-queer.appendix.char_subsets.subsetcreation}

To find character matches from the Fandom dataset in the Archetypes dataset (to utilize extant `\fandomqueer' coding), we perform entity resolution on characters' first names and stories between the two datasets and check this entire list for canonically queer labeling for the story contained in Archetypes; six matches are excluded because of character property mismatches (comic version to TV version) or uncertain, potentially non-canon, labeling. 

Next, we detail bootstrap sampling methodology to create subsets of straight- or queer-aligned characters. We create the subsets of `\textbf{\straightalign}', `\textbf{\queeralign}', `\textbf{\straightneigh}', `\textbf{\queeralign}' (alongside the extant \textbf{\fandomqueer} group). Originally we had sampled around 100 for each subset and grew the sizes to 120 because of the number of queer coded character matches from Fandom data (125) within Archetypes data. To test the stability of variance and determine stabilized sampling parameters, we perform bootstrap sampling and calculate radius of gyration ($R_g$) and coefficient of variation ($CV$). Simulations were over the parameters: top $N$ in trait, top $m$ seed characters in trait, and $k$ nearest neighbors.

\begin{figure}[!h]
    \centering
    \begin{minipage}{.45\textwidth}
        \centering
        \includegraphics[width=\textwidth]{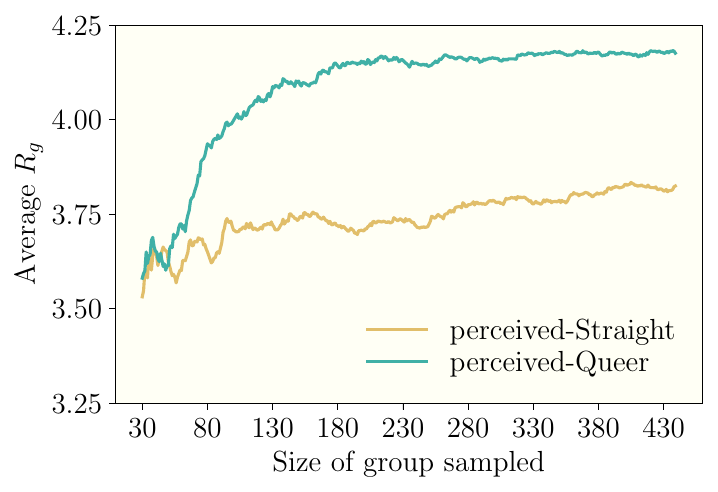}
    \end{minipage}
    \hspace{.2cm}
    \begin{minipage}{.45\textwidth}
        \centering
        \includegraphics[width=\textwidth]{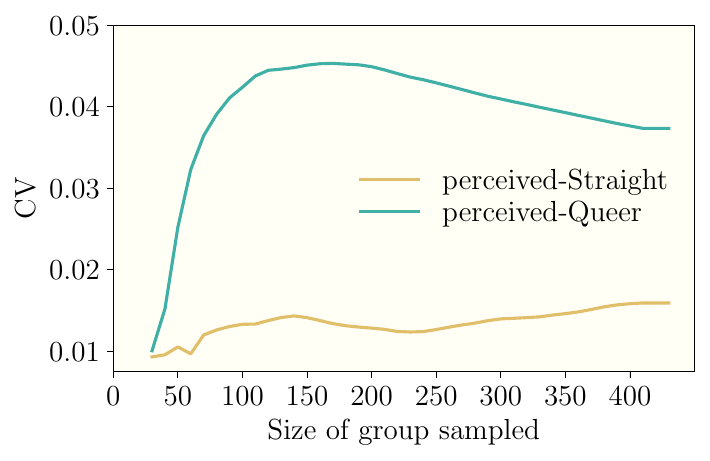}
    \end{minipage}
    \caption{Radii of gyration for \straightalign\ and \queeralign\ subsets along bootstrapped samples of $N$ (left). Coefficient of variation ($CV$) of $R_g$ for \straightalign\ and \queeralign\ groups along bootstrapped samples of $N$ (right).}
    \label{fig:radius-cv-straight-queer}
\end{figure}

The starting parameters differ for \straightalign and \queeralign\ versus neighbors. The parameter ranges were chosen to begin too low and too high in terms of final hypothetical sample size (given our original seed mentioned prior). For these high trait subsets, we sample over top $N$ in $[30, 440]$. For each $n$ along this top $N$ characters range, we sample $30$ characters $10,000$ times, calculating $R_g$, and taking the mean $R_g$ per $n$. Thus, as the sampled subset size grows, we have a bootstrapped radius of gyration (see Fig.~\ref{fig:radius-cv-straight-queer}).

\begin{figure}
    \centering
    \includegraphics[width=.77\textwidth]{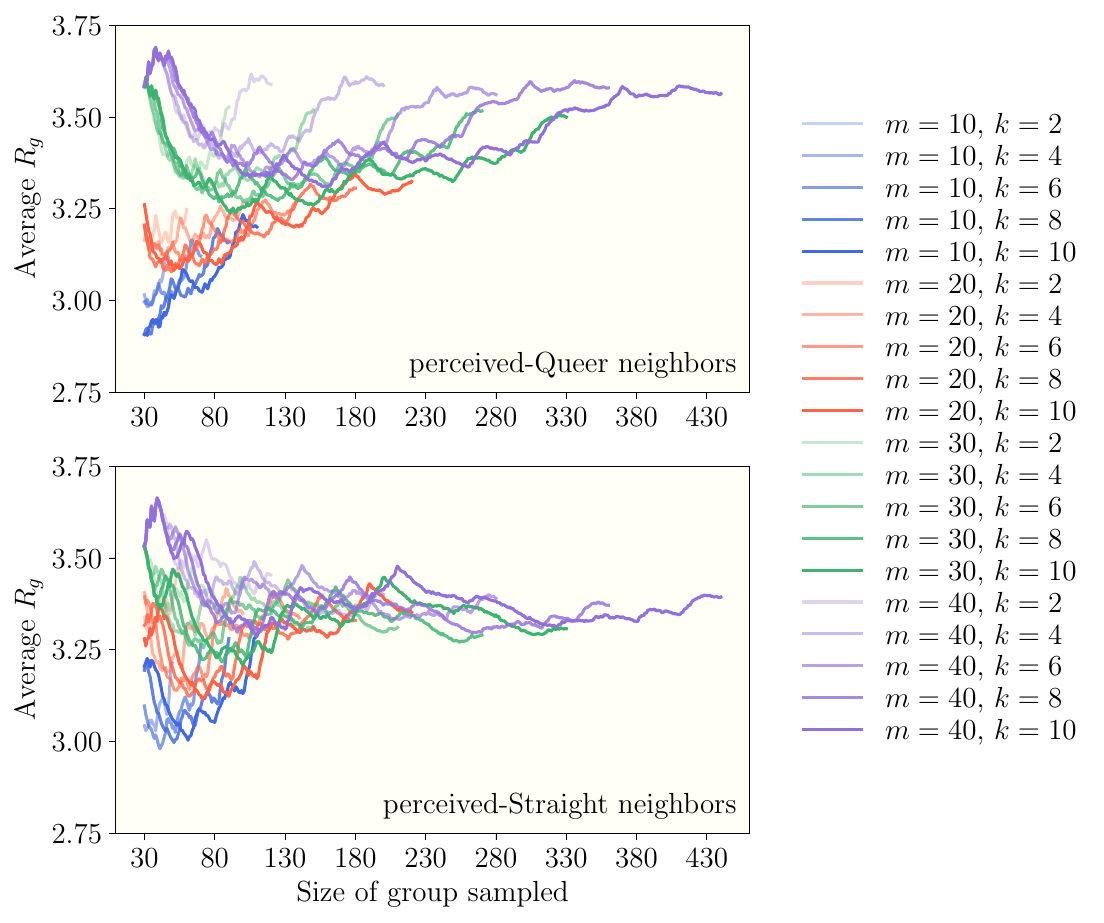}
    \caption{Radii of gyration for \queerneigh\ (upper) and \straightneigh\ (lower) along bootstrapped samples of $N$ using $m$ seed characters and $k$ neighbors.}
    \label{fig:radius-neighbors}
\end{figure}

\begin{figure*}
    \centering
    \includegraphics[width=.9\textwidth]{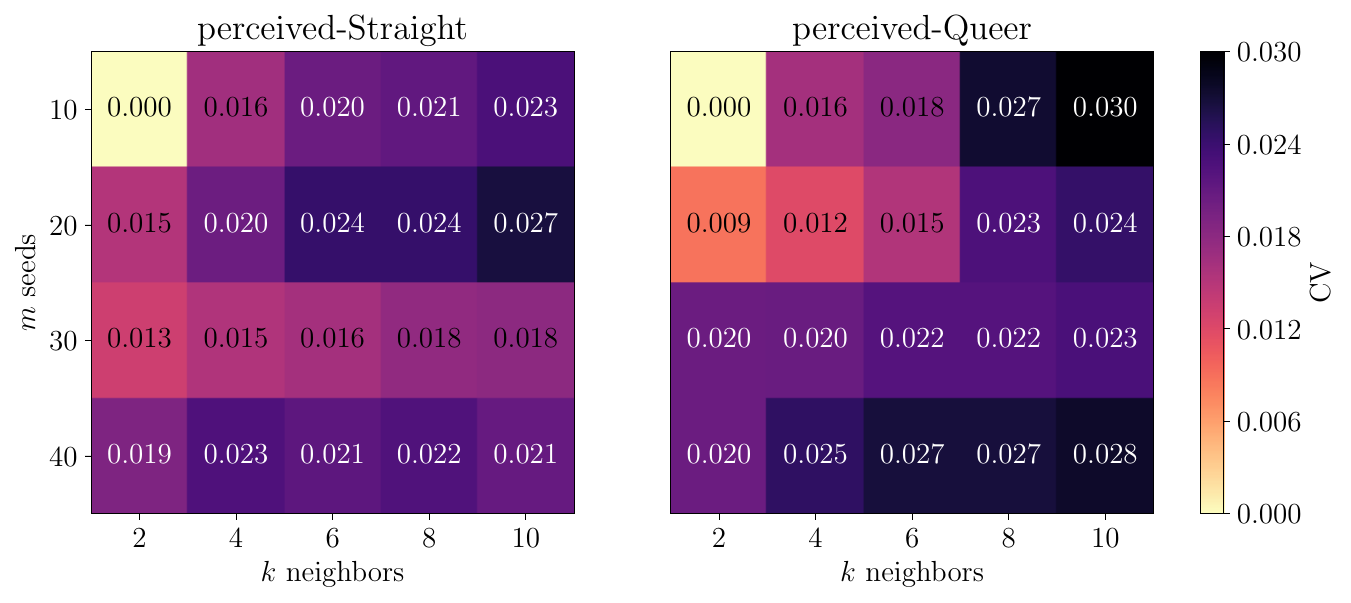}
    \caption{Coefficient of variation ($CV$) of $R_g$ for both neighbor subsets from bootstrapped sampling conditions of $m$ seed characters and $k$ neighbors.}
    \label{fig:radius-cv-neighbors}
\end{figure*}

For neighbor subsets, we sample over the product of $m$ top (seed) characters and $k$ neighbors ranges using $[10, 40]$ (every 10) and $[2, 10]$, respectively and inclusively. To build this list of characters, we use the seed characters and calculate their nearest neighbors by Euclidean distance across the full trait space, taking the $k$ closest in distance. 
The final list is comprised of the original $m$ plus each $m$'s $k$ neighbors; the $N$ built from this combination is $m \times k + m$. Using this $N$ per condition, we follow the same method as above by sampling for each $n$. These conditions present many more results to compare, hence the parameter comparison matrices (see Figs.~\ref{fig:radius-neighbors} and \ref{fig:radius-cv-neighbors}).

In our choice of final parameters, we look for approximate mean $R_g$ convergence among the groups that makes sense for both the high trait value and neighbor subsets using results in the figures. Note the simulations require moderately optimized code and multiprocessing (we use \texttt{numpy} and \texttt{multiprocessing} Python packages). Final parameters are $N=128$, $m=22$, and $k=6$. Final group sizes are: \straightalign\ $=128$, \queeralign\ $=128$, \queerneigh\ $=123$, and \straightneigh\ $=128$.

\vspace{1cm}

% 9July2026 comment: we don't reference these but am including until publication dictates otherwise.
\subsection{Character subset lists}
\label{sec:straight-queer.appendix.char_subsets.char_lists}
\newcommand{\queerneighlist}{
    \characterlinksimple{NCIS-Abby-Sciuto}{Abby Sciuto}, \characterlinksimple{Peaky-Blinders-Ada-Shelby}{Ada Shelby}, \characterlinksimple{The-Hangover-Alan}{Alan}, \characterlinksimple{Orange-is-the-New-Black-Alex-Vause}{Alex Vause}, \characterlinksimple{Schitts-Creek-Alexis-Rose}{Alexis Rose}, \characterlinksimple{Twilight-Alice-Cullen}{Alice Cullen}, \characterlinksimple{The-Notebook-Allison-Hamilton}{Allison Hamilton}, \characterlinksimple{A-Star-Is-Born-Ally-Maine}{Ally Maine}, \characterlinksimple{Sense8-Amanita-Caplan}{Amanita Caplan}, \characterlinksimple{Chilling-Adventures-of-Sabrina-Ambrose-Spellman}{Ambrose Spellman}, \characterlinksimple{Booksmart-Amy-Antsler}{Amy Antsler}, \characterlinksimple{Little-Women-Amy-March}{Amy March}, \characterlinksimple{Bones-Angela-Montenegro}{Angela Montenegro}, \characterlinksimple{Jennifers-Body-Anita-Needy-Lesnicki}{Anita 'Needy' Lesnicki}, \characterlinksimple{How-To-Get-Away-With-Murder-Annalise-Keating}{Annalise Keating}, \characterlinksimple{Inception-Ariadne}{Ariadne}, \characterlinksimple{Glee-Blaine-Anderson}{Blaine Anderson}, \characterlinksimple{The-Vampire-Diaries-Bonnie-Bennett}{Bonnie Bennett}, \characterlinksimple{Modern-Family-Cameron-Tucker}{Cameron Tucker}, \characterlinksimple{The-Vampire-Diaries-Caroline-Forbes}{Caroline Forbes}, \characterlinksimple{Sex-and-the-City-Carrie-Bradshaw}{Carrie Bradshaw}, \characterlinksimple{Smallville-Chloe-Sullivan}{Chloe Sullivan}, \characterlinksimple{Outlander-Claire-Randall}{Claire Randall}, \characterlinksimple{Eternal-Sunshine-of-the-Spotless-Mind-Clementine-Kruczynski}{Clementine Kruczynski}, \characterlinksimple{Before-Sunrise-Celine}{Céline}, \characterlinksimple{Agents-of-SHIELD-Daisy-Skye-Johnson}{Daisy 'Skye' Johnson}, \characterlinksimple{Mean-Girls-Damian-Leigh}{Damian Leigh}, \characterlinksimple{Mr-Robot-Darlene}{Darlene}, \characterlinksimple{Schitts-Creek-David-Rose}{David Rose}, \characterlinksimple{Twin-Peaks-Donna-Hayward}{Donna Hayward}, \characterlinksimple{That-70s-Show-Donna-Pinciotti}{Donna Pinciotti}, \characterlinksimple{Jurassic-Park-Dr-Ellie-Sattler}{Dr. Ellie Sattler}, \characterlinksimple{The-Hunger-Games-Effie-Trinket}{Effie Trinket}, \characterlinksimple{Seinfeld-Elaine-Benes}{Elaine Benes}, \characterlinksimple{The-Good-Place-Eleanor-Shellstrop}{Eleanor Shellstrop}, \characterlinksimple{Sex-Education-Eric-Effiong}{Eric Effiong}, \characterlinksimple{Orange-is-the-New-Black-Flaca-Gonzales}{Flaca Gonzales}, \characterlinksimple{Fleabag-Fleabag}{Fleabag}, \characterlinksimple{Tangled-Flynn-Rider}{Flynn Rider}, \characterlinksimple{Marvel-Cinematic-Universe-Gamora}{Gamora}, \characterlinksimple{Aladdin-Genie}{Genie}, \characterlinksimple{Harry-Potter-George-Weasley}{George Weasley}, \characterlinksimple{Harry-Potter-Ginny-Weasley}{Ginny Weasley}, \characterlinksimple{Crazy-Rich-Asians-Goh-Peik-Lin}{Goh Peik Lin}, \characterlinksimple{The-Fault-in-Our-Stars-Hazel-Grace-Lancaster}{Hazel Grace Lancaster}, \characterlinksimple{Portrait-of-a-Lady-on-Fire-Heloise}{Heloise}, \characterlinksimple{Breakfast-at-Tiffanys-Holly-Golightly}{Holly Golightly}, \characterlinksimple{Shameless-Ian-Gallagher}{Ian Gallagher}, \characterlinksimple{Greys-Anatomy-Izzie-Stevens}{Izzie Stevens}, \characterlinksimple{Brokeback-Mountain-Jack-Twist}{Jack Twist}, \characterlinksimple{Brooklyn-Nine-Nine-Jake-Peralta}{Jake Peralta}, \characterlinksimple{Breaking-Bad-Jane-Margolis}{Jane Margolis}, \characterlinksimple{Criminal-Minds-Jennifer-Jareau}{Jennifer Jareau}, \characterlinksimple{Forrest-Gump-Jenny-Curran}{Jenny Curran}, \characterlinksimple{Shadow-and-Bone-Jesper-Fahey}{Jesper Fahey}, \characterlinksimple{Before-Sunrise-Jesse}{Jesse}, \characterlinksimple{Outer-Banks-John-B}{John B}, \characterlinksimple{Euphoria-Jules-Vaughn}{Jules Vaughn}, \characterlinksimple{Emily-in-Paris-Julien}{Julien}, \characterlinksimple{The-Good-Wife-Kalinda-Sharma}{Kalinda Sharma}, \characterlinksimple{Parasite-Kim-Ki-jung}{Kim Ki-jung}, \characterlinksimple{Locke-and-Key-Kinsey-Locke}{Kinsey Locke}, \characterlinksimple{The-Umbrella-Academy-Klaus-Hargreeves}{Klaus Hargreeves}, \characterlinksimple{Glee-Kurt-Hummel}{Kurt Hummel}, \characterlinksimple{Downton-Abbey-Lady-Sybil-Crawley}{Lady Sybil Crawley}, \characterlinksimple{Supergirl-Lena-Luthor}{Lena Luthor}, \characterlinksimple{How-I-Met-Your-Mother-Lily-Aldrin}{Lily Aldrin}, \characterlinksimple{Bobs-Burgers-Linda-Belcher}{Linda Belcher}, \characterlinksimple{Sense8-Lito-Rodriguez}{Lito Rodriguez}, \characterlinksimple{Atomic-Blonde-Lorraine-Broughton}{Lorraine Broughton}, \characterlinksimple{Portrait-of-a-Lady-on-Fire-Marianne}{Marianne}, \characterlinksimple{Sense-and-Sensibility-Marianne-Dashwood}{Marianne Dashwood}, \characterlinksimple{Normal-People-Marianne-Sheridan}{Marianne Sheridan}, \characterlinksimple{Back-to-the-Future-Marty-McFly}{Marty McFly}, \characterlinksimple{Superstore-Mateo-Liwanag}{Mateo Liwanag}, \characterlinksimple{MASH-Maxwell-Klinger}{Maxwell Klinger}, \characterlinksimple{La-La-Land-Mia-Dolan}{Mia Dolan}, \characterlinksimple{Emily-in-Paris-Mindy-Chen}{Mindy Chen}, \characterlinksimple{Mean-Girls-Ms-Sharon-Norbury}{Ms. Sharon Norbury}, \characterlinksimple{Mulan-Mushu}{Mushu}, \characterlinksimple{X-Men-Mystique}{Mystique}, \characterlinksimple{Money-Heist-Nairobi}{Nairobi}, \characterlinksimple{The-Lion-King-Nala}{Nala}, \characterlinksimple{Orange-is-the-New-Black-Nicky-Nichols}{Nicky Nichols}, \characterlinksimple{Wynonna-Earp-Nicole-Haught}{Nicole Haught}, \characterlinksimple{The-Notebook-Noah-Calhoun}{Noah Calhoun}, \characterlinksimple{Sense8-Nomi-Marks}{Nomi Marks}, \characterlinksimple{Harry-Potter-Nymphadora-Tonks}{Nymphadora Tonks}, \characterlinksimple{How-To-Get-Away-With-Murder-Oliver-Hampton}{Oliver Hampton}, \characterlinksimple{The-Perks-of-Being-a-Wallflower-Patrick-Stewart}{Patrick Stewart}, \characterlinksimple{10-Things-I-Hate-About-You-Patrick-Verona}{Patrick Verona}, \characterlinksimple{Outer-Banks-Pope}{Pope}, \characterlinksimple{Anna-Karenina-Princess-Anna-Karenina}{Princess Anna Karenina}, \characterlinksimple{Shrek-Princess-Fiona}{Princess Fiona}, \characterlinksimple{One-Flew-Over-the-Cuckoos-Nest-Randle-McMurphy}{Randle McMurphy}, \characterlinksimple{My-Little-Pony-Friendship-Is-Magic-Rarity}{Rarity}, \characterlinksimple{Sense8-Riley-Blue}{Riley Blue}, \characterlinksimple{How-I-Met-Your-Mother-Robin-Scherbatsky}{Robin Scherbatsky}, \characterlinksimple{Jane-the-Virgin-Rogelio-De-La-Vega}{Rogelio De La Vega}, \characterlinksimple{Titanic-Rose-DeWitt-Bukater}{Rose DeWitt Bukater}, \characterlinksimple{Mamma-Mia-Rosie-Mulligan}{Rosie Mulligan}, \characterlinksimple{Arrow-Roy-Harper}{Roy Harper}, \characterlinksimple{Fullmetal-Alchemist-Brotherhood-Roy-Mustang}{Roy Mustang}, \characterlinksimple{High-School-Musical-Ryan-Evans}{Ryan Evans}, \characterlinksimple{Sailor-Moon-Sailor-Jupiter}{Sailor Jupiter}, \characterlinksimple{Sailor-Moon-Sailor-Venus}{Sailor Venus}, \characterlinksimple{The-Perks-of-Being-a-Wallflower-Sam-Button}{Sam Button}, \characterlinksimple{Outer-Banks-Sarah-Cameron}{Sarah Cameron}, \characterlinksimple{New-Girl-Schmidt}{Schmidt}, \characterlinksimple{The-Da-Vinci-Code-Sophie-Neveu}{Sophie Neveu}, \characterlinksimple{Stranger-Things-Steve-Harrington}{Steve Harrington}, \characterlinksimple{The-Good-Place-Tahani-Al-Jamil}{Tahani Al-Jamil}, \characterlinksimple{Yellowjackets-Taissa}{Taissa}, \characterlinksimple{Arrow-Thea-Queen}{Thea Queen}, \characterlinksimple{Little-Women-Theodore-Laurence}{Theodore Laurence}, \characterlinksimple{Unbreakable-Kimmy-Schmidt-Titus-Andromedon}{Titus Andromedon}, \characterlinksimple{Parks-and-Recreation-Tom-Haverford}{Tom Haverford}, \characterlinksimple{Divergent-Tris-Prior}{Tris Prior}, \characterlinksimple{WandaVision-Wanda-Maximoff}{Wanda Maximoff}, \characterlinksimple{Zombieland-Wichita}{Wichita}, \characterlinksimple{Buffy-the-Vampire-Slayer-Willow-Rosenberg}{Willow Rosenberg}, \characterlinksimple{New-Girl-Winston-Bishop}{Winston Bishop}, \characterlinksimple{Jane-the-Virgin-Xiomara-Villanueva}{Xiomara Villanueva}

}

\newcommand{\queeralignlist}{
    \characterlinksimple{Sex-Education-Adam-Groff}{Adam Groff}, \characterlinksimple{WandaVision-Agatha-Harkness}{Agatha Harkness}, \characterlinksimple{Lilo-and-Stitch-Agt-Wendell-Pleakley}{Agt. Wendell Pleakley}, \characterlinksimple{Supergirl-Alex-Danvers}{Alex Danvers}, \characterlinksimple{Orange-is-the-New-Black-Alex-Vause}{Alex Vause}, \characterlinksimple{The-Breakfast-Club-Allison-Reynolds}{Allison Reynolds}, \characterlinksimple{Sense8-Amanita-Caplan}{Amanita Caplan}, \characterlinksimple{Chilling-Adventures-of-Sabrina-Ambrose-Spellman}{Ambrose Spellman}, \characterlinksimple{Booksmart-Amy-Antsler}{Amy Antsler}, \characterlinksimple{My-Little-Pony-Friendship-Is-Magic-Applejack}{Applejack}, \characterlinksimple{Attack-on-Titan-Armin-Arlert}{Armin Arlert}, \characterlinksimple{Game-of-Thrones-Asha-Greyjoy}{Asha Greyjoy}, \characterlinksimple{Yellowjackets-Ben-Scott}{Ben Scott}, \characterlinksimple{Mulholland-Drive-Betty-Elms}{Betty Elms}, \characterlinksimple{Glee-Blaine-Anderson}{Blaine Anderson}, \characterlinksimple{Glee-Brittany-Pierce}{Brittany Pierce}, \characterlinksimple{The-OA-Buck-Vu}{Buck Vu}, \characterlinksimple{South-Park-Butters}{Butters}, \characterlinksimple{Arcane-Caitlyn-Kiramman}{Caitlyn Kiramman}, \characterlinksimple{Modern-Family-Cameron-Tucker}{Cameron Tucker}, \characterlinksimple{The-Land-Before-Time-Cera}{Cera}, \characterlinksimple{Riverdale-Cheryl-Blossom}{Cheryl Blossom}, \characterlinksimple{The-100-Clarke-Griffin}{Clarke Griffin}, \characterlinksimple{How-To-Get-Away-With-Murder-Connor-Walsh}{Connor Walsh}, \characterlinksimple{Community-Craig-Pelton}{Craig Pelton}, \characterlinksimple{Orange-is-the-New-Black-Crazy-Eyes}{Crazy Eyes}, \characterlinksimple{Arrow-Curtis-Holt}{Curtis Holt}, \characterlinksimple{Scandal-Cyrus-Beene}{Cyrus Beene}, \characterlinksimple{Mean-Girls-Damian-Leigh}{Damian Leigh}, \characterlinksimple{Crazy-Ex-Girlfriend-Darryl-Whitefeather}{Darryl Whitefeather}, \characterlinksimple{Schitts-Creek-David-Rose}{David Rose}, \characterlinksimple{White-Collar-Diana-Barrigan}{Diana Barrigan}, \characterlinksimple{Mr-Robot-Dominique-DiPierro}{Dominique DiPierro}, \characterlinksimple{Green-Book-Don-Shirley}{Don Shirley}, \characterlinksimple{Finding-Nemo-Dory}{Dory}, \characterlinksimple{Space-Force-Dr-Adrian-Mallory}{Dr. Adrian Mallory}, \characterlinksimple{Hannibal-Dr-Hannibal-Lecter}{Dr. Hannibal Lecter}, \characterlinksimple{Mindhunter-Dr-Wendy-Carr}{Dr. Wendy Carr}, \characterlinksimple{Cowboy-Bebop-Ed}{Ed}, \characterlinksimple{Pretty-Little-Liars-Emily-Fields}{Emily Fields}, \characterlinksimple{Brokeback-Mountain-Ennis-Del-Mar}{Ennis Del Mar}, \characterlinksimple{Sex-Education-Eric-Effiong}{Eric Effiong}, \characterlinksimple{Killing-Eve-Eve-Polastri}{Eve Polastri}, \characterlinksimple{Riverdale-Fangs-Fogarty}{Fangs Fogarty}, \characterlinksimple{Crazy-Rich-Asians-Goh-Peik-Lin}{Goh Peik Lin}, \characterlinksimple{Crazy-Ex-Girlfriend-Heather-Davis}{Heather Davis}, \characterlinksimple{Portrait-of-a-Lady-on-Fire-Heloise}{Heloise}, \characterlinksimple{Money-Heist-Helsinki}{Helsinki}, \characterlinksimple{Sense8-Hernando-Fuentes}{Hernando Fuentes}, \characterlinksimple{Hamlet-Horatio}{Horatio}, \characterlinksimple{Shameless-Ian-Gallagher}{Ian Gallagher}, \characterlinksimple{Broad-City-Ilana-Wexler}{Ilana Wexler}, \characterlinksimple{The-Princess-Bride-Inigo-Montoya}{Inigo Montoya}, \characterlinksimple{Brokeback-Mountain-Jack-Twist}{Jack Twist}, \characterlinksimple{Mean-Girls-Janis-Ian}{Janis Ian}, \characterlinksimple{Jennifers-Body-Jennifer-Check}{Jennifer Check}, \characterlinksimple{Wynonna-Earp-Jeremy-Chetri}{Jeremy Chetri}, \characterlinksimple{Shadow-and-Bone-Jesper-Fahey}{Jesper Fahey}, \characterlinksimple{Euphoria-Jules-Vaughn}{Jules Vaughn}, \characterlinksimple{Emily-in-Paris-Julien}{Julien}, \characterlinksimple{The-Good-Wife-Kalinda-Sharma}{Kalinda Sharma}, \characterlinksimple{Riverdale-Kevin-Keller}{Kevin Keller}, \characterlinksimple{The-Wire-Kima-Greggs}{Kima Greggs}, \characterlinksimple{Hamilton-King-George-III}{King George III}, \characterlinksimple{The-Umbrella-Academy-Klaus-Hargreeves}{Klaus Hargreeves}, \characterlinksimple{Glee-Kurt-Hummel}{Kurt Hummel}, \characterlinksimple{Death-Note-L}{L}, \characterlinksimple{Sex-Education-Lily-Iglehart}{Lily Iglehart}, \characterlinksimple{Sense8-Lito-Rodriguez}{Lito Rodriguez}, \characterlinksimple{Atomic-Blonde-Lorraine-Broughton}{Lorraine Broughton}, \characterlinksimple{Harry-Potter-Luna-Lovegood}{Luna Lovegood}, \characterlinksimple{Its-Always-Sunny-in-Philadelphia-Mac}{Mac}, \characterlinksimple{Portrait-of-a-Lady-on-Fire-Marianne}{Marianne}, \characterlinksimple{Superstore-Mateo-Liwanag}{Mateo Liwanag}, \characterlinksimple{Lucifer-Mazikeen}{Mazikeen}, \characterlinksimple{Gilmore-Girls-Michel-Gerard}{Michel Gerard}, \characterlinksimple{Modern-Family-Mitchell-Pritchett}{Mitchell Pritchett}, \characterlinksimple{Wheel-of-Time-Moiraine-Damodred}{Moiraine Damodred}, \characterlinksimple{The-100-Nathan-Miller}{Nathan Miller}, \characterlinksimple{Orange-is-the-New-Black-Nicky-Nichols}{Nicky Nichols}, \characterlinksimple{Wynonna-Earp-Nicole-Haught}{Nicole Haught}, \characterlinksimple{Sense8-Nomi-Marks}{Nomi Marks}, \characterlinksimple{Game-of-Thrones-Oberyn-Martell}{Oberyn Martell}, \characterlinksimple{Sex-Education-Ola-Nyman}{Ola Nyman}, \characterlinksimple{Frozen-Olaf}{Olaf}, \characterlinksimple{How-To-Get-Away-With-Murder-Oliver-Hampton}{Oliver Hampton}, \characterlinksimple{The-Wire-Omar-Little}{Omar Little}, \characterlinksimple{Gotham-Oswald-Cobblepot}{Oswald Cobblepot}, \characterlinksimple{SpongeBob-SquarePants-Patrick-Star}{Patrick Star}, \characterlinksimple{The-Perks-of-Being-a-Wallflower-Patrick-Stewart}{Patrick Stewart}, \characterlinksimple{The-Boondock-Saints-Paul-Smecker}{Paul Smecker}, \characterlinksimple{Dark-Peter-Doppler}{Peter Doppler}, \characterlinksimple{The-Land-Before-Time-Petrie}{Petrie}, \characterlinksimple{My-Little-Pony-Friendship-Is-Magic-Pinkie-Pie}{Pinkie Pie}, \characterlinksimple{Orange-is-the-New-Black-Piper-Chapman}{Piper Chapman}, \characterlinksimple{Arcane-Powder}{Powder}, \characterlinksimple{The-Boys-Queen-Maeve}{Queen Maeve}, \characterlinksimple{Monsters-Inc-Randall-Boggs}{Randall Boggs}, \characterlinksimple{Brooklyn-Nine-Nine-Raymond-Holt}{Raymond Holt}, \characterlinksimple{Mulholland-Drive-Rita}{Rita}, \characterlinksimple{Stranger-Things-Robin-Buckley}{Robin Buckley}, \characterlinksimple{Brooklyn-Nine-Nine-Rosa-Diaz}{Rosa Diaz}, \characterlinksimple{Ozark-Roy-Petty}{Roy Petty}, \characterlinksimple{Euphoria-Rue-Bennett}{Rue Bennett}, \characterlinksimple{High-School-Musical-Ryan-Evans}{Ryan Evans}, \characterlinksimple{Death-Note-Ryuk}{Ryuk}, \characterlinksimple{Mad-Men-Salvatore-Romano}{Salvatore Romano}, \characterlinksimple{Glee-Santana-Lopez}{Santana Lopez}, \characterlinksimple{Neon-Genesis-Evangelion-Shinji-Ikari}{Shinji Ikari}, \characterlinksimple{SpongeBob-SquarePants-SpongeBob-SquarePants}{SpongeBob SquarePants}, \characterlinksimple{The-Marvelous-Mrs-Maisel-Susie-Myerson}{Susie Myerson}, \characterlinksimple{Yellowjackets-Taissa}{Taissa}, \characterlinksimple{The-Wizard-of-Oz-The-Cowardly-Lion}{The Cowardly Lion}, \characterlinksimple{Downton-Abbey-Thomas-Barrow}{Thomas Barrow}, \characterlinksimple{Unbreakable-Kimmy-Schmidt-Titus-Andromedon}{Titus Andromedon}, \characterlinksimple{Arrested-Development-Tobias-Funke}{Tobias Funke}, \characterlinksimple{Riverdale-Toni-Topaz}{Toni Topaz}, \characterlinksimple{The-Little-Mermaid-Ursula}{Ursula}, \characterlinksimple{The-Umbrella-Academy-Vanya-Hargreeves}{Vanya Hargreeves}, \characterlinksimple{Game-of-Thrones-Varys}{Varys}, \characterlinksimple{Arcane-Vi}{Vi}, \characterlinksimple{Arcane-Viktor}{Viktor}, \characterlinksimple{Killing-Eve-Villanelle}{Villanelle}, \characterlinksimple{Wynonna-Earp-Waverly-Earp}{Waverly Earp}, \characterlinksimple{The-Simpsons-Waylon-Smithers}{Waylon Smithers}, \characterlinksimple{Stranger-Things-Will-Byers}{Will Byers}, \characterlinksimple{Buffy-the-Vampire-Slayer-Willow-Rosenberg}{Willow Rosenberg}, \characterlinksimple{Willy-Wonka-and-the-Chocolate-Factory-Willy-Wonka}{Willy Wonka}

}

\newcommand{\fandomqueerlist}{
    \characterlinksimple{Sex-Education-Adam-Groff}{Adam Groff}, \characterlinksimple{WandaVision-Agatha-Harkness}{Agatha Harkness}, \characterlinksimple{Harry-Potter-Albus-Dumbledore}{Albus Dumbledore}, \characterlinksimple{Supergirl-Alex-Danvers}{Alex Danvers}, \characterlinksimple{Orange-is-the-New-Black-Alex-Vause}{Alex Vause}, \characterlinksimple{Sense8-Amanita-Caplan}{Amanita Caplan}, \characterlinksimple{Chilling-Adventures-of-Sabrina-Ambrose-Spellman}{Ambrose Spellman}, \characterlinksimple{Booksmart-Amy-Antsler}{Amy Antsler}, \characterlinksimple{Futurama-Amy-Wong}{Amy Wong}, \characterlinksimple{Bones-Angela-Montenegro}{Angela Montenegro}, \characterlinksimple{Mr-Robot-Angela-Moss}{Angela Moss}, \characterlinksimple{Gotham-Barbara-Kean}{Barbara Kean}, \characterlinksimple{Community-Ben-Chang}{Ben Chang}, \characterlinksimple{Yellowjackets-Ben-Scott}{Ben Scott}, \characterlinksimple{The-Queens-Gambit-Beth-Harmon}{Beth Harmon}, \characterlinksimple{Riverdale-Betty-Cooper}{Betty Cooper}, \characterlinksimple{Mulholland-Drive-Betty-Elms}{Betty Elms}, \characterlinksimple{Glee-Blaine-Anderson}{Blaine Anderson}, \characterlinksimple{Bobs-Burgers-Bob-Belcher}{Bob Belcher}, \characterlinksimple{Glee-Brittany-Pierce}{Brittany Pierce}, \characterlinksimple{Buffy-the-Vampire-Slayer-Buffy-Summers}{Buffy Summers}, \characterlinksimple{Modern-Family-Cameron-Tucker}{Cameron Tucker}, \characterlinksimple{Psych-Carlton-Lassiter}{Carlton Lassiter}, \characterlinksimple{Supernatural-Castiel}{Castiel}, \characterlinksimple{New-Girl-Cece-Parekh}{Cece Parekh}, \characterlinksimple{Riverdale-Cheryl-Blossom}{Cheryl Blossom}, \characterlinksimple{Lucifer-Chloe-Decker}{Chloe Decker}, \characterlinksimple{The-Witcher-Ciri}{Ciri}, \characterlinksimple{The-100-Clarke-Griffin}{Clarke Griffin}, \characterlinksimple{Arrow-Curtis-Holt}{Curtis Holt}, \characterlinksimple{Mean-Girls-Damian-Leigh}{Damian Leigh}, \characterlinksimple{Mr-Robot-Darlene}{Darlene}, \characterlinksimple{Crazy-Ex-Girlfriend-Darryl-Whitefeather}{Darryl Whitefeather}, \characterlinksimple{Schitts-Creek-David-Rose}{David Rose}, \characterlinksimple{Its-Always-Sunny-in-Philadelphia-Dennis-Reynolds}{Dennis Reynolds}, \characterlinksimple{Mr-Robot-Dominique-DiPierro}{Dominique DiPierro}, \characterlinksimple{Hannibal-Dr-Alana-Bloom}{Dr. Alana Bloom}, \characterlinksimple{Hannibal-Dr-Hannibal-Lecter}{Dr. Hannibal Lecter}, \characterlinksimple{Mindhunter-Dr-Wendy-Carr}{Dr. Wendy Carr}, \characterlinksimple{The-Good-Place-Eleanor-Shellstrop}{Eleanor Shellstrop}, \characterlinksimple{Star-Trek-Deep-Space-Nine-Elim-Garak}{Elim Garak}, \characterlinksimple{Brokeback-Mountain-Ennis-Del-Mar}{Ennis Del Mar}, \characterlinksimple{Sex-Education-Eric-Effiong}{Eric Effiong}, \characterlinksimple{Killing-Eve-Eve-Polastri}{Eve Polastri}, \characterlinksimple{Two-and-a-Half-Men-Evelyn-Harper}{Evelyn Harper}, \characterlinksimple{Riverdale-Fangs-Fogarty}{Fangs Fogarty}, \characterlinksimple{Fleabag-Fleabag}{Fleabag}, \characterlinksimple{The-Boys-Frenchie}{Frenchie}, \characterlinksimple{Breaking-Bad-Gus-Fring}{Gus Fring}, \characterlinksimple{Attack-on-Titan-Hange-Zoe}{Hange Zoe}, \characterlinksimple{Sense8-Hernando-Fuentes}{Hernando Fuentes}, \characterlinksimple{Brokeback-Mountain-Jack-Twist}{Jack Twist}, \characterlinksimple{Star-Trek-Deep-Space-Nine-Jadzia-Dax}{Jadzia Dax}, \characterlinksimple{Jennifers-Body-Jennifer-Check}{Jennifer Check}, \characterlinksimple{Wynonna-Earp-Jeremy-Chetri}{Jeremy Chetri}, \characterlinksimple{Shadow-and-Bone-Jesper-Fahey}{Jesper Fahey}, \characterlinksimple{Euphoria-Jules-Vaughn}{Jules Vaughn}, \characterlinksimple{Emily-in-Paris-Julien}{Julien}, \characterlinksimple{The-Good-Wife-Kalinda-Sharma}{Kalinda Sharma}, \characterlinksimple{Ted-Lasso-Keeley-Jones}{Keeley Jones}, \characterlinksimple{Riverdale-Kevin-Keller}{Kevin Keller}, \characterlinksimple{Star-Trek-Deep-Space-Nine-Kira-Nerys}{Kira Nerys}, \characterlinksimple{The-Umbrella-Academy-Klaus-Hargreeves}{Klaus Hargreeves}, \characterlinksimple{Glee-Kurt-Hummel}{Kurt Hummel}, \characterlinksimple{Vikings-Lagertha}{Lagertha}, \characterlinksimple{Sex-Education-Lily-Iglehart}{Lily Iglehart}, \characterlinksimple{The-Simpsons-Lisa-Simpson}{Lisa Simpson}, \characterlinksimple{Sense8-Lito-Rodriguez}{Lito Rodriguez}, \characterlinksimple{Marvel-Cinematic-Universe-Loki}{Loki}, \characterlinksimple{Orange-is-the-New-Black-Lorna-Morello}{Lorna Morello}, \characterlinksimple{Atomic-Blonde-Lorraine-Broughton}{Lorraine Broughton}, \characterlinksimple{Lucifer-Lucifer-Morningstar}{Lucifer Morningstar}, \characterlinksimple{Its-Always-Sunny-in-Philadelphia-Mac}{Mac}, \characterlinksimple{Archer-Malory-Archer}{Malory Archer}, \characterlinksimple{Superstore-Mateo-Liwanag}{Mateo Liwanag}, \characterlinksimple{Lucifer-Mazikeen}{Mazikeen}, \characterlinksimple{Gilmore-Girls-Michel-Gerard}{Michel Gerard}, \characterlinksimple{Modern-Family-Mitchell-Pritchett}{Mitchell Pritchett}, \characterlinksimple{Wheel-of-Time-Moiraine-Damodred}{Moiraine Damodred}, \characterlinksimple{Euphoria-Nate-Jacobs}{Nate Jacobs}, \characterlinksimple{The-100-Nathan-Miller}{Nathan Miller}, \characterlinksimple{Orange-is-the-New-Black-Nicky-Nichols}{Nicky Nichols}, \characterlinksimple{Wynonna-Earp-Nicole-Haught}{Nicole Haught}, \characterlinksimple{Black-Swan-Nina-Sayers}{Nina Sayers}, \characterlinksimple{Sense8-Nomi-Marks}{Nomi Marks}, \characterlinksimple{The-100-Octavia-Blake}{Octavia Blake}, \characterlinksimple{Sex-Education-Ola-Nyman}{Ola Nyman}, \characterlinksimple{The-Wire-Omar-Little}{Omar Little}, \characterlinksimple{Gotham-Oswald-Cobblepot}{Oswald Cobblepot}, \characterlinksimple{Archer-Pamela-Poovey}{Pamela Poovey}, \characterlinksimple{SpongeBob-SquarePants-Patrick-Star}{Patrick Star}, \characterlinksimple{The-Perks-of-Being-a-Wallflower-Patrick-Stewart}{Patrick Stewart}, \characterlinksimple{Jane-the-Virgin-Petra-Solano}{Petra Solano}, \characterlinksimple{Orange-is-the-New-Black-Piper-Chapman}{Piper Chapman}, \characterlinksimple{The-Boys-Queen-Maeve}{Queen Maeve}, \characterlinksimple{Brooklyn-Nine-Nine-Raymond-Holt}{Raymond Holt}, \characterlinksimple{Rick-and-Morty-Rick-Sanchez}{Rick Sanchez}, \characterlinksimple{Mulholland-Drive-Rita}{Rita}, \characterlinksimple{The-Office-Robert-California}{Robert California}, \characterlinksimple{Stranger-Things-Robin-Buckley}{Robin Buckley}, \characterlinksimple{Brooklyn-Nine-Nine-Rosa-Diaz}{Rosa Diaz}, \characterlinksimple{Euphoria-Rue-Bennett}{Rue Bennett}, \characterlinksimple{High-School-Musical-Ryan-Evans}{Ryan Evans}, \characterlinksimple{Glee-Santana-Lopez}{Santana Lopez}, \characterlinksimple{Star-Trek-Voyager-Seven-of-Nine}{Seven of Nine}, \characterlinksimple{Yellowjackets-Shauna}{Shauna}, \characterlinksimple{Neon-Genesis-Evangelion-Shinji-Ikari}{Shinji Ikari}, \characterlinksimple{SpongeBob-SquarePants-SpongeBob-SquarePants}{SpongeBob SquarePants}, \characterlinksimple{Archer-Sterling-Archer}{Sterling Archer}, \characterlinksimple{Rick-and-Morty-Summer-Smith}{Summer Smith}, \characterlinksimple{Yellowjackets-Taissa}{Taissa}, \characterlinksimple{Downton-Abbey-Thomas-Barrow}{Thomas Barrow}, \characterlinksimple{Unbreakable-Kimmy-Schmidt-Titus-Andromedon}{Titus Andromedon}, \characterlinksimple{Riverdale-Toni-Topaz}{Toni Topaz}, \characterlinksimple{Mr-Robot-Tyrell-Wellick}{Tyrell Wellick}, \characterlinksimple{Crazy-Ex-Girlfriend-Valencia-Perez}{Valencia Perez}, \characterlinksimple{Riverdale-Veronica-Lodge}{Veronica Lodge}, \characterlinksimple{Killing-Eve-Villanelle}{Villanelle}, \characterlinksimple{Wynonna-Earp-Waverly-Earp}{Waverly Earp}, \characterlinksimple{The-Simpsons-Waylon-Smithers}{Waylon Smithers}, \characterlinksimple{Stranger-Things-Will-Byers}{Will Byers}, \characterlinksimple{Hannibal-Will-Graham}{Will Graham}, \characterlinksimple{Buffy-the-Vampire-Slayer-Willow-Rosenberg}{Willow Rosenberg}, \characterlinksimple{Sense8-Wolfgang-Bogdanow}{Wolfgang Bogdanow}, \characterlinksimple{Chilling-Adventures-of-Sabrina-Zelda-Spellman}{Zelda Spellman}

}

\newcommand{\straightneighlist}{
    \characterlinksimple{Criminal-Minds-Aaron-Hotchner}{Aaron Hotchner}, \characterlinksimple{Mean-Girls-Aaron-Samuels}{Aaron Samuels}, \characterlinksimple{The-West-Wing-Abbey-Bartlet}{Abbey Bartlet}, \characterlinksimple{Greys-Anatomy-Alex-Karev}{Alex Karev}, \characterlinksimple{Riverdale-Archie-Andrews}{Archie Andrews}, \characterlinksimple{To-Kill-a-Mockingbird-Atticus-Finch}{Atticus Finch}, \characterlinksimple{The-Wire-Beadie-Russell}{Beadie Russell}, \characterlinksimple{The-100-Bellamy-Blake}{Bellamy Blake}, \characterlinksimple{Star-Trek-Deep-Space-Nine-Benjamin-Sisko}{Benjamin Sisko}, \characterlinksimple{This-Is-Us-Beth-Pearson}{Beth Pearson}, \characterlinksimple{Riverdale-Betty-Cooper}{Betty Cooper}, \characterlinksimple{Mindhunter-Bill-Tench}{Bill Tench}, \characterlinksimple{Lord-of-the-Rings-Boromir}{Boromir}, \characterlinksimple{Marvel-Cinematic-Universe-Captain-America}{Captain America}, \characterlinksimple{CSI-Crime-Scene-Investigation-Captain-Jim-Brass}{Captain Jim Brass}, \characterlinksimple{Shameless-Carl-Gallagher}{Carl Gallagher}, \characterlinksimple{Desperate-Housewives-Carlos-Solis}{Carlos Solis}, \characterlinksimple{Psych-Carlton-Lassiter}{Carlton Lassiter}, \characterlinksimple{Lucifer-Chloe-Decker}{Chloe Decker}, \characterlinksimple{American-Sniper-Chris-Kyle}{Chris Kyle}, \characterlinksimple{The-100-Clarke-Griffin}{Clarke Griffin}, \characterlinksimple{The-Boondock-Saints-Connor-MacManus}{Connor MacManus}, \characterlinksimple{The-Cabin-in-the-Woods-Curt-Vaughan}{Curt Vaughan}, \characterlinksimple{The-Walking-Dead-Dale-Horvath}{Dale Horvath}, \characterlinksimple{Forrest-Gump-Dan-Taylor}{Dan Taylor}, \characterlinksimple{The-Walking-Dead-Daryl-Dixon}{Daryl Dixon}, \characterlinksimple{Supernatural-Dean-Winchester}{Dean Winchester}, \characterlinksimple{Fast-and-Furious-Dominic-Toretto}{Dominic Toretto}, \characterlinksimple{Law-and-Order-SVU-Donald-Cragen}{Donald Cragen}, \characterlinksimple{A-Nightmare-on-Elm-Street-Donald-Thompson}{Donald Thompson}, \characterlinksimple{Jurassic-Park-Dr-Alan-Grant}{Dr. Alan Grant}, \characterlinksimple{Sherlock-Dr-John-Watson}{Dr. John Watson}, \characterlinksimple{The-Good-Doctor-Dr-Marcus-Andrews}{Dr. Marcus Andrews}, \characterlinksimple{Dune-Duke-Leto-Atreides}{Duke Leto Atreides}, \characterlinksimple{White-Collar-Elizabeth-Burke}{Elizabeth Burke}, \characterlinksimple{The-Blacklist-Elizabeth-Keen}{Elizabeth Keen}, \characterlinksimple{Law-and-Order-SVU-Elliot-Stabler}{Elliot Stabler}, \characterlinksimple{After-Life-Emma}{Emma}, \characterlinksimple{Twilight-Emmett-Cullen}{Emmett Cullen}, \characterlinksimple{Brokeback-Mountain-Ennis-Del-Mar}{Ennis Del Mar}, \characterlinksimple{Friday-Night-Lights-Eric-Taylor}{Eric Taylor}, \characterlinksimple{Riverdale-FP-Jones}{FP Jones}, \characterlinksimple{Glee-Finn-Hudson}{Finn Hudson}, \characterlinksimple{Blue-Bloods-Frank-Reagan}{Frank Reagan}, \characterlinksimple{Hamilton-George-Washington}{George Washington}, \characterlinksimple{Breaking-Bad-Hank-Schrader}{Hank Schrader}, \characterlinksimple{The-Blacklist-Harold-Cooper}{Harold Cooper}, \characterlinksimple{Dexter-Harry-Morgan}{Harry Morgan}, \characterlinksimple{Gotham-Harvey-Bullock}{Harvey Bullock}, \characterlinksimple{Chilling-Adventures-of-Sabrina-Harvey-Kinkle}{Harvey Kinkle}, \characterlinksimple{Psych-Henry-Spencer}{Henry Spencer}, \characterlinksimple{A-Star-Is-Born-Jack-Maine}{Jack Maine}, \characterlinksimple{LOST-Jack-Shephard}{Jack Shephard}, \characterlinksimple{Sons-of-Anarchy-Jackson-Jax-Teller}{Jackson 'Jax' Teller}, \characterlinksimple{Gotham-James-Gordon}{James Gordon}, \characterlinksimple{Modern-Family-Jay-Pritchett}{Jay Pritchett}, \characterlinksimple{Criminal-Minds-Jennifer-Jareau}{Jennifer Jareau}, \characterlinksimple{Cowboy-Bebop-Jet-Black}{Jet Black}, \characterlinksimple{Stranger-Things-Jim-Hopper}{Jim Hopper}, \characterlinksimple{The-Flash-Joe-West}{Joe West}, \characterlinksimple{Dexter-Joey-Quinn}{Joey Quinn}, \characterlinksimple{The-Breakfast-Club-John-Bender}{John Bender}, \characterlinksimple{Arrow-John-Diggle}{John Diggle}, \characterlinksimple{Yellowstone-John-Dutton}{John Dutton}, \characterlinksimple{Dirty-Dancing-Johnny-Castle}{Johnny Castle}, \characterlinksimple{Cobra-Kai-Johnny-Lawrence}{Johnny Lawrence}, \characterlinksimple{The-West-Wing-Josiah-Bartlet}{Josiah Bartlet}, \characterlinksimple{Psych-Juliet-OHara}{Juliet O'Hara}, \characterlinksimple{LOST-Kate-Austen}{Kate Austen}, \characterlinksimple{Castle-Kate-Beckett}{Kate Beckett}, \characterlinksimple{Battlestar-Galactica-Laura-Roslin}{Laura Roslin}, \characterlinksimple{How-To-Get-Away-With-Murder-Laurel-Castillo}{Laurel Castillo}, \characterlinksimple{Arrow-Laurel-Lance}{Laurel Lance}, \characterlinksimple{Battlestar-Galactica-Lee-Apollo-Adama}{Lee 'Apollo' Adama}, \characterlinksimple{NCIS-Leroy-Jethro-Gibbs}{Leroy Jethro Gibbs}, \characterlinksimple{Fast-and-Furious-Letty-Ortiz}{Letty Ortiz}, \characterlinksimple{West-Side-Story-Lieutenant-Schrank}{Lieutenant Schrank}, \characterlinksimple{Prison-Break-Lincoln-Burrows}{Lincoln Burrows}, \characterlinksimple{Shameless-Lip-Gallagher}{Lip Gallagher}, \characterlinksimple{Gilmore-Girls-Luke-Danes}{Luke Danes}, \characterlinksimple{Raiders-of-the-Lost-Ark-Marion-Ravenwood}{Marion Ravenwood}, \characterlinksimple{Little-Women-Marmee-March}{Marmee March}, \characterlinksimple{Jaws-Martin-Brody}{Martin Brody}, \characterlinksimple{True-Detective-Marty-Hart}{Marty Hart}, \characterlinksimple{Chicago-Fire-Matthew-Casey}{Matthew Casey}, \characterlinksimple{Gladiator-Maximus}{Maximus}, \characterlinksimple{Agents-of-SHIELD-Melinda-May}{Melinda May}, \characterlinksimple{Better-Call-Saul-Mike-Ehrmantraut}{Mike Ehrmantraut}, \characterlinksimple{Breaking-Bad-Mike-Ehrmantraut}{Mike Ehrmantraut}, \characterlinksimple{Baywatch-Mitch-Buchannon}{Mitch Buchannon}, \characterlinksimple{Sex-and-the-City-Mr-Big}{Mr. Big}, \characterlinksimple{Inception-Mr-Saito}{Mr. Saito}, \characterlinksimple{The-Lion-King-Nala}{Nala}, \characterlinksimple{CSI-Crime-Scene-Investigation-Nick-Stokes}{Nick Stokes}, \characterlinksimple{Twin-Peaks-Norma-Jennings}{Norma Jennings}, \characterlinksimple{Arrow-Oliver-Queen}{Oliver Queen}, \characterlinksimple{Parasite-Park-Dong-ik}{Park Dong-ik}, \characterlinksimple{The-Hangover-Phil}{Phil}, \characterlinksimple{Agents-of-SHIELD-Phil-Coulson}{Phil Coulson}, \characterlinksimple{The-Crown-Philip-Duke-of-Edinburgh}{Philip, Duke of Edinburgh}, \characterlinksimple{Star-Wars-A-New-Hope-Princess-Leia}{Princess Leia}, \characterlinksimple{Arrow-Quentin-Lance}{Quentin Lance}, \characterlinksimple{Vikings-Ragnar-Lothbrok}{Ragnar Lothbrok}, \characterlinksimple{Money-Heist-Raquel-Murillo}{Raquel Murillo}, \characterlinksimple{This-Is-Us-Rebecca-Pearson}{Rebecca Pearson}, \characterlinksimple{That-70s-Show-Red-Forman}{Red Forman}, \characterlinksimple{Orange-is-the-New-Black-Red-Reznikov}{Red Reznikov}, \characterlinksimple{The-Walking-Dead-Rick-Grimes}{Rick Grimes}, \characterlinksimple{Yellowstone-Rip-Wheeler}{Rip Wheeler}, \characterlinksimple{Parks-and-Recreation-Ron-Swanson}{Ron Swanson}, \characterlinksimple{Harry-Potter-Ron-Weasley}{Ron Weasley}, \characterlinksimple{Ted-Lasso-Roy-Kent}{Roy Kent}, \characterlinksimple{Glee-Sam-Evans}{Sam Evans}, \characterlinksimple{Battlestar-Galactica-Samuel-Longshot-Anders}{Samuel 'Longshot' Anders}, \characterlinksimple{Battlestar-Galactica-Saul-Tigh}{Saul Tigh}, \characterlinksimple{LOST-Sawyer-Ford}{Sawyer Ford}, \characterlinksimple{This-Is-Us-Sophie}{Sophie}, \characterlinksimple{Schitts-Creek-Stevie-Budd}{Stevie Budd}, \characterlinksimple{Friday-Night-Lights-Tami-Taylor}{Tami Taylor}, \characterlinksimple{Independence-Day-Thomas-J-Whitmore}{Thomas J. Whitmore}, \characterlinksimple{Dexter-Thomas-Matthews}{Thomas Matthews}, \characterlinksimple{Friday-Night-Lights-Tim-Riggins}{Tim Riggins}, \characterlinksimple{Divergent-Tobias-Four-Eaton}{Tobias 'Four' Eaton}, \characterlinksimple{Desperate-Housewives-Tom-Scavo}{Tom Scavo}, \characterlinksimple{The-Godfather-Vito-Corleone}{Vito Corleone}, \characterlinksimple{Chicago-Fire-Wallace-Boden}{Wallace Boden}, \characterlinksimple{The-Good-Wife-Will-Gardner}{Will Gardner}, \characterlinksimple{Battlestar-Galactica-William-Adama}{William Adama}

}

\newcommand{\straightalignlist}{
    \characterlinksimple{Criminal-Minds-Aaron-Hotchner}{Aaron Hotchner}, \characterlinksimple{Greys-Anatomy-Alex-Karev}{Alex Karev}, \characterlinksimple{Veep-Amy-Brookheimer}{Amy Brookheimer}, \characterlinksimple{The-Office-Angela-Martin}{Angela Martin}, \characterlinksimple{Riverdale-Archie-Andrews}{Archie Andrews}, \characterlinksimple{Entourage-Ari-Gold}{Ari Gold}, \characterlinksimple{Mindhunter-Bill-Tench}{Bill Tench}, \characterlinksimple{The-Wire-Bunk-Moreland}{Bunk Moreland}, \characterlinksimple{Shameless-Carl-Gallagher}{Carl Gallagher}, \characterlinksimple{Game-of-Thrones-Catelyn-Stark}{Catelyn Stark}, \characterlinksimple{Twilight-Charlie-Swan}{Charlie Swan}, \characterlinksimple{Sex-and-the-City-Charlotte-York}{Charlotte York}, \characterlinksimple{American-Sniper-Chris-Kyle}{Chris Kyle}, \characterlinksimple{The-Cabin-in-the-Woods-Curt-Vaughan}{Curt Vaughan}, \characterlinksimple{The-Wire-Cutty-Wise}{Cutty Wise}, \characterlinksimple{Blue-Bloods-Danny-Reagan}{Danny Reagan}, \characterlinksimple{Greys-Anatomy-Derek-Shepherd}{Derek Shepherd}, \characterlinksimple{Fast-and-Furious-Dominic-Toretto}{Dominic Toretto}, \characterlinksimple{Mad-Men-Don-Draper}{Don Draper}, \characterlinksimple{Law-and-Order-SVU-Donald-Cragen}{Donald Cragen}, \characterlinksimple{The-West-Wing-Donna-Moss}{Donna Moss}, \characterlinksimple{Game-of-Thrones-Eddard-Stark}{Eddard Stark}, \characterlinksimple{Crazy-Rich-Asians-Eleanor-Sung-Young}{Eleanor Sung-Young}, \characterlinksimple{The-Vampire-Diaries-Elena-Gilbert}{Elena Gilbert}, \characterlinksimple{The-Blacklist-Elizabeth-Keen}{Elizabeth Keen}, \characterlinksimple{Law-and-Order-SVU-Elliot-Stabler}{Elliot Stabler}, \characterlinksimple{Emily-in-Paris-Emily-Cooper}{Emily Cooper}, \characterlinksimple{Gilmore-Girls-Emily-Gilmore}{Emily Gilmore}, \characterlinksimple{After-Life-Emma}{Emma}, \characterlinksimple{Friday-Night-Lights-Eric-Taylor}{Eric Taylor}, \characterlinksimple{The-Wire-Ervin-Burrell}{Ervin Burrell}, \characterlinksimple{Riverdale-FP-Jones}{FP Jones}, \characterlinksimple{Glee-Finn-Hudson}{Finn Hudson}, \characterlinksimple{Avatar-The-Last-Airbender-Firelord-Ozai}{Firelord Ozai}, \characterlinksimple{Blue-Bloods-Frank-Reagan}{Frank Reagan}, \characterlinksimple{The-Wire-Frank-Sobotka}{Frank Sobotka}, \characterlinksimple{The-Hunger-Games-Gale-Hawthorne}{Gale Hawthorne}, \characterlinksimple{Stargate-SG-1-George-S-Hammond}{George S. Hammond}, \characterlinksimple{Breaking-Bad-Hank-Schrader}{Hank Schrader}, \characterlinksimple{Dark-Hannah-Kahnwald}{Hannah Kahnwald}, \characterlinksimple{Dexter-Harry-Morgan}{Harry Morgan}, \characterlinksimple{Seinfeld-Helen-Seinfeld}{Helen Seinfeld}, \characterlinksimple{Psych-Henry-Spencer}{Henry Spencer}, \characterlinksimple{Raiders-of-the-Lost-Ark-Indiana-Jones}{Indiana Jones}, \characterlinksimple{Hannibal-Jack-Crawford}{Jack Crawford}, \characterlinksimple{30-Rock-Jack-Donaghy}{Jack Donaghy}, \characterlinksimple{Stargate-SG-1-Jack-ONeill}{Jack O'Neill}, \characterlinksimple{This-Is-Us-Jack-Pearson}{Jack Pearson}, \characterlinksimple{LOST-Jack-Shephard}{Jack Shephard}, \characterlinksimple{Speed-Jack-Traven}{Jack Traven}, \characterlinksimple{Sons-of-Anarchy-Jackson-Jax-Teller}{Jackson 'Jax' Teller}, \characterlinksimple{Tommorrow-Never-Dies-James-Bond}{James Bond}, \characterlinksimple{Modern-Family-Jay-Pritchett}{Jay Pritchett}, \characterlinksimple{The-Fall-Jim-Burns}{Jim Burns}, \characterlinksimple{The-Office-Jim-Halpert}{Jim Halpert}, \characterlinksimple{8-Mile-Jimmy-B-Rabbit-Smith-Jr}{Jimmy 'B-Rabbit' Smith Jr.}, \characterlinksimple{The-Flash-Joe-West}{Joe West}, \characterlinksimple{Yellowstone-John-Dutton}{John Dutton}, \characterlinksimple{Bloodline-John-Rayburn}{John Rayburn}, \characterlinksimple{The-Man-in-the-High-Castle-John-Smith}{John Smith}, \characterlinksimple{The-West-Wing-Josiah-Bartlet}{Josiah Bartlet}, \characterlinksimple{Friday-Night-Lights-Julie-Taylor}{Julie Taylor}, \characterlinksimple{Yellowstone-Kayce-Dutton}{Kayce Dutton}, \characterlinksimple{Chicago-Fire-Kelly-Severide}{Kelly Severide}, \characterlinksimple{This-Is-Us-Kevin-Pearson}{Kevin Pearson}, \characterlinksimple{Parasite-Kim-Ki-taek}{Kim Ki-taek}, \characterlinksimple{The-West-Wing-Leo-McGarry}{Leo McGarry}, \characterlinksimple{NCIS-Leroy-Jethro-Gibbs}{Leroy Jethro Gibbs}, \characterlinksimple{Crouching-Tiger-Hidden-Dragon-Li-Mu-Bai}{Li Mu Bai}, \characterlinksimple{Star-Trek-Deep-Space-Nine-Lieutenant-Commander-Worf}{Lieutenant Commander Worf}, \characterlinksimple{West-Side-Story-Lieutenant-Schrank}{Lieutenant Schrank}, \characterlinksimple{Shameless-Lip-Gallagher}{Lip Gallagher}, \characterlinksimple{The-Walking-Dead-Lori-Grimes}{Lori Grimes}, \characterlinksimple{ER-Luka-Kovac}{Luka Kovac}, \characterlinksimple{Gilmore-Girls-Luke-Danes}{Luke Danes}, \characterlinksimple{The-Umbrella-Academy-Luther-Hargreeves}{Luther Hargreeves}, \characterlinksimple{Parks-and-Recreation-Mark-Brendanawicz}{Mark Brendanawicz}, \characterlinksimple{Girls-Marnie-Michaels}{Marnie Michaels}, \characterlinksimple{Jaws-Martin-Brody}{Martin Brody}, \characterlinksimple{Ozark-Marty-Byrde}{Marty Byrde}, \characterlinksimple{True-Detective-Marty-Hart}{Marty Hart}, \characterlinksimple{The-Big-Bang-Theory-Mary-Cooper}{Mary Cooper}, \characterlinksimple{Friday-Night-Lights-Matt-Saracen}{Matt Saracen}, \characterlinksimple{Chicago-Fire-Matthew-Casey}{Matthew Casey}, \characterlinksimple{The-Sopranos-Meadow-Soprano}{Meadow Soprano}, \characterlinksimple{Scandal-Mellie-Grant}{Mellie Grant}, \characterlinksimple{Arrested-Development-Michael-Bluth}{Michael Bluth}, \characterlinksimple{Desperate-Housewives-Mike-Delfino}{Mike Delfino}, \characterlinksimple{Better-Call-Saul-Mike-Ehrmantraut}{Mike Ehrmantraut}, \characterlinksimple{Sex-and-the-City-Mr-Big}{Mr. Big}, \characterlinksimple{The-Lion-King-Mufasa}{Mufasa}, \characterlinksimple{The-Last-of-the-Mohicans-Nathaniel-Hawkeye-Poe}{Nathaniel 'Hawkeye' Poe}, \characterlinksimple{Gone-Girl-Nick-Dunne}{Nick Dunne}, \characterlinksimple{Law-and-Order-SVU-Odafin-Tutuola}{Odafin Tutuola}, \characterlinksimple{The-Office-Pam-Beesly}{Pam Beesly}, \characterlinksimple{Parasite-Park-Dong-ik}{Park Dong-ik}, \characterlinksimple{White-Collar-Peter-Burke}{Peter Burke}, \characterlinksimple{The-Hangover-Phil}{Phil}, \characterlinksimple{Groundhog-Day-Phil-Connors}{Phil Connors}, \characterlinksimple{The-Crown-Queen-Elizabeth-II}{Queen Elizabeth II}, \characterlinksimple{This-Is-Us-Randall-Pearson}{Randall Pearson}, \characterlinksimple{That-70s-Show-Red-Forman}{Red Forman}, \characterlinksimple{Gilmore-Girls-Richard-Gilmore}{Richard Gilmore}, \characterlinksimple{Greys-Anatomy-Richard-Webber}{Richard Webber}, \characterlinksimple{Casablanca-Rick-Blaine}{Rick Blaine}, \characterlinksimple{The-Walking-Dead-Rick-Grimes}{Rick Grimes}, \characterlinksimple{Yellowstone-Rip-Wheeler}{Rip Wheeler}, \characterlinksimple{Parks-and-Recreation-Ron-Swanson}{Ron Swanson}, \characterlinksimple{The-Marvelous-Mrs-Maisel-Rose-Weissman}{Rose Weissman}, \characterlinksimple{Ted-Lasso-Roy-Kent}{Roy Kent}, \characterlinksimple{LOST-Sawyer-Ford}{Sawyer Ford}, \characterlinksimple{Bones-Seeley-Booth}{Seeley Booth}, \characterlinksimple{The-Walking-Dead-Shane-Walsh}{Shane Walsh}, \characterlinksimple{Community-Shirley-Bennett}{Shirley Bennett}, \characterlinksimple{The-Office-Stanley-Hudson}{Stanley Hudson}, \characterlinksimple{The-Wire-Stringer-Bell}{Stringer Bell}, \characterlinksimple{The-Spectacular-Now-Sutter-Keely}{Sutter Keely}, \characterlinksimple{Friday-Night-Lights-Tami-Taylor}{Tami Taylor}, \characterlinksimple{Independence-Day-Thomas-J-Whitmore}{Thomas J. Whitmore}, \characterlinksimple{Peaky-Blinders-Thomas-Shelby}{Thomas Shelby}, \characterlinksimple{Friday-Night-Lights-Tim-Riggins}{Tim Riggins}, \characterlinksimple{The-Sopranos-Tony-Soprano}{Tony Soprano}, \characterlinksimple{Entourage-Turtle}{Turtle}, \characterlinksimple{Star-Trek-Voyager-Tuvok}{Tuvok}, \characterlinksimple{The-Godfather-Vito-Corleone}{Vito Corleone}, \characterlinksimple{Breaking-Bad-Walter-White}{Walter White}, \characterlinksimple{The-Good-Wife-Will-Gardner}{Will Gardner}, \characterlinksimple{Battlestar-Galactica-William-Adama}{William Adama}

}

\begin{itemize}[label={},itemsep=.25cm]
    \item \textbf{\queeralign}: \queeralignlist
    \item \textbf{\queerneigh}: \queerneighlist
    \item \textbf{\fandomqueer}: \fandomqueerlist
    \item \textbf{\straightneigh}: \straightneighlist
    \item \textbf{\straightalign}: \straightalignlist
\end{itemize}

\clearpage
\section{Traits and archetype membership counts}
\label{sec:straight-queer.appendix.traits_and_arch_membership_props}

\subsection{Trait counts}

Each character has a top trait, and some of these traits surface exclusively within one subset. We detail the number of unique (set) of traits and the number of traits exclusively appearing in each subset. For top traits within groups,
\queeralign\ shows $69$ unique and $7$ exclusive,
\queerneigh\ shows $52$ unique and $11$ exclusive,
\fandomqueer\ shows $66$ unique and $3$ exclusive.
\straightneigh\ shows $61$ unique and $14$ exclusive, and
\straightalign\ shows $72$ unique and $18$ exclusive.

Table~\ref{tab:top_group_traits} describes the most occurring top character traits by group. Trait labels are in the queer-aligned direction. The following list contains the exclusive top traits within each group:
\begin{itemize}[label={}]
    \item \textbf{\queeralign} (7): \traitlinksimple{resistant}{resigned}, \traitlinksimple{harsh}{gentle}, \traitlinksimple{anxious}{calm}, \traitlinksimple{protagonist}{antagonist}, \traitlinksimple{scheduled}{spontaneous}, \traitlinksimple{macho}{metrosexual}, \traitlinksimple{extrovert}{introvert}

    \item \textbf{\queerneigh} (11): \traitlinksimple{feeler}{thinker}, \traitlinksimple{fresh}{stinky}, \traitlinksimple{goal-oriented}{experience-oriented}, \traitlinksimple{stereotypical}{boundary breaking}, \traitlinksimple{English}{German}, \traitlinksimple{pronatalist}{child free}, \traitlinksimple{unannoying}{annoying}, \traitlinksimple{wolf}{bear}, \traitlinksimple{stoic}{expressive}, \traitlinksimple{Italian}{Swedish}, \traitlinksimple{soulful}{soulless}

    \item \textbf{\fandomqueer} (3): \traitlinksimple{cheesy}{chic}, \traitlinksimple{cocky}{timid}, \traitlinksimple{chosen one}{everyman}

    \item \textbf{\straightneigh} (14): \traitlinksimple{lawyerly}{engineerial}, \traitlinksimple{scientific}{artistic}, \traitlinksimple{coarse}{delicate}, \traitlinksimple{manic}{mild}, \traitlinksimple{go-getter}{slugabed}, \traitlinksimple{lion}{zebra}, \traitlinksimple{corporate}{freelance}, \traitlinksimple{spartan}{glamorous}, \traitlinksimple{boy/girl-next-door}{celebrity}, \traitlinksimple{flower child}{goth}, \traitlinksimple{main character}{side character}, \traitlinksimple{physicist}{photographer}, \traitlinksimple{French}{Russian}, \traitlinksimple{rock}{rap}

    \item \textbf{\straightalign} (18): \traitlinksimple{reasonable}{deranged}, \traitlinksimple{vintage}{trendy}, \traitlinksimple{non-gamer}{gamer}, \traitlinksimple{chronically single}{serial dater}, \traitlinksimple{no-nonsense}{dramatic}, \traitlinksimple{generous}{stingy}, \traitlinksimple{dramatic}{comedic}, \traitlinksimple{practical}{imaginative}, \traitlinksimple{earthly}{divine}, \traitlinksimple{preppy}{punk rock}, \traitlinksimple{obsessed}{aloof}, \traitlinksimple{uptight}{easy}, \traitlinksimple{pointed}{random}, \traitlinksimple{friendly}{unfriendly}, \traitlinksimple{unstirring}{quivering}, \traitlinksimple{cursed}{blessed}, \traitlinksimple{honorable}{cunning}, \traitlinksimple{routine}{innovative}

\end{itemize}

\begin{table*}[h]
\centering
\begin{tabular}{@{}c@{\hskip 0.01\textwidth}c@{}}
\begin{minipage}{0.38\textwidth}
\centering
\textbf{\straightalign}\\[4pt]
\begin{tabular}{p{6cm}r}
\hline
\textbf{Trait} & \textbf{$n$} \\
\hline
\traitlinksimple{handshakes}{hugs} & 8 \\
\traitlinksimple{gendered}{androgynous} & 7 \\
\traitlinksimple{hygienic}{gross} & 6 \\
\traitlinksimple{pro}{noob} & 5 \\
\traitlinksimple{militaristic}{hippie} & 5 \\
\traitlinksimple{master}{apprentice} & 4 \\
\traitlinksimple{f***-the-police}{tattle-tale} & 4 \\
\traitlinksimple{straight edge}{junkie} & 4 \\
\traitlinksimple{strong identity}{social chameleon} & 4 \\
\traitlinksimple{driven}{unambitious} & 3 \\
\hline
\end{tabular}
\\[4pt]\small 72 unique $|$ 18 exclusive
\end{minipage}
&
\begin{minipage}{0.38\textwidth}
\centering
\textbf{\queeralign}\\[4pt]
\begin{tabular}{p{6cm}r}
\hline
\textbf{Trait} & \textbf{$n$} \\
\hline
\traitlinksimple{straight}{queer} & 25 \\
\traitlinksimple{f***-the-police}{tattle-tale} & 5 \\
\traitlinksimple{hygienic}{gross} & 4 \\
\traitlinksimple{rebellious}{obedient} & 3 \\
\traitlinksimple{expressive}{monotone} & 3 \\
\traitlinksimple{big-vocabulary}{small-vocabulary} & 3 \\
\traitlinksimple{sassy}{chill} & 3 \\
\traitlinksimple{bold}{shy} & 2 \\
\traitlinksimple{extrovert}{introvert} & 2 \\
\traitlinksimple{anxious}{calm} & 2 \\
\hline
\end{tabular}
\\[4pt]\small 69 unique $|$ 7 exclusive
\end{minipage}
\\
\noalign{\vskip 20pt}
\begin{minipage}{0.38\textwidth}
\centering
\textbf{\fandomqueer}\\[4pt]
\begin{tabular}{p{6cm}r}
\hline
\textbf{Trait} & \textbf{$n$} \\
\hline
\traitlinksimple{straight}{queer} & 18 \\
\traitlinksimple{hygienic}{gross} & 10 \\
\traitlinksimple{bold}{shy} & 5 \\
\traitlinksimple{f***-the-police}{tattle-tale} & 5 \\
\traitlinksimple{stable}{unstable} & 5 \\
\traitlinksimple{rebellious}{obedient} & 4 \\
\traitlinksimple{self-disciplined}{disorganized} & 3 \\
\traitlinksimple{egalitarian}{racist} & 3 \\
\traitlinksimple{cannibal}{vegan} & 3 \\
\traitlinksimple{masculine}{feminine} & 2 \\
\hline
\end{tabular}
\\[4pt]\small 66 unique $|$ 3 exclusive
\end{minipage}
&
\begin{minipage}{0.38\textwidth}
\centering
\textbf{\queerneigh}\\[4pt]
\begin{tabular}{p{6cm}r}
\hline
\textbf{Trait} & \textbf{$n$} \\
\hline
\traitlinksimple{straight}{queer} & 18 \\
\traitlinksimple{egalitarian}{racist} & 7 \\
\traitlinksimple{hygienic}{gross} & 6 \\
\traitlinksimple{rebellious}{obedient} & 5 \\
\traitlinksimple{bold}{shy} & 4 \\
\traitlinksimple{masculine}{feminine} & 4 \\
\traitlinksimple{bookish}{sporty} & 4 \\
\traitlinksimple{stylish}{slovenly} & 3 \\
\traitlinksimple{complicated}{simple} & 3 \\
\traitlinksimple{manicured}{scruffy} & 3 \\
\hline
\end{tabular}
\\[4pt]\small 52 unique $|$ 11 exclusive
\end{minipage}
\\
\noalign{\vskip 20pt}
\begin{minipage}{0.38\textwidth}
\centering
\textbf{\straightneigh}\\[4pt]
\begin{tabular}{p{6cm}r}
\hline
\textbf{Trait} & \textbf{$n$} \\
\hline
\traitlinksimple{hygienic}{gross} & 12 \\
\traitlinksimple{handshakes}{hugs} & 8 \\
\traitlinksimple{militaristic}{hippie} & 7 \\
\traitlinksimple{jock}{nerd} & 5 \\
\traitlinksimple{self-disciplined}{disorganized} & 4 \\
\traitlinksimple{driven}{unambitious} & 4 \\
\traitlinksimple{gendered}{androgynous} & 4 \\
\traitlinksimple{strong identity}{social chameleon} & 4 \\
\traitlinksimple{bold}{shy} & 3 \\
\traitlinksimple{rebellious}{obedient} & 3 \\
\hline
\end{tabular}
\\[4pt]\small 61 unique $|$ 14 exclusive
\end{minipage}
\end{tabular}
\vspace{1cm}
\caption{Top character traits by group. Trait labels are flipped according to their relationships in euclidean distance with \traitlinksimple{straight}{queer} from Tab.~\ref{tab:trait_closeness}. Left-hand trait labels correspond to the straight-aligned direction while right-hand correspond to the queer-aligned direction.}
\label{tab:top_group_traits}
\end{table*}

\subsection{Archetype membership counts}

% top primary archetypes
Characters' assigned archetype labels~\cite{dodds2025ArchetypometricsPragmateiaEmpirical} are collected in Tab.~\ref{tab:archetype_proportions} where we report all top proportions using a cut-off of $17$ traits so proportions for any group only reflect $n>1$ for archetypes. Across subsets, the top 2 archetypes are from strong primary dimensions:
\\
\textbf{\archetype{Hero}} (\straightneigh\ (0.48), \straightalign\ (0.26), \queeralign\ (0.15), \fandomqueer\ (0.14), \ldots, \queeralign\ (0.08 but below top 2))
\\
and
\\
\textbf{\archetype{Adventurer}} (\queeralign\ (0.48), \queeralign\ (0.15), \fandomqueer\ (0.11), with \straightalign\ and \straightneigh\ not having adventurer in their top lists).

The strictly \archetype{Demon} archetype appears less frequently (\fandomqueer\ (0.06), \queeralign\ (0.05), \straightalign\ (0.03)).
% secondary archetypes
No secondary archetypes occur alone (e.g., Brute-Lone Wolf-Hero as a triple variant versus \archetype{Brute} alone).
Lone Wolf variants occur most frequently in top lists of \straightalign\ and \straightneigh\ while Diva variants tend to occur more in \fandomqueer\ and \queerneigh.
Outcast variants seldom appear, and Sophisticate variants do not appear in the top list.
Brute variants only appear in the top lists of \straightalign\ and \straightneigh\ while Geek variants likewise only appear in top lists of \queeralign, \queerneigh, and \fandomqueer\ (consist with the quartile trend).

% in python, use the new function using:
% #1 is what to SHOW (ordered), #2 is what to SILENTLY LINK (by-class)
% \newcommand{\archetypedisplayordered}[2]{\archetypelinkbaseordered{#1}{#2}}
% waiting on PSD response bc 9 characters' have erroneous matches between these 2 orderings, which makes it impossible to hyperlink them here
\begin{table*}
\centering
    \begin{minipage}{0.45\textwidth}
    \centering
    \begin{tabular}{l}
    \hline
    \textbf{\straightalign} \\
    \hline
    Hero (0.26) \\
    Traditionalist-Hero (0.14) \\
    Angel-Hero (0.07) \\
    Brute-Hero (0.05) \\
    Brute-Lone Wolf-Hero (0.04) \\
    Lone Wolf-Hero (0.04) \\
    Brute-Traditionalist-Hero (0.04) \\
    Hero-Demon (0.04) \\
    Demon-Hero-Traditionalist (0.03) \\
    Lone Wolf-Adventurer-Hero (0.03) \\
    Traditionalist-Lone Wolf-Hero (0.03) \\
    Traditionalist-Demon-Hero (0.03) \\
    Demon (0.03) \\
    Lone Wolf-Brute-Hero (0.03) \\
    Demon-Diva-Hero (0.03) \\
    Demon-Hero (0.03) \\
    Brute-Demon-Traditionalist (0.03) \\
    \hline
    \end{tabular}
    \end{minipage}
\hfill
    \begin{minipage}{0.45\textwidth}
    \centering
    \begin{tabular}{l}
    \hline
    \textbf{\queeralign} \\
    \hline
    Adventurer (0.15) \\
    Hero (0.14) \\
    Hero-Adventurer-Diva (0.05) \\
    Demon-Hero (0.05) \\
    Hero-Adventurer-Demon (0.05) \\
    Demon (0.05) \\
    Geek-Hero-Angel (0.04) \\
    Geek-Adventurer-Hero (0.04) \\
    Diva-Adventurer (0.04) \\
    Lone Wolf-Adventurer-Hero (0.04) \\
    Adventurer-Hero-Demon (0.04) \\
    Angel-Adventurer (0.04) \\
    Hero-Adventurer (0.03) \\
    Hero-Angel (0.03) \\
    Hero-Geek-Angel (0.03) \\
    Hero-Demon (0.03) \\
    Geek-Traditionalist-Outcast (0.03) \\
    \hline
    \end{tabular}
    \end{minipage}

\vspace{1cm}

    \begin{minipage}{0.3\textwidth}
    \centering
    \begin{tabular}{l}
    \hline
    \textbf{\fandomqueer} \\
    \hline
    Hero (0.14) \\
    Adventurer (0.11) \\
    Demon-Hero (0.07) \\
    Hero-Adventurer-Demon (0.06) \\
    Demon (0.06) \\
    Hero-Adventurer (0.06) \\
    Diva-Hero-Demon (0.04) \\
    Hero-Demon (0.04) \\
    Hero-Adventurer-Diva (0.04) \\
    Demon-Adventurer (0.04) \\
    Diva-Adventurer (0.04) \\
    Hero-Angel-Adventurer (0.03) \\
    Demon-Traditionalist-Hero (0.03) \\
    Geek-Demon-Hero (0.03) \\
    Adventurer-Demon-Hero (0.03) \\
    Adventurer-Hero-Demon (0.03) \\
    Lone Wolf-Adventurer-Hero (0.03) \\
    \hline
    \end{tabular}
    \end{minipage}
\hfill
    \begin{minipage}{0.3\textwidth}
    \centering
    \begin{tabular}{l}
    \hline
    \textbf{\queerneigh} \\
    \hline
    Adventurer (0.48) \\
    Diva-Adventurer (0.15) \\
    Demon-Hero (0.15) \\
    Adventurer-Hero (0.12) \\
    Demon-Adventurer (0.10) \\
    Angel-Hero (0.10) \\
    Hero-Adventurer (0.08) \\
    Geek-Adventurer-Hero (0.08) \\
    Angel-Hero-Adventurer (0.08) \\
    Angel-Adventurer-Hero (0.08) \\
    Hero (0.08) \\
    Hero-Adventurer-Diva (0.06) \\
    Adventurer-Diva-Hero (0.06) \\
    Outcast-Hero-Adventurer (0.04) \\
    Geek-Hero-Angel (0.04) \\
    Adventurer-Angel-Hero (0.04) \\
    Demon-Lone Wolf-Adventurer (0.04) \\
    \hline
    \end{tabular}
    \end{minipage}
\hfill
    \begin{minipage}{0.3\textwidth}
    \centering
    \begin{tabular}{l}
    \hline
    \textbf{\straightneigh} \\
    \hline
    Hero (0.48) \\
    Angel-Hero (0.13) \\
    Traditionalist-Hero (0.08) \\
    Hero-Angel (0.07) \\
    Lone Wolf-Hero (0.07) \\
    Brute-Hero (0.05) \\
    Lone Wolf-Adventurer-Hero (0.05) \\
    Traditionalist-Lone Wolf-Hero (0.05) \\
    Brute-Traditionalist-Hero (0.05) \\
    Demon-Hero-Traditionalist (0.03) \\
    Lone Wolf-Outcast-Hero (0.03) \\
    Hero-Adventurer-Brute (0.03) \\
    Brute-Lone Wolf-Hero (0.03) \\
    Adventurer-Angel-Hero (0.03) \\
    Outcast-Angel-Brute (0.03) \\
    Demon-Hero (0.03) \\
    Demon-Traditionalist-Hero (0.03) \\
    \hline
    \end{tabular}
    \end{minipage}
\caption{Most populous archetypes per group by proportions. The cut-off applied here is used to populate only archetypes with count $>1$ in any group.}
\label{tab:archetype_proportions}
\end{table*}

\clearpage
\section{Trait closeness}
\label{sec:straight-queer.appendix.traitdistance}

We measure closeness of other traits to \semdiff{straight}{queer} using two calculations: Euclidean distance and cosine similarity. For distance, we measure the distance from \semdiff{straight}{queer} to every other trait and to every other trait's flipped version (its inverse vector scores). We apply the flipped score and ordering of the label if a trait's flipped distance is closer to \semdiff{straight}{queer} than its unflipped distance. For cosine, we flip the reverse the score and flip label for negative values of \semdiff{straight}{queer}. Flipping the labels aligns the interpretation between these two measures with the left- and right-hand labels of the \semdiff{straight}{queer} trait. For example, in \traitlinksimple{gendered}{androgynous}, `gendered' shows lower distance to the straight direction and `androgynous' shows lower distance to the queer direction.

The $30$ closest traits from each metric overlap by $20\%$ and grows to $50\%$ when taking the top $100$ closest traits. For just the top $30$ closest traits, overlapping ones include, in no order, \traitlinksimple{English}{German}, \traitlinksimple{basic}{hipster}, \traitlinksimple{believable}{poorly-written}, \semdiff{gendered}{androgynous}, \traitlinksimple{neurotypical}{autistic}, and \traitlinksimple{patriotic}{unpatriotic}. 

\vspace{1cm}
\begin{table*}[!h]
    \centering
    \begin{tabular}{p{1.2cm}ll}
    \hline
    \textbf{Rank} & \textbf{Distance} & \textbf{Inner Product} \\
    \hline
    1 & \traitlinksimple{gendered}{androgynous} & \traitlinksimple{gendered}{androgynous} \\
2 & \traitlinksimple{neurotypical}{autistic} & \traitlinksimple{persistent}{quitter} \\
3 & \traitlinksimple{basic}{hipster} & \traitlinksimple{diligent}{lazy} \\
4 & \traitlinksimple{patriotic}{unpatriotic} & \traitlinksimple{motivated}{unmotivated} \\
5 & \traitlinksimple{real}{philosophical} & \traitlinksimple{driven}{unambitious} \\
6 & \traitlinksimple{earth}{air} & \traitlinksimple{bold}{shy} \\
7 & \traitlinksimple{concrete}{abstract} & \traitlinksimple{neurotypical}{autistic} \\
8 & \traitlinksimple{prideful}{envious} & \traitlinksimple{opinionated}{neutral} \\
9 & \traitlinksimple{western}{eastern} & \traitlinksimple{hygienic}{gross} \\
10 & \traitlinksimple{literal}{metaphorical} & \traitlinksimple{straight edge}{junkie} \\
11 & \traitlinksimple{empirical}{theoretical} & \traitlinksimple{go-getter}{slugabed} \\
12 & \traitlinksimple{libertarian}{socialist} & \traitlinksimple{English}{German} \\
13 & \traitlinksimple{repetitive}{varied} & \traitlinksimple{active}{slothful} \\
14 & \traitlinksimple{pensive}{serene} & \traitlinksimple{workaholic}{slacker} \\
15 & \traitlinksimple{Roman}{Greek} & \traitlinksimple{resourceful}{helpless} \\
16 & \traitlinksimple{classical}{avant-garde} & \traitlinksimple{important}{irrelevant} \\
17 & \traitlinksimple{straightforward}{cryptic} & \traitlinksimple{pointed}{random} \\
18 & \traitlinksimple{wooden}{plastic} & \traitlinksimple{competent}{incompetent} \\
19 & \traitlinksimple{efficient}{overprepared} & \traitlinksimple{believable}{poorly-written} \\
20 & \traitlinksimple{English}{German} & \traitlinksimple{devoted}{unfaithful} \\
21 & \traitlinksimple{believable}{poorly-written} & \traitlinksimple{focused}{absentminded} \\
22 & \traitlinksimple{permanent}{transient} & \traitlinksimple{rock}{rap} \\
23 & \traitlinksimple{biased}{impartial} & \traitlinksimple{self-disciplined}{disorganized} \\
24 & \traitlinksimple{resolute}{wavering} & \traitlinksimple{proud}{apologetic} \\
25 & \traitlinksimple{chortling}{giggling} & \traitlinksimple{basic}{hipster} \\
26 & \traitlinksimple{Coke}{Pepsi} & \traitlinksimple{loyal}{traitorous} \\
27 & \traitlinksimple{triggered}{trolling} & \traitlinksimple{high IQ}{low IQ} \\
28 & \traitlinksimple{resistant}{resigned} & \traitlinksimple{guarded}{open} \\
29 & \traitlinksimple{stoic}{hypochondriac} & \traitlinksimple{patriotic}{unpatriotic} \\
30 & \traitlinksimple{hard-work}{natural-talent} & \traitlinksimple{demanding}{unchallenging} \\
    \hline
    \end{tabular}
    \caption{Trait closeness in Euclidean distance and by inner product, ordered by their closeness (Euclidean) or similarity (inner product). Left-hand labels are in the \textbf{straight} direction while right-hand labels are in the \textbf{queer} direction.}
    \label{tab:trait_closeness}
    \end{table*}

\end{document}